\documentclass[11pt,a4paper]{article}

\usepackage[a4paper]{geometry} 
\usepackage{titling} 
\usepackage{setspace} 
\usepackage{titlesec}

\usepackage[T1]{fontenc}
\usepackage{lmodern}
\usepackage[dvipsnames]{xcolor}
\usepackage{microtype}
\usepackage{dsfont}
\usepackage{parskip}

\usepackage{amsmath}
\usepackage{tabularx}
\usepackage{amssymb}
\usepackage{physics}
\usepackage{tensor}
\usepackage{braket}
\usepackage{slashed}
\usepackage{accents}

\usepackage{appendix}
\usepackage{comment}
\usepackage[colorlinks=true
  ,urlcolor=blue
  ,anchorcolor=blue
  ,citecolor=blue
  ,filecolor=blue
  ,linkcolor=blue
  ,menucolor=blue
  ,linktocpage=true
  ,pdfproducer=medialab
  ,pdfa=true
]{hyperref}
\usepackage{cleveref}

\usepackage{import}
\usepackage{standalone}

\usepackage{tikz} 
\usepackage{graphicx}
\usetikzlibrary{patterns}

\usepackage{array}   

\usepackage{cite}

\newcolumntype{L}{>{$}c<{$}} 

\usepackage[width=.8\textwidth]{caption} 

\numberwithin{equation}{section}
\numberwithin{table}{section}

\title{Quantum fields in two-dimensional de Sitter space}
\author{Ben, Dio, Tarek, Gizem}

\begin{document}\spacing{1.2}

\vspace*{2.5cm}
\begin{center}
{  \Large \textsc{The Discreet Charm of the Discrete Series in DS$_2$}} \\ \vspace*{1.3cm}

\end{center}

\begin{center}
Dionysios Anninos,$^1$ Tarek Anous,$^2$ Ben Pethybridge,$^1$ and  Gizem \c{S}eng\"or$^3$\\ 
\end{center}
\begin{center}
{
\footnotesize
$^1$ Department of Mathematics, King's College London, Strand, London WC2R 2LS, UK \\
$^2$ School of Mathematical Sciences, Queen Mary University of London, Mile End Road, London, E1 4NS, UK \\
$^3$  Department of Physics, Bo\u{g}azi\c{c}i University, 34342 Bebek, Istanbul, Turkey

}
\end{center}
\begin{center}
{\footnotesize\textsf{\href{mailto:dionysios.anninos@kcl.ac.uk}{dionysios.anninos@kcl.ac.uk}, \href{mailto:t.anous@qmul.ac.uk}{t.anous@qmul.ac.uk}, \href{mailto:ben.pethybridge@kcl.ac.uk}{ben.pethybridge@kcl.ac.uk}, \href{mailto:gizem.sengor@boun.edu.tr}{gizem.sengor@boun.edu.tr}} } 
\end{center}

\vspace*{0.5cm}

\vspace*{1.5cm}
\begin{abstract}
\noindent We study quantum field theories placed on a two-dimensional de Sitter spacetime (dS$_2$) with an eye on the group-theoretic organisation of single and multi-particle states.  We explore the distinguished role of the discrete series unitary irreducible representation (UIR) in the Hilbert space. By employing previous attempts to realise these states in free tachyonic scalar field theories, we propose how the discrete series may contribute to the K\"all\'en-Lehmann decomposition of an interacting scalar two-point function. We also study BF gauge theories with $SL(N,\mathbb{R})$ gauge group in dS$_2$ and establish a relation between the discrete series UIRs and the operator content of these theories. Although present at the level of the operators, states carrying discrete series quantum numbers are projected out of the gauge-invariant Hilbert space. This projection is reminiscent of what happens for quantum field theories coupled to semiclassical de Sitter gravity, where we must project onto the subspace of de Sitter invariant states. We discuss how to impose the diffeomorphism constraints on local field-theory operators coupled to two-dimensional gravity in de Sitter, with particular emphasis on the role of contact terms. Finally, we discuss an SYK-type model with a random two-body interaction that encodes an infinite tower of discrete series operators. We speculate on its potential microscopic connection to the $SL(N,\mathbb{R})$ BF theory in the large-$N$ limit.  


\end{abstract}

\newpage
\tableofcontents

\section{Introduction}
In Minkowski space, Wigner's classification \cite{Wigner:1939cj} links the very tangible concept of a particle---a tiny streaming packet of energy and momentum---to the abstract notion of a unitary irreducible representation (UIR) of the Poincar\'e group. This principle is so powerful at constraining the observable physics \cite{Weinberg:1995mt} that one is lead to apply it in other maximally symmetric spacetimes. In de Sitter space, this has been advocated either directly or indirectly in several papers \cite{Dirac:1935zz,nachtmann1967quantum,borner1969classical,Tagirov:1972vv, Deser:1983mm,Deser:2001us,Deser:2001pe,Deser:2003gw,Higuchi:2010xt, Vasiliev:1990en, Bros:1990cu, Bros:1995js, Bros:2009bz, Bros:2010wa, Bros:2010rku, Epstein:2012zz, Epstein:2014jaa, Epstein:2018wfh,Joung:2006gj,Joung:2007je, Marolf:2008it, Marolf:2010nz, Basile:2016aen,
Marolf:2010zp, Marolf:2011sh,Anninos:2019oka,Anous:2020nxu,Sengor:2019mbz, Sengor:2021zlc, Sun:2021thf,Letsios:2023qzq,Enayati:2022hed,Takook:2023yum,Loparco:2023rug,Higuchi:1986wu,barutfronsdal,mukunda}. Relatedly,   dS/CFT considerations \cite{Witten:2001kn,Maldacena:2002vr,Strominger:2001pn, Anninos:2011ui, Anninos:2017eib,Isono:2020qew} and the bootstrap and $S$-matrix methods \cite{Anninos:2014lwa, Anninos:2014hia,Arkani-Hamed:2015bza, Sleight:2019hfp,Sleight:2020obc,Goodhew:2020hob,Melville:2021lst,Arkani-Hamed:2018kmz, Baumann:2019oyu,Baumann:2020dch,Hogervorst:2021uvp, Penedones:2023uqc,Arkani-Hamed:2017fdk,Benincasa:2022omn,Albayrak:2023hie,Benincasa:2018ssx}  exploit the de Sitter isometries to constrain physical phenomena. In this paper, we explore the logic of Wigner's classification on particle dynamics in the setting of quantum fields propagating on a two-dimensional de Sitter spacetime (dS$_2$).\footnote{Reviews on quantum aspects of de Sitter spacetimes include \cite{Galante:2023uyf,Spradlin:2001pw,Anninos:2012qw,Bousso:2002fq}.} 

We choose to work in dS$_2$, rather than its four-dimensional counterpart, so as to have a simplified playground in which to explore the relevant physics. Indeed, the dS$_2$ group of isometries, $SO(1,2)$ (or its double cover $SL(2,\mathbb{R})$) shares many features in common with the $SO(1,4)$ isometry group of dS$_4$. Of particular interest to us is that both groups contain discrete series UIRs, in addition to the more generic principal series representations, associated with heavy fields in de Sitter, as was first established in the works of Bargmann and Harish-Chandra \cite{Bargmann:1946me,Harish-Chandra.1952,Thomas:1941}. In dS$_4$, the discrete series UIRs appear in the single-particle Hilbert space of higher-spin gauge fields, both massless and partially massless \cite{Deser:2001us,Deser:2001pe,Deser:2003gw,Higuchi:1986py}, including, as a special case, the linearized graviton. However, in dS$_2$, healthy models with discrete series UIRs have been more elusive. For example, they have been shown to arise in the single-particle Hilbert space in free tachyonic scalar models \cite{Bros:2010wa,Epstein:2014jaa}.

One reason to concern ourselves with discrete series UIRs is that they generically appear in the multiparticle (tensor-product) Hilbert space of heavy fields \cite{Dobrev:1977qv, Repka1978TensorR}. This statement  deserves  scrutiny. Since de Sitter is a time-dependent spacetime, energy is not conserved, allowing for interesting phenomenology. For example \cite{nachtmann1968dynamische, Bros:2010rku,Epstein:2012zz}, a  scalar particle of mass $m_0$ in de Sitter space, coupled via a cubic interaction to particles with masses $m_1$ and $m_2$, can decay into these particles even if $m_0<m_1+m_2$---a process forbidden in flat space by energy conservation. A similarly counter-intuitive fact is that the two-particle Hilbert space of a single heavy field in de Sitter carries the discrete series UIR, which, as we just mentioned, is generally constructed as a scalar tachyon in dS$_2$.  We are thus motivated to explore how these discrete series UIRs can arise in different QFT constructions on a rigid dS$_2$ background. We also investigate how the discrete series  contribute in a K\"all\'en-Lehmann \cite{Kallen:1952zz,Lehmann:1954xi} spectral decomposition of the two-point function for general interacting scalar fields in dS$_2$ (see  \cite{Bros:1990cu,Bros:1995js,Bros:2009bz,Hollands:2011we,Epstein:2012zz,Hogervorst:2021uvp,DiPietro:2021sjt,Loparco:2023rug} for related discussions).

We find that we can construct operators furnishing the discrete series representations in BF-type gauge theories on dS$_2$. However, such BF gauge theories are topological and as a result these UIRs eventually are projected out of the gauge-invariant Hilbert space. In a sense, they only exist in the pre-Hilbert space of the theory, meaning they may come alive if we break the gauge invariance or do something  to alter the structure of the spacetime. At least semiclassically, BF gauge theories can be thought of as higher-spin fields in two-dimensions \cite{Alkalaev:2013fsa,Alkalaev:2019xuv}, thus providing a link between the two- and four-dimensional constructions of these UIRs. 

The simplest example is a BF-theory with $SL(2,\mathbb{R})$ gauge group, which can be mapped to de Sitter JT gravity \cite{Isler:1989hq,Chamseddine:1989yz}. The discrete series operators are built from the Weyl mode of the two-dimensional metric. But since one must further impose the diffeomorphism constraints of the theory, the Hilbert space is severely reduced. This is reminiscent of the need to gauge the dS$_4$ isometry group in semiclassical quantum gravity near a dS$_4$ vacuum, which restricts the Hilbert space to just the de Sitter invariant states \cite{Higuchi:1991tk,Higuchi:1991tm,Marolf:2008it,Marolf.2009yal}. This naturally leads us to ask how to construct gravitational observables. In particular, we discuss how contact terms in $n$-point functions of local operators in dS$_2$ \cite{Sengor:2021zlc} propagate into the appropriate gauge invariant observables, suitably constructed.

In anticipation of future work we examine the appearance of the discrete series in a microphysical SYK model endowed with a two-body random interaction and discuss how it may connect to the aforementioned BF theories at large-$N$ \cite{Grumiller:2013swa,Alkalaev:2020kut}. More generally, the parallels between dS$_2$ and dS$_4$ motivate us to explore the space of integrable/solvable quantum field theories in dS$_2$ and cosmological analogues of the Yang-Baxter equation. This is in the spirit of the Schwinger model as a toy model for various phenomena in four-dimensional quantum field theory, and rhymes with  recent efforts on quantum gravity in dS$_2$ \cite{Martinec:2014uva,Bautista:2015wqy,Bautista:2019jau,Anninos:2021ene,Muhlmann:2021clm,Muhlmann:2022duj,Maldacena:2019cbz,Cotler:2019nbi,Anninos:2017hhn,Anninos:2018svg}.

The paper is structured as follows. In section \ref{sec:geom} we discuss the basic geometric aspects of dS$_2$ and introduce various relevant coordinate systems. In section \ref{sec:UIRs} we consider the unitary irreducible representations of $SL(2,\mathbb{R})$, the free Green's function for particles in the principal series UIR and a simple model furnishing the $\Delta=1$ discrete series UIR. In section \ref{sec:spectroscopic decompositions} we elaborate on the K\"all\'en-Lehman decomposition of a de Sitter invariant two-point function of interacting scalar fields. In  \cref{discretesec} we describe how the operator content of BF-type gauge theories furnish the discrete series UIR, at the level of the pre-Hilbert space. In section \ref{sec:contactterms} we discuss how the constraints from gauging the dS$_2$ isometry group in a theory of gravity affect observables and correlation functions, bearing in mind the role of potential contact terms. In the outlook, section \ref{sec:outlook}, we discuss a microphysical SYK model that encodes an infinite tower of discrete series particles and speculate on its relation to a higher-spin theory of dS$_2$. In \cref{App:Repthy} we review the unitary irreducible representations of the $SO(1,2)$ isometry group of dS$_2$. In \cref{app:circle} we construct a discrete series UIR via a Clebsch-Gordan analysis at the level of a quantum mechanics of two degrees of freedom each furnishing a principal series UIR.

\section{Geometry of \texorpdfstring{dS$_2$}{dS2}}\label{sec:geom}
In this section we discuss the geometry of a two-dimensional de Sitter spacetime, referred to as dS$_2$.  We also discuss some group theoretic properties of its isometry group $SO(1,2)  \cong SL(2,\mathbb{R}) /\mathbb{Z}_2$.

\subsection{Geometry}
Let us begin by reviewing the geometry of dS$_2$. The usual starting point is to view this spacetime as a Lorentzian hypersurface embedded in a three-dimensional ambient Minkowski spacetime, satisfying the equation
\begin{equation}\label{eq:hyperboloid}
-\left(X^0\right)^2 + \left(X^1\right)^2 + \left(X^2\right)^2 = \ell^2~, \quad\quad X^A \equiv \left(X^0,X^1,X^2\right) \in \mathbb{R}^{3}~. 
\end{equation}
The metric of dS$_2$ is induced from the flat metric of the ambient spacetime
\begin{equation}\label{eq:ambientmetric}
ds^2 =\eta_{AB}\,dX^AdX^B= -\left( d X^0\right)^2 + \left( d X^1\right)^2 + \left(d X^2\right)^2~,
\end{equation}
by solving \eqref{eq:hyperboloid}.
It follows from the above construction that the isometry group of dS$_2$ is the Lorentz group in three-dimensions, $SO(1,2)$. We will also permit fermionic fields on this spacetime, meaning we should actually consider the double cover of $SO(1,2)$, namely $G \equiv SL(2,\mathbb{R})$. The generators of $G$ are constructed as follows, consider the differential operators:  
\begin{equation}
     {\mathcal{L}}_{AB}= -i\left(X_A\partial_{X^B}- X_B\partial_{X^A}\right) \, ,
     \label{def:dsgens}
\end{equation}
which under commutation satisfy
\begin{equation}\label{eq:lorentzalg}
    \left[ {\mathcal{L}}_{AB}, {\mathcal{L}}_{CD}\right] = -i\left(\eta_{AD}{\mathcal{L}}_{BC}-\eta_{AC} {\mathcal{L}}_{BD}+\eta_{BC}{\mathcal{L}}_{AD}-\eta_{BD} {\mathcal{L}}_{AC}\right) \, .
\end{equation}
In this paper, we will use a hat $\widehat{\ }$ to denote abstract operators. Whenever an operator appears without a hat, we mean the representation as a differential operator.  
If we define 
\begin{equation}
    {L}_0\equiv{\mathcal{L}}_{12}~,\qquad\qquad {L}_{\pm1}\equiv-\left({\mathcal{L}}_{02}\pm i{\mathcal{L}}_{01}\right)~,
    \label{def:sl2r}
\end{equation}
 we observe that the above algebra is isomorphic to that of $SL(2,\mathbb{R})$, 
\begin{equation}\label{eq:sl2ralgebra}
    [L_n, L_m]=(n-m)L_{n+m}~.
\end{equation}
The quadratic Casimir is given by 
\begin{equation}\label{eq:cas}
{\mathcal{C}} \equiv  \frac{1}{2} {\mathcal{L}}_{AB}{\mathcal{L}}^{AB} ={L}_0^2-\frac{1}{2}\left({L}_{-1}{L}_{1}+{L}_{1}{L}_{-1}\right)\, . 
\end{equation}
From here, we will use the shorthand $L_\pm\equiv L_{\pm1}$.
The maximal compact subgroup of $G$ is  $K \equiv SO(2)$, generated by $L_{0}$. It is worth noting that dS$_2$ shares its isometry group with the Poincar\'e disk, also known as Euclidean AdS$_2$. 

Correlation functions on the spacetime will be given as functions of de Sitter invariant quantities. For example, the two-point function will depend on the following de Sitter invariant distance, defined in terms of coordinates on the hyperboloid: 
\begin{equation}\label{eq:invariantdistance}
 u(X,Y)\equiv \frac{ \eta_{AB}\left(X^A-Y^A\right)\left(X^B-Y^B\right)}{2\ell^2}=1-\frac{\eta_{AB}\,X^A\,Y^B}{\ell^2}~,
\end{equation}
where contractions are made using the ambient Minkowski metric \eqref{eq:ambientmetric}. From this formula we conclude that $u(X,X)=0$, as expected, whereas $u(X,-X)=2$ for antipodally separated points. Thus points are spacelike separated for $u>0$, null separated when $u=0$, and timelike separated for $u<0$~. Lastly, for $u>2$, there exist no spacelike geodesic paths connecting the two points.  

We can select from a variety of parameterizations of the induced metric on the hypersurface (\ref{eq:hyperboloid}). Here we list a few.

\paragraph{Global coordinates on dS$_2$.}  The global chart:
\begin{equation}\label{eq:globalcoords}
	X^{A} =\ell  \, (\sinh \tau,\, \cos\vartheta \cosh \tau ,\, \sin \vartheta \cosh \tau ) \, ,
\end{equation}
covers the entire manifold, resulting in the two-dimensional line element
\begin{equation}\label{globalds2}
\frac{ds^2}{\ell^2} = -d\tau^2 + \cosh^2\tau \, d\vartheta^2~, \quad\quad \tau \in \mathbb{R}~, \quad \vartheta \sim \vartheta +2\pi~.
\end{equation}
In this coordinate system, the invariant distance $u(X,Y)$ defined in \eqref{eq:invariantdistance} can be expressed simply as
\begin{equation}
    u=1+\sinh \tau\sinh \tau'-\cos(\vartheta-\vartheta')\cosh \tau \cosh \tau'~.
\end{equation}
The Killing vector fields of dS$_2$ in the global chart are given explicitly by
\begin{align}
   &{\mathcal{L}}_{12} = -i\partial_\vartheta~, \label{def:killing0} \\
    &{\mathcal{L}}_{01} = i\left(\cos\vartheta\, \partial_\tau-\sin\vartheta\,\tanh \tau\,  \partial_\vartheta\right)~, \label{def:killing1}\\
    &{\mathcal{L}}_{02} = i\left(\sin\vartheta\, \partial_\tau+\cos\vartheta\,\tanh \tau \, \partial_\vartheta\right) \, , \label{def:killing-1}
\end{align}
which respectively generate the rotational and two boost symmetries of dS$_2$. 
Similarly, we have
\begin{equation}\label{eq:confopsglobal}
L_0=-i\partial_\vartheta~,\qquad L_{\pm}=e^{\mp i\vartheta}\left(-i\tanh\tau\partial_\vartheta \pm \partial_\tau\right)~.
\end{equation}
The quadratic Casimir is then given as a differential operator on scalar fields, by
\begin{equation}\label{eq:casimirlaplacianrel}
{\mathcal{C}} =  -{\ell^2}\square_{{\text{dS}}}\, ,
\end{equation}
where $\square_{{\text{dS}}}$ is the scalar Laplacian on dS$_2$. 

One can unwrap the spatial $S^1$ by taking $\vartheta \in \mathbb{R}$ in (\ref{globalds2}), as considered in \cite{Epstein:2018wfh,Anninos:2019oka}. The resulting spacetime is smooth and has an isometry group given by the universal cover of $G$, namely $\widetilde{SL(2,\mathbb{R})}$, which is also the isometry group of Lorentzian AdS$_2$.    
\paragraph{Conformal compactification of dS$_2$.} It will occasionally be convenient to work in a global coordinate system with the infinite coordinate time compactified to a finite interval, given by
\begin{equation}\label{confmetric}
\frac{ds^2}{\ell^2} = \frac{-dT^2 + d\vartheta^2}{\sin^2 T}~, \quad\quad T \in (-\pi,0)~,
\end{equation}
and obtained from \eqref{eq:globalcoords} by the identification
\begin{equation}\label{eq:golbaltoconfcomp}
    \cosh \tau= -\frac{1}{\sin T}~. 
\end{equation}
In this coordinate system, the invariant distance $u(X,Y)$ takes the following form
\begin{equation}
    u=\frac{\cos(T-T')-\cos(\vartheta-\vartheta')}{\sin T\sin T'}=\frac{2\sin\left(\vartheta^{-}-\vartheta'^{-}\right)\sin\left(\vartheta^{+}-\vartheta'^{+}\right)}{\sin \left(\vartheta^{+}-\vartheta^{-}\right)\sin \left(\vartheta'^{+}-\vartheta'^{-}\right)}~,
\end{equation}
where
\begin{equation}
    \vartheta^{\pm}\equiv\frac{\vartheta\pm T}{2}
\end{equation}
are, respectively, the instantaneous left and right moving coordinates on global dS$_2$.
The metric (\ref{confmetric}) is conformally equivalent to the Lorentzian cylinder over a finite time-interval. In this case we have Killing vectors given by 
\begin{align}
   {\mathcal{L}}_{12} &= -i\partial_\vartheta~, \\
    {\mathcal{L}}_{01} &= -i\left( \cos \vartheta  \,\sin T\,\partial_T+ \sin\vartheta\,  \cos T\,\partial_\vartheta\right)~, \\
    {\mathcal{L}}_{02} &= -i\left( \sin \vartheta \, \sin T\,\partial_T- \cos\vartheta \, \cos T\,\partial_\vartheta\right) \, .
\end{align}
These give us: 
\begin{equation}
L_0=-i\partial_\vartheta~,\qquad L_{\pm}=-e^{\mp i\vartheta}\left(i\cos T\,\partial_\vartheta \pm \sin T\,\partial_T\right)~.
\end{equation}
\paragraph{Planar coordinates on dS$_2$.} We also make use of the planar coordinate system when discussing phenomena that are localised near a boundary point. This coordinate patch is analogous to the Poincar\'e patch in AdS. The metric is
\begin{equation}\label{planar}
\frac{ds^2}{\ell^2} = \frac{-d\eta^2 + dx^2}{\eta^2}~, \quad\quad \eta \in (-\infty,0)~, \quad\quad x \in \mathbb{R}~,
\end{equation}
and covers half of the global dS manifold. It contains the late time slice with a single point removed. It is obtained from the embedding coordinates
\begin{equation}\label{eq:planarcoords}
    X^{A} =\frac{\ell}{2\eta}  \, \left(\eta^2-1-{x^2},\,{2x},\,\eta^2+1-{x^2} \right) \, . 
\end{equation}
\paragraph{Euclidean continuation to $S^2$.} The global patch of dS$_2$ can be Wick rotated to the standard metric on the two-sphere. In the coordinates (\ref{globalds2}) one takes $\tau \to  i\left(\psi-\tfrac{\pi}{2}\right)$ and restricts the range of $\psi \in (0,\pi)$ to obtain the smooth geometry. In the coordinates (\ref{confmetric}) one takes $T \to -i X -\tfrac{\pi}{2}$ to obtain the metric 
\begin{equation}
\frac{ds^2}{\ell^2} = \frac{dX^2 + d\vartheta^2}{\cosh^2 X}~,
\label{eq:eglobal}
\end{equation}
with $X\in \mathbb{R}$. Again this metric is that of the two-sphere. Similarly to \eqref{eq:golbaltoconfcomp}, the relationship between $X$ and the polar angle $\psi$ is
\begin{equation}
    \cosh X=\frac{1}{\sin\psi}~.
\end{equation}
At least in the absence of gravity,  quantum field theory on the Euclidean sphere plays an important role in the axiomatic formulation of quantum field theory on de Sitter space (see for instance \cite{Schlingemann:1999mk,Higuchi:2010xt}). In particular, correlation functions of local operators on the Euclidean sphere can be continued to dS$_2$ correlation functions in the Euclidean vacuum.

\subsubsection*{Discrete symmetries}
In addition to the continuous isometries described above, the dS$_2$ spacetime also enjoys discrete antipodal (A), parity (P), and time reversal (T) symmetries. 
\begin{center}
\begin{tabular}{| L || L | L | L ||} 
 \hline
   & P & T & A \\ [0.5ex] 
 \hline\hline
 \text{Global} & \vartheta \rightarrow 2 \pi -\vartheta  & t\rightarrow -t & \vartheta\rightarrow\vartheta + \pi \text{ , } t\rightarrow -t \\ 
 \hline
 \text{Planar} & x\rightarrow-x & \times & \eta\rightarrow-\eta \\
 \hline
 \text{Conformal} & \vartheta \rightarrow 2 \pi -\vartheta  & T\rightarrow -\pi-T & \vartheta\rightarrow\vartheta + \pi \text{ , } T\rightarrow -\pi-T\\
 \hline
\end{tabular}
\end{center}
The above table  summarizes the action of these symmetries in terms of the corresponding coordinates of each coordinate patch.

\section{Unitary irreducible representations} \label{sec:UIRs}
\subsection{General theory}
The unitary irreducible representations (UIRs) of $SO(1,2)$ are well known \cite{Dobrev:1977qv} (see e.g. \cite{Sengor:2019mbz,Anous:2020nxu,Anninos:2020hfj,Sun:2021thf,Hogervorst:2021uvp,Penedones:2023uqc} for recent literature). Here we will be brief. Let us label the eigenvalue of the  quadratic Casimir \eqref{eq:cas} as $\Delta(\Delta-1)$.
The quantity $\Delta$ is known as the \emph{conformal weight} and labels a particular representation. Only a few possible choices of $\Delta$ lead to unitary irreducible representations, which we review in \cref{App:Repthy}. 

To set the stage, we remind the reader that a state $|n,\Delta\rangle$ in a UIR is labeled by its eigenvalues under the maximal commuting subgroup of $SL(2,\mathbb{R})$:
\begin{equation}\label{eq:state}
    \widehat{\mathcal{C}}|\Delta,n\rangle=\Delta(\Delta-1)|\Delta,n\rangle~, \qquad \widehat{L}_0|\Delta,n\rangle=-n |\Delta,n\rangle~,\quad \widehat{L}_\pm|\Delta,n\rangle=-(n\pm \Delta)|\Delta,n\pm1\rangle~. 
\end{equation}
Recall that operators denoted with a ~$\widehat{\ }$ ~signify abstract matrix (not necessarily differential) operators.
Because $\widehat{L}_0$ is a compact generator, which acts by rotating  the de Sitter hyperboloid, its eigenvalues must be integers (or half integers for the double cover). 

To have a UIR means we have a positive, semi-definite inner product on the space of states defined above. The various distinct UIRs for which this is possible are: 
\begin{itemize}
    \item The \emph{principal series}, $\pi_\nu$, for which $\Delta=\frac{1}{2}(1+i\nu)$ with $\nu\in \mathbb{R}$~.
    \item The \emph{complementary series}, ${\gamma}_\Delta$, for which $0<\Delta<1$~. 
    \item The \emph{discrete series}, $D^\pm_\Delta$, for which $\Delta$ is either a positive integer or half integer. The $(+)$ refers to the highest-weight module, which has an element annihilated by $L_+$. The $(-)$ refers to a lowest weight module which contains an element annihilated by $L_-$. 
\end{itemize}
Moreover, as  suggested by the Casimir eigenvalue, there is an isomorphism between $\pi_\nu$ and $\pi_{-\nu}$ and $\gamma_\Delta$ and $\gamma_{1-\Delta}$. The isomorphism does not hold for the $D^\pm_{\Delta}$, as unitarity restricts $\Delta\geq 1$ for $D_\Delta^\pm$.  The ranges of $n$ differ across the various representations, so we provide a handy summary in \cref{tab:reptab} for the scalar representations. For both scalar and fermionic representations, see \cref{App:Repthy}.

\begin{table}\renewcommand\arraystretch{1.3}\begin{center}
\begin{tabular}{|c|c|c|c|}
\hline
\textbf{Rep.} & \textbf{Range of $\Delta$} & \textbf{Range of} $n$ & \textbf{Scalar} $m^2 \ell^2$\\
\hline
$\pi_\nu$ & $\Delta= \tfrac{1}{2}(1+i \mathbb{R})$  & $n\in \mathbb{Z}$ & $m^2 \ell^2>\tfrac{1}{4}$\\
$\gamma_\Delta$ &$0<\Delta<1$  & $n\in \mathbb{Z}$ & $0 < m^2 \ell^2 <\tfrac{1}{4}$\\
$D^\pm_\Delta$ & $\Delta\in\mathbb{Z^+}$ & $n=\mp\Delta,\mp(\Delta+1),\dots$& $m^2\ell^2= -t(t+1)$ with $t\in \mathbb{N}_0$ \\
\hline
\end{tabular}\end{center}\caption{Summary of the various scalar representations and their eigenvalues under the $SL(2,\mathbb{R})$ algebra. For more details see \cref{App:Repthy}. }\label{tab:reptab}
\end{table}

Much like states can be organised into UIRs, we can also discuss the transformation properties of certain operators under $SL(2,\mathbb{R})$. In particular, an operator $\mathcal{O}_{\Delta,n}$ satisfying 
\begin{equation}\label{confOps}
\left[ \widehat{L}_0, \mathcal{O}_{\Delta,n}\right] = n \mathcal{O}_{\Delta,n}~, \quad\quad \left[\widehat{L}_\pm, \mathcal{O}_{\Delta,n}\right] = \left( n \pm \Delta \right) \mathcal{O}_{\Delta,n\pm1}~, 
\end{equation}
is said to be a conformal operator of weight $\Delta$. Although we have states and operators furnishing UIRs of the de Sitter group, there is no state-operator correspondence as is usual for conformal field theory \cite{DiPietro:2021sjt}. Acting with a local conformal operator at the Euclidean de Sitter boundary creates a non-normalizable state due to coincident point singularities. Relatedly, the future boundary of de Sitter space, though Euclidean, arises as the end point of the bulk Lorentzian spacetime's time evolution. As such, the imprints of a Lorentzian structure such as a non-commuting operator algebra and standard Hermiticity conditions must be obeyed at $\mathcal{I}^+$.

Recall from \eqref{eq:casimirlaplacianrel} that the Casimir operator $\mathcal{C}$ can be represented by the Laplacian on dS$_2$ as $\mathcal{C}=-\ell^2\square_{\text{dS}}$. The equation of motion for a massive scalar field is
\begin{equation}\label{eq:massiveeom}
    \square_{{\text{dS}}}\,\phi=m^2\phi~,
\end{equation}
implying the relationship
\begin{equation}\label{eq:masstodelta}
    \Delta(\Delta-1)=-m^2\ell^2 \quad \implies \quad \Delta=\frac{1}{2}\left(1\pm\sqrt{1-4m^2\ell^2}\right)~. 
\end{equation}
From here we determine that states in the complementary series are faithfully represented by scalars whose mass squared satisfies: $0<m^2\ell^2<\frac{1}{4}$, whereas states in the principal series are given by scalars with $m^2\ell^2>\frac{1}{4}$. In order to obtain a state in the discrete series, for $\Delta=1+t$ with  $t\in \mathbb{N}_0$, we would need
\begin{equation}\label{tachyon}
    m^2\ell^2= -t(t+1)~.
\end{equation}
Namely, for $t=0$, the scalar is massless, otherwise the scalar must be tachyonic. Perhaps, then, discrete series states are unphysical in dS$_2$, and can be rightfully ignored. However, recalling \cite{Repka1978TensorR} the tensor product of scalar representations: 
\begin{equation}
    \pi_\nu \otimes \pi_{\nu'}
    = 
    \left(\mathop{\bigoplus}_{
    \Delta} D^\pm_\Delta\right)\oplus 
    \left(2\int_{\mathbb{R}^+}d\omega\, \pi_\omega\right) ~,
    \label{res:principal}
\end{equation}
one notes that $D^\pm_\Delta$ invariably makes an appearance in the two-particle Hilbert space of massive scalar fields on dS$_2$.\footnote{We elaborate on the two-particle Hilbert space from the perspective of a Hilbert space built from wavefunctions on $S^1$ in \cref{app:circle}.} Given Wigner's interpretation of UIRs as single particle states in quantum field theory, we are led to ask if these discrete series bound states can be interpreted as asymptotic free particle states under some suitable definition. In fact, much of this paper will concern itself with how the discrete series appears in different guises in the context of dS$_2$. In section \ref{sec:spectroscopic decompositions} we will explore the consequences of $D_{\Delta}^\pm$ appearing in the multi-particle Hilbert space by studying the spectral decomposition (or K\"all\'en-Lehmann representation) of the two-point function of a heavy interacting scalar on dS$_2$.  

An interesting consequence of our analysis is that a QFT whose classical equations of motion on dS$_2$ are given by \eqref{eq:massiveeom} is not  guaranteed to have a state of dimension $\Delta$ (related to the mass via the relation \eqref{eq:masstodelta}) in its Hilbert space. This is most obvious, whenever the  the metric degrees of freedom are dynamical. Suitably gauging the ambient $SL(2,\mathbb{R})$ isometry of dS$_2$ will result in a reduction of the physical Hilbert space---a fact we will explore in a few examples in \cref{discretesec}.

%

%

\subsection{Free Fock space: principal series}\label{sec:freefock}
So far our exposition has been quite abstract, so as an instructive aid to the reader, we will show how to build the principal series UIR, as in \eqref{eq:state}, using single-particle excitations of a free quantum field theory on dS$_2$. We start with the action: 
\begin{equation}
 	 S=-\frac{1}{2}\int d^2x\sqrt{-g}\left[g^{\mu\nu}\partial_\mu\phi\partial_\nu\phi+m^2\phi^2\right]~,
 \end{equation}
 from which we can derive the equation of motion \eqref{eq:massiveeom}:
 \begin{equation}
    \square_{{\text{dS}}}\,\phi=m^2\phi~. 
\end{equation}
In this section, we will insist that $m^2\ell^2>\tfrac{1}{4}$ such that $\Delta=\tfrac{1}{2}(1+i\nu)$. There are several ways to decompose our classical field into modes. Working in global coordinates \eqref{eq:globalcoords}, we choose to expand our field in modes that are regular near the pole of the lower half-sphere $\left({\tau}\rightarrow-\frac{i\pi}{2}\right)$ in the Euclidean continuation of the global coordinates defined above \eqref{eq:eglobal}. Being regular on the south pole, these modes define a Hadamard state, as we will come to see \cite{Schlingemann:1999mk,Higuchi:2010xt}. That is, we write
\begin{equation}\label{eq:phiprincipaldef}
	\phi(\tau,\vartheta)=\sum_{n=-\infty}^\infty a^\Delta_n\,\phi^{E,\Delta}_n(\tau,\vartheta)+a_n^{\Delta\dagger} \phi_n^{*E,\Delta}(\tau,\vartheta)
\end{equation}
where \cite{Mottola:1984ar,Bousso:2001mw}: 
\begin{equation}\label{eq:eucmode}
 \phi_n^{E,\Delta}(\tau,\vartheta)=f_n^{E,\Delta}(\tau)\frac{e^{-in\vartheta}}{\sqrt{2\pi}}~,
\end{equation}
and the time dependent factor is:
\begin{equation}\label{eq:EucmodeTimeDep}
f_n^{E,\Delta}(\tau)\equiv e^{i\alpha_n}\sqrt{\frac{\Gamma(\Delta-|n|)\Gamma(1-\Delta-|n|)}{2}}\,P_{-\Delta}^{|n|}\left(i\sinh\tau\right)~,
\end{equation}
where $P_a^b(x)$ is an associated Legendre function. We have chosen a particular phase factor
\begin{equation}\label{eq:phaseeuclideanmodes}
e^{2i\alpha_n}=e^{i\pi\left(|n|+\frac{1}{2}\right)}\frac{\Gamma(|n|+1-\Delta)\Gamma(1+\Delta)}{\Gamma(|n|+\Delta)\Gamma(2-\Delta)}~,
\end{equation}
which will play an important role in what follows.
These modes are normalized such that 
\begin{equation}
	(\phi_n^{E,\Delta},\phi_m^{E,\Delta})=\delta_{nm}~,\quad\quad\quad (\phi_n^{E,\Delta},\phi^{*E,\Delta}_m)=0~,
\end{equation}
where the bracket $(\cdot,\cdot)$ denotes Klein-Gordon inner-product: 
  \begin{equation}\label{eq:kgnormwf}
 	(\phi_1,\phi_2)=-i\int_\Sigma d\Sigma^\mu\left(\phi_1\partial_\mu\phi_2^*-\phi_2^*\partial_\mu\phi_1\right)=-i\cosh\tau\int_0^{2\pi} d\vartheta \left(\phi_1\partial_\tau\phi_2^*-\phi_2^*\partial_\tau\phi_1\right)~.
 \end{equation}
The phase factor \eqref{eq:phaseeuclideanmodes}  ensures that these modes transform nicely under the generators of $SL(2,\mathbb{R})$ given in \eqref{eq:confopsglobal}: 
\begin{equation}
	L_0\phi^{E,\Delta}_n=-n\phi^{E,\Delta}_n~,\quad\quad L_\pm \phi^{E,\Delta}_n=-(n\pm\Delta)\phi^{E,\Delta}_{n\pm1}~.
\end{equation}
One may find the above equation surprising, given that the Euclidean modes have an admixture of falloffs ($e^{-\Delta\tau}$ and $e^{-(1-\Delta)\tau}$) at late times, but it is nevertheless possible to express them as transforming properly under the conformal algebra.

Canonical quantization proceeds by promoting $\phi(\tau,\vartheta)$ and its canonical conjugate $\pi(\tau,\vartheta)$ to operators, 
where
\begin{equation}
\pi(\tau,\vartheta) \equiv\frac{\delta\mathcal{L}}{\delta(\partial_\tau\phi(\tau,\vartheta))}  =\cosh\tau\,\partial_\tau\phi~, 
\end{equation}
and demanding
\begin{equation}
[\phi(\tau,\vartheta),\pi(\tau,\vartheta')]=i\delta(\vartheta-\vartheta')~.
\end{equation}
This can be achieved by promoting $(a^\Delta_n,a_n^{\Delta\dagger})$  to operators that satisfy:
\begin{equation}\label{eq:canonicalcomm}
	[a^\Delta_n,a_m^{\Delta\dagger}]=\delta_{nm}~,\qquad\qquad [a^\Delta_n,a^\Delta_m]=[a_n^{\Delta\dagger},a_m^{\Delta\dagger}]=0~.
\end{equation}
We must also choose a state on top of which we build our Fock space. The Euclidean vacuum $|\Omega\rangle$, is defined such that
\begin{equation}
a^\Delta_n|\Omega\rangle=0 ~, \qquad \forall n~.
\end{equation}
What remains is to identify the basis states of the principal series UIR: $\ket{\Delta,n}$. A natural expectation is the following:
\begin{equation}
    \ket{\Delta,n}\equiv a^{\Delta\dagger}_n\ket{\Omega}
\end{equation}
at least at the single particle level \cite{Guijosa:2005qi}. To check that this is indeed correct, we must write down the conformal generators in the basis of creation and annihilation operators:
\begin{equation}
\widehat{L}_{n}=-\sum_{k=-\infty}^\infty (k+n\Delta)\, a_{k+n}^{\Delta\dagger} a_k^\Delta~,
\end{equation}
for $n=\{-1,0,1\}$.
Using the canonical commutation relations \eqref{eq:canonicalcomm}, we find 
\begin{equation}
\left[\widehat{L}_n,\widehat{L}_m\right]=(n-m)\widehat{L}_{n+m}~,
\end{equation}
as required. It is also straightforward to check that 
\begin{equation}
\left[\widehat{L}_n,\phi(\tau,\vartheta)\right]=-L_n\phi(\tau,\vartheta)~,
\end{equation}
where the operators on the right hand side are the differential representation of the algebra given in \eqref{eq:confopsglobal}. Finally given the definitions, a short computation yields: 
\begin{equation}
  \widehat{L}_0\,a^{\Delta\dagger}_n|\Omega\rangle=-n\, a^{\Delta\dagger}_n |\Omega\rangle~,\quad \widehat{L}_\pm\, a_n^{\Delta\dagger}|\Omega\rangle=-(n\pm \Delta)a^{\Delta\dagger}_{n\pm1}|\Omega\rangle~.
\end{equation}
Hence we see that we have correctly identified $\ket{\Delta,n}\equiv a_n^{\Delta\dagger}\ket{\Omega}$. Given these definitions, the mode functions may be expressed as overlaps of the field operator $\phi$ and the state $\ket{\Delta,n}$: 
\begin{equation}\label{eq:fieldmodeoverlaps}
\phi_n^{E,\Delta}(\tau,\vartheta)=\bra{\Omega}\phi(\tau,\vartheta)\ket{\Delta,n}~,\qquad\qquad \phi_n^{*E,\Delta}(\tau,\vartheta)=\bra{\Delta,n}\phi(\tau,\vartheta)\ket{\Omega}~.
\end{equation}

\subsection{Free two-point function \label{sec:2ptfunc}}
We now turn to the two-point function, or propagator, of a free, minimally-coupled,  massive scalar field $\phi$ on dS$_2$
\begin{equation}
G_{\rm f}(X,Y)\equiv \langle\Omega|\phi(X)\phi(Y)|\Omega\rangle~.  
\end{equation}
The subscript ``${\rm f}$'' refers to the fact that it is free. 
This satisfies 
\begin{equation}\label{eq:wightman}
\left(\square_{\rm dS} -m^2\right)G_{\rm f}(X,Y)=0~,
\end{equation}
with $\Delta$ related to $m$ through \eqref{eq:masstodelta}. Here we are choosing to study the Wightman function, but we could just as well study the retarded, advanced or Feynman propagator by replacing the right hand side of \eqref{eq:wightman} with $\delta(X,Y)/\sqrt{-g}$ and choosing suitable boundary conditions. 
The above differential equation can be expressed as an ODE of the de Sitter invariant distance $u(X,Y)$ defined in \eqref{eq:invariantdistance}
\begin{equation}
   u(2-u)G_{\rm f}''(u)+2(1-u)G_{\rm f}'(u)+\Delta(\Delta-1)G_{\rm f}(u)=0~. 
   \label{eq:inv2point}
\end{equation}
Being a second order differential equation, there are two linearly independent solutions 
\begin{equation}
    G_{\rm f}(u)=
    c_1\, {}_2F_1\left(\Delta,1-\Delta,1,1-\tfrac{u}{2}\right)+c_2\, {}_2F_1\left(\Delta,1-\Delta,1,\tfrac{u}{2}\right)~.
\end{equation}
 For $\Delta$ in the principal or complementary series, the term proportional to $c_1$ has the appropriate lightcone singularity in the limit $u\rightarrow 0$, while the term proportional to $c_2$ has an antipodal singularity in the limit $u\rightarrow 2$. Both behaviors are allowed in a de Sitter invariant state, however, we will further demand that the state $|\Omega\rangle$, upon which we build our Fock space, be Hadamard, which disallows any field singularities at spacelike-separated points. This fixes: 
 \begin{equation}
    G^\Delta_{\rm f}(u)\equiv
\frac{\Gamma(\Delta)\Gamma(1-\Delta)}{4\pi}{}_2F_1\left(\Delta,1-\Delta,1,1-\tfrac{u}{2}\right)  ~, \qquad \Delta\in \pi_\nu \quad \text{ or } \quad \gamma_\Delta ~,
\label{eq:massivefreeg}
 \end{equation}
 where the coefficient is set by demanding that we match onto the flat space answer in the limit $u\rightarrow 0$~, which, in this case, is 
 \begin{equation}\label{eq:flatspacesing}
     G_{\rm f}(u)\underset{u\rightarrow0}{\approx}-\frac{1}{4\pi}\log \frac{u}{2}~.
 \end{equation}
Moreover, for $\pi_\nu$,  we can show the Hadamard two-point function admits a Fourier decomposition in terms of the Euclidean modes constructed in \cref{sec:freefock}: 
\begin{equation}\label{eq:2pteuclideandecom}
G^\Delta_{\rm f}(u)=\sum_{n=-\infty}^\infty \phi_n^{E,\Delta}(\tau,\vartheta)\phi_n^{*E,\Delta}(\tau',\vartheta')=\frac{\Gamma(\Delta)\Gamma(1-\Delta)}{4\pi}{}_2F_1\left(\Delta,1-\Delta,1,1-\tfrac{u}{2}\right)~, \quad \Delta\in \pi_\nu~.
\end{equation}
 
 The case of $\Delta$ in the discrete series is subtle and requires some care. Let us parametrize $\Delta=1+t$ for $t=0,1,2,\dots$. Note that for these values of $\Delta$, the coefficient of \eqref{eq:massivefreeg} diverges as $\Gamma(-t)$. This divergence has a physical origin \cite{Folacci:1992xc}, as we will discuss shortly. For now, let us repeat the exercise and try and solve \eqref{eq:inv2point} for these values of $\Delta$. The two independent solutions are: 
 \begin{equation}\label{eq:indepdiscreteseriessols}
      G_{\rm f}(u)=c_1\, P_t(1-u)+c_2\, Q_t (1-u)~, 
 \end{equation}
 where $P_t$ and $Q_t$ are Legendre functions of order $t$. A peculiarity: The term proportional to $c_1$ is a polynomial of order $t$ in $(1-u)$ and thus has no lightcone divergence in the  limit $u\rightarrow 0$. Alternatively, the term proportional to $c_2$ has both lightcone and antipodal divergences. If we demand that $|\Omega\rangle$ be Hadamard, we are required to set $c_2=0$, throwing away both the coincident-point and antipodal singularities, \emph{together}. This leaves a correlator free of divergences, or branch cuts---which is certainly not expectated for a local quantum field. It would seem, then, that there is no room for the discrete series to contribute to the two-point function, at least if we are to have a standard coincident point singularity. 
 
 As we will discuss in the next section, this conclusion is not quite correct. Following the work of \cite{Allen:1987tz,Folacci:1992xc,Bros:2010wa,Bonifacio:2018zex}, we will show that a more delicate treatment indeed leads to a contribution from the discrete series UIR.

\subsubsection{Revisiting the discrete series two-point function} \label{sec:discreterevesit}
We now proceed to explain the physical origin behind the divergence in \eqref{eq:massivefreeg} when $\Delta=1+t$, following \cite{Folacci:1992xc}. For this, let us recall that the Hadamard Wightman function on dS$_2$ can be obtained via analytic continuation of the two-point function on an $S^2$ of radius $\ell$ \cite{Schlingemann:1999mk,Higuchi:2010xt} . Thus, we should compute the Euclidean path integral
\begin{equation}
    G^\Delta_{\rm E}(\Omega,\Omega')=\frac{\int \mathcal{D} \phi\,\phi(
    \Omega)\phi(\Omega')\, e^{-S_{\rm E}[\phi]}}{\int \mathcal{D} \phi\, e^{-S_{\rm E}[\phi]}}~,
\end{equation}
where the Euclidean action is given by 
\begin{equation}
    S_{\rm E}[\phi]=\frac{1}{2}\int_{S^2} d^2x\sqrt{g}\,\phi(\Omega)\left[-\square_{S^2}+m^2\right]\phi(\Omega)~.
\end{equation}
As usual, to evaluate the path integral it is convenient to expand the field $\phi(\Omega)$ in a basis of eigenfunctions of the two-sphere Laplacian, as
\begin{equation}
    \phi(\Omega)=\sum_{L=0}^\infty\sum_{M=-L}^L c_{LM}\, Y_{L}^M(\Omega)~,\qquad\qquad \mathcal{D} \phi=\prod_{L=0}^\infty\prod_{M=-L}^Ldc_{LM}~.
\end{equation}
We have chosen the $Y_L^M(\Omega)$ to be real-valued, such that the $c_{LM}$ are themselves real-valued. The $Y_L^M(\Omega)$ satisfy the standard orthonormality conditions
\begin{equation}\label{eq:ylmproperties}
    \square_{S^2}Y_{L}^M=-\frac{L(L+1)}{\ell^2}Y_L^M~,\qquad\qquad \int_{S^2} d^2x\sqrt{g}\,Y_{L}^M Y_{L'}^{M'}=\ell^2\,\delta_{LL'}\delta_{MM'}~,
\end{equation}
as well as the completeness relation
\begin{equation}\label{eq:ylmcompleteness}
 \sum_{L=0}^\infty \sum_{M= -L}^{L}  Y_{L}^M(\Omega)   Y_L^M(\Omega') =  \delta(\Omega,\Omega')~.
\end{equation}
Performing the remaining Gaussian integrals over the $c_{LM}$ leads to the Euclidean Green's function in momentum space
\begin{equation}
    G^\Delta_{\rm E}(\Omega,\Omega')=\sum_{L=0}^\infty\sum_{M=-L}^L \frac{Y_L^M(\Omega) Y_L^M(\Omega')}{L(L+1)+m^2\ell^2}~.
\end{equation}
Upon employing the addition theorem
\begin{equation}\label{eq:additionylm}
    \sum_{M= -L}^{L}  Y_{L}^M(\Omega)   Y_L^M(\Omega')=\frac{2L+1}{4\pi}P_L(1-u(\Omega,\Omega'))~, 
\end{equation}
we can further express the Euclidean Green's function in the following form
\begin{equation}\label{eq:EuclideanRep}
    G^\Delta_{\rm E}(\Omega,\Omega')=\frac{1}{4\pi}\sum_{L=0}^\infty \frac{2L+1}{L(L+1)+m^2\ell^2}P_L(1-u(\Omega,\Omega'))~.
\end{equation}
In turn, for generic $m^2\ell^2$, the above sum can be performed explicitly, resulting in an expression involving the hypergeometric function. Namely, 
 \begin{equation}
   G^\Delta_{\rm E}(\Omega,\Omega')=
\frac{\Gamma(\Delta)\Gamma(1-\Delta)}{4\pi}{}_2F_1\left(\Delta,1-\Delta,1,1-\tfrac{u(\Omega,\Omega')}{2}\right)~,\label{eq:massivefreegsphere}
 \end{equation}
 as expected from \eqref{eq:massivefreeg}, and where we have used \eqref{eq:masstodelta}, which relates $m^2\ell^2$ with $\Delta$. Here $u(\Omega,\Omega')$ is the appropriate geodesic distance on the $S^2$, equivalent to the analytic continuation of \eqref{eq:invariantdistance} to Euclidean signature. 
 In this case, it is straightforward to verify that 
 \begin{equation}
1-\frac{u(\Omega,\Omega')}{2}=\cos^2\left(\frac{\theta(\Omega,\Omega')}{2}\right)\implies u(\Omega,\Omega')=2\sin^2\left(\frac{\theta(\Omega,\Omega')}{2}\right)~,
 \end{equation}
 where $\theta(\Omega,\Omega')$ is the angle subtended by a geodesic arc connecting $\Omega$ to $\Omega'$.
 Notice, however, that when $m^2\ell^2=-t(t+1)$, with $t\in \mathbb{N}_0$, there are a collection of $(2t+1)$ modes, precisely those with $L=t$, with vanishing Euclidean action. The integrals over these modes necessarily lead to divergences which must be dealt with.\footnote{The modes with $L<t$ have negative Euclidean action, and may seem even more problematic. One way to deal with these is  by analytically continuing the contour of integration for the offending $c_{LM}$'s.}  Thus, the Lorentzian discrete-series divergence originates from the fact that this theory suffers from a Euclidean vacuum state which is non-normalizable, precisely due to these problematic modes \cite{Ford:1984hs,Allen:1985ux}. 
 
 One can now proceed to try and define an appropriate Euclidean two-point function for the discrete series UIR \cite{Folacci:1992xc}. The idea is to eliminate the problematic modes from the sum altogether:
 \begin{equation}\label{eq:zeromodesremoved}
    H^{\Delta=1+t}_{\rm f}(\Omega,\Omega')=\sum_{\substack{L=0\\L\neq t}}^\infty\sum_{M=-L}^L \frac{Y_L^M(\Omega) Y_L^M(\Omega')}{L(L+1)-t(t+1)}~,
\end{equation}
and try to give this function a Lorentzian Hilbert space interpretation, as in \cite{Bros:2010wa,Epstein:2014jaa}. From here on, we will label free propagators on the discrete series as $H_{\rm f}$, the ``${\rm f}$'' again referring to the fact that it is free, so as to distinguish it from the typical two-point function of the principal and complementary series. This procedure is inherently ad-hoc, and the final answer will necessarily be ambiguous, moreover, it is difficult to reconcile with local quantum field theory, although in the next section we will give an example of how to proceed when $\Delta=1$. 

Note that the completeness relation \eqref{eq:ylmcompleteness} implies that 
\begin{equation}\label{eq:inhomogeneous}
    \left[-\ell^2\square_{ S^2}-t(t+1)\right]H^{\Delta=1+t}_{\rm f}(\Omega,\Omega')= \delta(\Omega,\Omega')-\frac{2t+1}{4\pi}P_t(1-u(\Omega,\Omega'))~,
\end{equation}
where we have used the addition theorem \eqref{eq:additionylm}.
We see that the Klein-Gordon operator acting on the two-point function (with the problematic zero-modes removed) isn't sourced by a local $\delta$-function disturbance, but rather, by a function supported on the entire $S^2$---evidence of some tension with locality. Moreover, this right hand side implies that the notion of the identity operator on the Hilbert space needs modification whenever the discrete series is concerned. Since all we've done is remove an entire $SO(3)$ representation from the sum, the final answer remains $SO(3)$-invariant, and the analytically continued result will therefore be de Sitter invariant.  

It is possible to solve this inhomogeneous Klein-Gordon equation outright, giving: 
\begin{equation}\label{eq:full-sol}
   H^{\Delta=1+t}_{\rm f}(u;\alpha)=-\frac{1}{4\pi}P_t(1-u)\left(\log\frac{u}{2}+\alpha\right)-\frac{1}{2\pi}\sum_{s=0}^{t-1}\frac{2s+1}{t(t+1)-s(s+1)}P_s(1-u)~,
\end{equation}
where the constant $\alpha$ is an ambiguity proportional to a homogeneous solution to  \eqref{eq:indepdiscreteseriessols}. The Green's function defined in \eqref{eq:zeromodesremoved} is equivalent to \eqref{eq:full-sol} with $\alpha=0$, but we have included the $\alpha\neq0$ term in order to provide the general solution to \eqref{eq:inhomogeneous}, which reflects the ambiguity in defining the procedure for removing the zero-modes. Indeed, in \cite{Folacci:1992xc}, the parameter $\alpha$ is related to a BRST gauge-fixing procedure.
The formula \eqref{eq:full-sol} has the appropriate short-distance singularity as in \eqref{eq:flatspacesing}, as expected for a two-point function in a Hadamard state, but the coefficient $\alpha$ can't be fixed by any local requirement. 
We will also make use of the following definition
\begin{equation}
H^{\Delta=1+t}_{\rm f}(u)\equiv H^{\Delta=1+t}_{\rm f}(u;0)~.
\end{equation}
Suffice it to say: if we ever encounter the Klein-Gordon operator with $m^2\ell^2=-t(t+1)$ under any circumstance in dS$_2$, we should exercise care. In what follows, we will provide some examples where such equations arise.

\paragraph{Comment on positivity of $H_{\rm f}$ for $0<u\leq2$:} In the spacelike separated regime ($0<u\leq 2$), Euclidean and Lorentzian correlators agree. On the sphere, we typically interpret the two-point function at antipodally-separated points as the norm of a state. Based on this intuition, we expect the two-point function in this regime to be positive definite. However \eqref{eq:full-sol} is oscillatory in this regime, and is not sign definite at antipodal points for every $t$. This is unlike the principal and complementary series correlators \eqref{eq:massivefreegsphere}, which are positive definite for $0<u\leq 2$. 

\paragraph{Comment on late time behavior of $H_{\rm f}$:} We now turn to the late time behavior of these correlation functions. Choosing the global coordinate system \eqref{globalds2}, and taking $\tau=\tau'\rightarrow \infty$ we find the following late time behavior for a discrete series two-point function: 
\begin{equation}\label{eq:latetimediscreteseries2pt}
    \lim_{\tau\rightarrow\infty} H^{\Delta=1+k}_{\rm f}(u;\alpha)=e^{2k\tau}\frac{(-1)^{k+1}}{4\pi}\binom{k-\tfrac{1}{2}}{k}\sin^{2k}\left(\frac{\vartheta-\vartheta'}{2}\right)\left\lbrace2\tau+\alpha+\log\left[ \frac{1}{4}\sin^{2}\left(\frac{\vartheta-\vartheta'}{2}\right) \right]\right\rbrace
\end{equation}
where we have used $k$ in place of $t$ to disambiguate it from the time coordinate $\tau$. This piece of the correlation function grows at late times. Note that the dependence on the ambiguous parameter $\alpha$ is subleading at the future boundary, albeit only polynomially in the global time $\tau$. On the other hand, the late time behavior of the free principal series correlator is 
\begin{equation}\label{eq:latetimeprincipalseries2pt}
    \lim_{\tau\rightarrow\infty} G^{\Delta}_{\rm f}(u)= e^{-2\Delta\tau}\frac{\Gamma(\Delta)\Gamma\left(\tfrac{1}{2}-\Delta\right)}{4\pi^{3/2}}\left\lvert\sin\left(\frac{\vartheta-\vartheta'}{2}\right)\right\rvert^{-2\Delta}+\text{c.c.}~, \qquad \Delta\in\pi_\nu~,
\end{equation}
meaning that, at late times, the discrete series contributions, if present, will wash out the imprint of the principal series on the conformal boundary.
 
Before moving on, let us briefly comment on a familiar example of a discrete series theory: the case of the free, massless scalar. The massless free boson has an action invariant under constant shifts of the field $\phi(x)\rightarrow \phi(x)+c$, and integrating over the constant mode of $\phi$ leads to a divergence since this mode is not Gaussian suppressed in Euclidean signature, again rendering the Laplacian operator  non-invertible. There are two familiar remedies: we can either compactify the zero-mode by identifying $\phi\sim\phi+R$ 
in which case the field $\phi$ is no longer well-defined as a local scalar operator on Hilbert space and we must instead consider operators such as $:e^{2\pi i \phi/R}:$,\footnote{Even in the case of the compact free scalar, global constraints arise when the zero-mode is treated carefully, see Exercise 9.2 of  \cite{DiFrancesco:1997nk}. For example, shifts of $\phi$ act as $U(1)$ phase rotations of the vertex operator $:e^{2\pi i \phi/R}:$, and the only non-vanishing vacuum correlators are those of charge-neutral strings of vertex operators.} or we can gauge the shift symmetry, in which case $\phi$ is not a gauge-invariant operator on Hilbert space. In both cases, the bare field $\phi(x)$ loses its status as a well-defined quantum field acting on Hilbert space. 

It is of crucial importance that we acknowledge that the discrete series can not be relegated as an easy-to-ignore curiosity. In \cref{app:circle}, we solve a quantum mechanical model that may be thought of as the late-time single-particle Hilbert space of two principal series fields propagating in dS$_2$. We show that the discrete series arises in the the two-particle Hilbert space of this quantum mechanical example, via a simple Clebsch-Gordan analysis. The projecting-out of the problematic modes happens simply by demanding normalizability of the two-body wavefunctions.

In the following section, we will examine the case of a free massless scalar with a gauged shift symmetry and show that the correlators of this theory can be derived starting from \eqref{eq:full-sol}.

\subsection{Scalar with a gauged shift symmetry and  the \texorpdfstring{$\Delta=1$}{} discrete series}\label{Delta1}
%
With the general discussion of the previous section now behind us, let us provide a simple example where the Green's function \eqref{eq:full-sol} (for $\Delta=1$) makes an indirect appearance. Consider a massless scalar coupled to a gauge field:
\begin{equation}\label{BFmatter}
S = -\frac{1}{2} \int d^2 x\sqrt{-g} g^{\mu\nu} \left( \partial_\mu \phi - A_\mu  \right) \ \left( \partial_\nu \phi - A_\nu  \right)  + k \int d^2x \sqrt{-g} \, B \, \epsilon^{\mu\nu} F_{\mu\nu}~.
\end{equation}
Here, $F_{\mu\nu} =\partial_\mu A_\nu-\partial_\nu A_\mu$, $B$ is a real-valued  scalar and $\epsilon^{\mu\nu}$ is the antisymmetric  Levi-Civita tensor with $\epsilon^{T\vartheta}=+{1}/{\sqrt{-g}}$.\footnote{We will distinguish between the Levi-Civita tensor and symbol by denoting the latter as $\tilde{\epsilon}^{\mu\nu}$ such that $\tilde{\epsilon}^{T\vartheta}=+1$.} The model's global shift symmetry $\phi(x)\rightarrow\phi(x)+c$, absent the gauge field, is promoted to a local symmetry. We take the shift symmetry to be non-compact such that  $k\in \mathbb{R}$. Explicitly, the model's  Abelian gauge invariance is $\phi(x) \to \phi(x) + \omega(x)$ and $A_\mu \to A_\mu + \partial_\mu \omega$ with $\omega$ a smooth real-valued function. 

Had we not gauged the shift symmetry, the model would suffer from a pathological zero mode, as explained in the previous section. 
In \cite{Folacci:1992xc}, the constant shift mode is gauged via a non-local condition on the $S^2$. The model \eqref{BFmatter} follows the spirit of \cite{Folacci:1992xc}, but the advantage of this setup is that we are always in the realm of local quantum field theory. 

It is convenient to consider the model in the global coordinate system (\ref{confmetric}), for which the Weyl factor drops out altogether. Gauge invariant operators are given by
\begin{equation}
 B~, \quad\quad F_{\mu\nu}~, \quad\quad \mathcal{O}_\mu(T,\vartheta) = \partial_\mu \phi(T,\vartheta) - A_\mu(T,\vartheta)~, \quad\quad \chi = \oint_{\mathcal{C}} A_\mu dx^\mu~, 
\end{equation}
and combinations thereof. The curve $\mathcal{C}$ is taken to be a closed spacelike curve. In addition, the dressed operators
\begin{equation}
\mathcal{O}_q(T,\vartheta) = e^{-i q \phi(T,\vartheta) } \exp \, i q {\int_{\mathcal{L}} A_\mu dx^\mu}~,
\end{equation}
where $\mathcal{L}$ is a curve beginning at some reference point and ending at $(T,\vartheta)$, can be arranged into gauge-invariant combinations by taking products for which the sum of the $q$ vanishes.  

We can construct a Hilbert space by acting on the vacuum state $|\Omega\rangle$ with  suitable combinations or distributions of the gauge-invariant operators. Working in the $A_T = 0$ gauge,  the ensuing constraint is given by 
\begin{equation}
2k \, \partial_\vartheta B - \partial_T \phi = 0~.
\end{equation}
This fixes the non-constant spatial modes of $B$, leaving only the constant mode $b \equiv \frac{1}{2\pi}\oint d\vartheta B$ as an independent gauge-invariant operator. We must further ensure invariance under residual gauge transformations given when $\omega$ is purely a function of $\vartheta$. This can be used to gauge away the spatial non-zero modes of $A_\vartheta$, again leaving $\chi$ as the gauge-invariant operator.

Thus, we land on the non-gauge invariant operator algebra
\begin{equation}
[ \phi(T,\vartheta), \partial_T \phi(T,\vartheta') ] = i  \delta(\vartheta-\vartheta')~, \quad\quad [ b , \chi ] = -\frac{ i}{2k}~.
\end{equation}
 To create single-particle states, we can build a creation operator out of the gauge-invariant operators. Creation and annihilation operators which stem from gauge invariant operators, expressed in the $A_T = 0$ gauge, read as follows 
\begin{eqnarray}
a_n - a_{-n}^\dag &\equiv& i\sqrt{\frac{2}{|n|}} \int {d\vartheta} e^{-i n \vartheta} \partial_T \phi(T,\vartheta)|_{T=0}~, \\
a_n + a_{-n}^\dag &\equiv&  - i \, \text{sgn} \, n\,\sqrt{\frac{2}{|n|}} \int {d\vartheta} e^{-i n \vartheta} \left( \partial_\vartheta \phi(0,\vartheta) - A_\vartheta(0,\vartheta) \right) ~,
\end{eqnarray}
where $n \in \mathbb{Z}/\{0\}$. We thus define the vacuum $|\Omega\rangle$ as the state annihilated by the $a_n$, while acting with the $a_n^\dag$ for either $n > 0$ (or $n<0$) furnishes the lowest (highest) weight $\Delta=1$ UIR. Additional states are created by acting with the operators $\chi$ and $\mathcal{O}_q(T,\vartheta)$. 

One can also consider the model in Euclidean signature, on the two-sphere, whose metric is given by \eqref{eq:eglobal}. The path-integral of interest is now
\begin{equation}
\mathcal{Z}_{\text{BF}} = \int \frac{\mathcal{D} \phi  \mathcal{D} A_\mu  \mathcal{D} B}{\text{vol} \, {\mathcal{G}}}  \, e^{-S_E[ \phi, A_\mu ]} \, e^{i k \int \sqrt{g} B \epsilon^{\mu\nu} F_{\mu\nu}}~,
\end{equation}
where
\begin{equation}
S_E[ \phi, A_\mu ] = \frac{1}{2} \int d^2 x\sqrt{g} g^{\mu\nu} \left( \partial_\mu \phi - A_\mu  \right) \ \left( \partial_\nu \phi - A_\nu  \right)~,
\end{equation}
and $\text{vol}\, {\mathcal{G}}$ is the volume of the gauge group. Path-integrating over $B$ imposes that $A_\mu = \partial_\mu \xi$ is locally pure gauge, and $\xi$ is a non-constant function which we use to parameterize the entire field configuration space of flat-curvature connections. Up to a Jacobian, this imposes $\mathcal{D}A_\mu\rightarrow \mathcal{D}\xi$. One can subsequently eliminate any $\xi$ dependence from the action by a shift in $\phi \to \phi + \xi$. The path-integral over $\xi$ then cancels against $\text{vol}\, {\mathcal{G}}$, save for the zero-mode corresponding to the constant part of the gauge group. This remaining zero-mode is cancelled by the integral over the constant mode of $\phi$. One subsequently computes expectation values of gauge-invariant operators. Employing the results in \cref{sec:discreterevesit} 
we have the Euclidean two-point function
\begin{equation}\label{2ptD1}
\langle   \mathcal{O}_\mu(\Omega)  \mathcal{O}_\mu(\Omega') \rangle = - \frac{1}{4 \pi}\partial_\mu  \partial_{\mu'} \,  \left(\log \frac{u(\Omega,\Omega')}{2}+\alpha\right)~,
\end{equation}
where $\Omega$ and $\Omega'$ are points on the two-sphere, and 
\begin{equation}
u(\Omega,\Omega') = \frac{\cosh(X-X')-\cos(\vartheta-\vartheta')}{\cosh X\cosh X'}~,
\end{equation}
is the invariant length on the two-sphere, in analogy with (\ref{eq:invariantdistance}). The result can be Wick rotated back to dS$_2$, producing an $SL(2,\mathbb{R})$ covariant result. Importantly \eqref{2ptD1} is obtained by taking derivatives of $H_{\text{f}}^{\Delta=1}(u;\alpha)$ written in \eqref{eq:full-sol}, and the necessity to compute a correlation function of gauge-invariant operators kills any dependence on the ambiguous coefficient $\alpha$.

It may be possible to repeat this exercise and gauge the non-constant global shift-symmetries of the scalar in the case of the $t\neq0$ discrete series.

\subsubsection{The \texorpdfstring{$\Delta=2$}{d=2} discrete-series equation}
 Now we briefly comment on an example where the $\Delta=2$ UIR makes an appearance. We will consider a different example in \cref{sec:BFsl2r}, which shares some features with this one. Recall, following \eqref{eq:massiveeom}, that the $\Delta=2$ Casimir equation is: 
 \begin{equation}
     \ell^2\square_{\rm dS}\phi=-2\phi~. 
 \end{equation}
 This equation can be derived in a setting where we couple 2d quantum gravity to a two-dimensional conformal-matter field theory with large positive central charge. Upon integrating out the matter-CFT, the fluctuations of the Weyl factor $\omega$ of the physical metric in the Weyl gauge take the form of a tachyonic scalar in de Sitter (as noted in footnote 6 of \cite{Anninos:2021ene}, see also \cite{Folacci:1996dv}). To be explicit, parameterize the metric $g_{\mu\nu}=e^{2\omega(T,\vartheta)} \tilde{g}_{\mu\nu}$ with $\tilde{g}_{\mu\nu}$ given by \eqref{confmetric}. The constant-curvature equation of motion $R[g]=2/\ell^2$ can be written as: 
\begin{equation}\label{Omega}
\ell^2 \square_{\text{dS}} \omega(T,\vartheta) \approx - 2 \omega(T,\vartheta)~,
\end{equation}
for small $\omega(T,\vartheta)$. As we discuss in section \ref{sec:BFsl2r}, in a somewhat different example, the conformal factor $\omega(T,\vartheta)$ is subject to the residual diffeomorphism constraints, which, in turn, remove the three Euclidean zero-modes of (\ref{Omega}). 

\subsection{Discrete series analogues in \texorpdfstring{dS$_4$}{dS4}}\label{sec:ds4discrete}

Let us now discuss the analogues of the discrete series in four-dimensional de Sitter space. The isometry group of dS$_4$ is $SO(1,4)$, so in addition to conformal dimension $\Delta$, UIRs in this setting are also labeled by a spin $s$ quantum number associated to the $SO(3)$-rotation subgroup of $SO(1,4)$.  A spin $s$ field on dS$_{4}$ has Casimir eigenvalue, directly generalizing the $SL(2,\mathbb{R})$ case: 
\begin{equation}
    \mathcal{C}=\Delta(\Delta-3)+s(s+1)~. 
\end{equation}
There are two possible discrete series analogs in dS$_4$ (see Section 4 of \cite{Sun:2021thf}):
\begin{itemize}
\item\textbf{Exceptional Type I:} In the $s=0$ sector, one finds a collection of discrete scalar UIRs, known as the Exceptional Type I representations, labeled by a discrete conformal dimension $\Delta=3+k$ where $k$, again, is a non-negative integer. One proposal for a free-field-theoretic construction of these representations is the following: Consider the action
\begin{equation}\label{eq:d4tachyon}
    S=-\frac{1}{2}\int d^4 x\,\sqrt{-g}\left[g^{\mu\nu}\partial_\mu \phi\partial_\nu \phi +m_k^2\,\phi^2\right]
\end{equation}
where the mass is tachyonic and satisfies: 
\begin{equation}
    m_k^2\ell^2=-k(k+3)~,
\end{equation}
analogous to \eqref{tachyon}. This field is massless for $k=0$, but is otherwise tachyonic for larger values of $k$. As noted in \cite{Folacci:1992xc}, this theory suffers from a set of unsuppressed Euclidean zero-modes associated to the following symmetry of the action \eqref{eq:d4tachyon} \cite{Bonifacio:2018zex}: 
\begin{equation}\label{eq:shiftsymm}
\phi \to \phi + \lambda_k~, \quad\quad \lambda_k = s_{A_1 A_2 \ldots A_{k}} X^{A_1} X^{A_2} \ldots X^{A_{k}}~,
\end{equation}
where $s_{A_1 A_2 \ldots A_{k}}$ is a real traceless and symmetric constant tensor and the $X^A$ are coordinates on the ambient hyperboloid, as in \eqref{eq:hyperboloid}. For $k=0$, this is the familiar shift symmetry of the free, massless scalar. 
If this symmetry is gauged \cite{Folacci:1992xc,Folacci:1996dv,Bros:2010wa}, as in the $\Delta=1$ example of \cref{Delta1}, then the theory might be amenable to quantization. So far, no one has yet attempted the exercise.

\item \textbf{Exceptional Type II:} For $s\neq 0$, one finds an additional family of discrete UIRs realised as free spin-$s$ (partially) massless gauge fields in dS$_4$. These fields have:
\begin{equation}\label{eq:weightpartialmassless}
    \Delta=2+t~, \qquad\qquad t=0,1,\dots s-1~. 
\end{equation}
The quantity $t$ is called the \emph{depth}. To realize these exceptional series on de Sitter, consider the following field theory of a fully-symmetrized, transverse, traceless, spin-$s$ field which satisfies the following equations of motion \cite{Hinterbichler:2016fgl,Brust:2016zns}: 
\begin{align}\label{eq:massivespinseom1}
   & \left[\square_{\rm dS}-m^2+\frac{s(s-2)-2}{\ell^2}\right]\phi_{\mu_1\dots\mu_s}=0~,
    &\nabla^\nu\phi_{\nu\mu_2\dots\mu_s}&=0~,
    &\phi^\nu_{~~\nu\mu_3\dots\mu_s}&=0~.
\end{align}
At generic values of the mass, this equation propagates $2s+1$ degrees of freedom. In dS$_{4}$, the Higuchi bound for such a spin-$s$ field is given by: 
\begin{equation}
    m^2\ell^2\geq s(s-1)~. 
\end{equation}
Below this value of the mass, one of the St\"{u}ckelberg fields that implement the transverse-tracelessness conditions obtains a ghost-like kinetic term, rendering the theory non-unitary. However, there are a set of special masses, all at or below the Higuchi bound where the theory develops a gauge symmetry that removes the ghosts. These masses are: 
\begin{equation}
    m^2_{s,t}\ell^2=(s-1-t)(s+t)~, \qquad\qquad t=0,1,\dots,s-1~.
\end{equation}
At these special points, known as the \emph{partially massless} points, the equations of motion \eqref{eq:massivespinseom1} develop a symmetry under $\phi\rightarrow\phi+\delta\phi$, where
\begin{equation}
    \delta\phi_{\mu_1\dots\mu_s}=\nabla_{(\mu_{t+1}}\nabla_{\mu_{t+2}}\dots\nabla_{\mu_s}\lambda_{\mu_1\dots\mu_t)}+\dots
\end{equation}
and the additional dots indicate terms with fewer derivatives.\footnote{As we will not need it, we do not provide the full expression for the gauge invariance of \eqref{eq:massivespinseom1}. The interested reader can find it in equation (2.5) of \cite{Hinterbichler:2016fgl}.} The gauge parameter must itself satisfy: 
\begin{align}\label{eq:gaugepinseom}
   & \left[\square_{\rm dS}+\frac{(s-1)(s+2)-t}{\ell^2}\right]\lambda_{\mu_1\dots\mu_t}=0~,
    &\nabla^\nu\lambda_{\nu\mu_2\dots\mu_t}&=0~,
    &\lambda^\nu_{~~\nu\mu_3\dots\mu_t}&=0~.
\end{align}
In total, this gauge symmetry amounts to removing a massive spin-$t$ field's worth of propagating degrees of freedom. Thus the partially massless spin-$s$ field propagates a total of $2(s-t)$
degrees of freedom. Note that the maximal-depth field with $t=s-1$, the field is massless and propagates two polarizations, just like the graviton. At the time of writing, besides Vasiliev theory on dS$_4$ (which has an infinite tower of massless fields), no consistent interacting theory with partially massless fields is known. Vasiliev theory possesses an enormous higher-spin gauge symmetry \cite{Vasiliev:1990en}, one which encompasses the underlying $SO(1,4)$ global symmetry of the dS$_4$ spacetime. So while we might naively think the Vasiliev fields carry conformal dimensions labeled by $\Delta=1+s$, the gauge-invariant Hilbert space only consists of $\Delta=0$ states.
\end{itemize}
In dS$_2$ there is no intrinsic spin -- the $SO(3)$ rotation group is replaced by $SO(1) \cong \mathbb{Z}_2$ with respect to which states can be graded. We may therefore wonder if the discrete series states in dS$_2$ admit  realisations that share features with either the Exceptional Series I or II of dS$_4$, as described above. Although the tachyonic scalar theory seems the most natural to consider -- due to the absence of spin in two-dimensions -- in section \ref{discretesec} we will describe a scenario more closely related to the $s\neq 0$ partially-massless fields of $SO(1,4)$.

\section{Spectral Decomposition}
\label{sec:spectroscopic decompositions}
 In this section, we discuss the K\"all\'en-Lehmann \cite{Kallen:1952zz,Lehmann:1954xi} spectral decomposition of the two-point function for general interacting scalar fields in dS$_2$. This has been discussed in many works before us \cite{Bros:1990cu,Bros:1995js,Bros:2009bz,Hollands:2011we,Epstein:2012zz,Hogervorst:2021uvp,DiPietro:2021sjt,Loparco:2023rug}. Our goal here is to focus on the 
  contributions from  discrete series states $D^\pm_\Delta$.  

 To set the stage, let us quickly review the spectral decomposition of the two-point function of an interacting scalar field $\phi$ in $d$-dimensional flat space:
 \begin{equation}
     G_{\rm i}(x-y)=\bra{\Omega}\phi(x)\phi(y)\ket{\Omega}~.
 \end{equation}
 The subscript ``$\rm{i}$'' denotes that it is ``interacting.''
Under very general assumptions,  this correlation function can be expressed as \cite{Kallen:1952zz,Lehmann:1954xi,Weinberg:1995mt}
\begin{equation}\label{eq:KallenLehmanFlat}
    G_{\rm i}(x-y)=\int_0^\infty d\mu^2 \rho_{\phi}(\mu^2) \int \frac{d^d p}{(2\pi)^d} \frac{e^{ip\cdot (x-y)}}{p^2+\mu^2}~.
\end{equation}
In this form,  unitarity of the two-point function demands $\rho_\phi(\mu^2)>0$ and,
\begin{equation}
    \int_0^\infty d\mu^2\rho_\phi(\mu^2)=1~.
\end{equation}
The usefulness of this representation stems from the fact that the interacting correlation function $G_{\rm i}(x-y)$ can be expressed, up to an undetermined function, out of free-field propagators, under any circumstance; allowing us to anchor our intuition on free-field theory, and free-field excitations.  
The quantity $\rho_\phi(\mu^2)$ is called the \emph{spectral density} of the interacting field $\phi$, and is chock-full of information about the underlying theory---namely it has within it data about the overlap between $\phi(x)\ket{\Omega}$ and any state in the Hilbert space. Hence, for example, it can tell us if the field $\phi(x)$ creates single-particle excitations, or if it is composite. 

To see this, note that spectral function governs the analytic properties of the two-point amplitude $G_{\rm i}(p)=\int_0^\infty d\mu^2  \frac{\rho_\phi(\mu^2)}{p^2+\mu^2}$~, which will have poles at the single-particle states created by $\phi(x)$ and a branch cut starting at the first multi-particle state. Hence if $G_{\rm i}(p)$ only has a single pole, we have an invariant notion of a `free-field.' Alternatively we define $\phi(x)$ as composite if $G_{\rm i}(p)$ only has branch cuts and no poles. 

On general curved backgrounds, we can't use analyticity arguments, as we do in flat space, to give an invariant notion to the meaning of single-particle (poles) and multi-particle (branch-cuts) states. But hope is not lost. de Sitter is a maximally-symmetric spacetime, and we have the constraining power of group theory to guide us---so while we may not have a notion of multi-\emph{particles}, we do have tensor product representations, as we discuss now. 

One might then think, following \eqref{eq:EuclideanRep}, that the natural generalization of \eqref{eq:KallenLehmanFlat} to Euclidean dS$_2$ is: 
\begin{equation}
    G_{\rm i}(u)=\frac{1}{4\pi}\int_{\frac{1}{4\ell^2}}^\infty dm^2\rho_\phi(m^2) \sum_{L=0}^{\infty}\frac{2L+1}{L(L+1)+m^2\ell^2}P_L(1-u)~,
\end{equation}
but, this formula, for the subtleties described in \cref{sec:discreterevesit}, misses contributions from states in the discrete series $D_\Delta^\pm$, which are required to appear by group theoretic considerations \cite{Dobrev:1977qv}.

\subsection{Spectral decomposition: a proposal}
To derive the spectral decomposition, we start with the identity operator on the field-theoretic Hilbert space  of a general interacting scalar field theory: 
\begin{equation}\label{identity}
\mathds{1}= \ket{\Omega}\bra{\Omega} + \sum_{\Delta,n} \frac{\ket{\Delta,n}\bra{\Delta,n}}{\braket{\Delta,n|\Delta,n}} \, . 
\end{equation}

The state $|\Omega\rangle$ is the (Hadamard) Bunch-Davies vacuum, and the states $|\Delta,n\rangle$ carry quantum numbers, respectively, under the quadratic Casismir $\mathcal{C}$ and the action of rotation $L_0$, as in \eqref{eq:state}.\footnote{More generally, there is also a sum over states transforming with either positive (even) or negative (odd) action under the operator $e^{2\pi i L_0}$. The scalar case we are studying leaves only states with even action under this transformation. The odd states are relevant to the case of fermionic fields as considered in \cite{Pethybridge:2021rwf,Schaub:2023scu}, to which our analysis can be extended. We note that the complementary series are non-unitary in the odd case, and so would not appear in the decomposition of the identity.} The normalization factor must be included for a generic UIR, since only the principal series can simultaneously be made unit normalized while also faithfully transforming under the action of the ladder operators \eqref{eq:state}, see \cref{App:Repthy}.

Let us now consider the two-point function of a general interacting scalar field in dS$_2$:
\begin{equation}
G_{\rm i}(X,Y)\equiv \langle\Omega|\phi(X)\phi(Y)|\Omega\rangle~.
\end{equation}
Inserting the identity operator, this implies: 
\begin{equation}
G_{\rm i}(X,Y) =\langle\phi\rangle^2+ \sum_{\Delta,n} \frac{\bra{\Omega}\phi(X)\ket{\Delta,n}\bra{\Delta,n} \phi(Y)\ket{\Omega}}{\braket{\Delta,n|\Delta,n}} \, .
\label{eqn:scalar2pt}
\end{equation}
Although we have included the possibility of a vacuum expectation value for the field $\phi$, from now on, we will assume that $\langle\phi\rangle\equiv\bra{\Omega}\phi(X)\ket{\Omega}=0$ in the Bunch-Davies state. We will use an additional fact about $SL(2,R)$ representation theory, namely that the operator $\mathds{1}$ \emph{excludes the complementary series} \cite{Dobrev:1977qv}. Using this:
\begin{multline}
G_{\rm i}(X,Y)=\sum_{t=0}^\infty\left\lbrace\sum_{ n\neq\{-t,..,t\}}\frac{\Gamma(|n|+1+t)}{\Gamma(|n|-t)}\bra{\Omega}\phi(X)\ket{1+t,n}\bra{1+t,n}\phi(Y)\ket{\Omega}\right\rbrace\\
+\int_{-\infty}^{+\infty} d\nu\left\lbrace\sum_{n\in \mathbb{Z}}\left\langle\Omega\right\vert\phi(X)\left\lvert\tfrac{1}{2}\left(1 + i \nu\right), n\right\rangle\left\langle\tfrac{1}{2}\left(1 + i \nu\right), n\right\vert\phi(Y)\left\vert \Omega\right\rangle \right\rbrace~.\label{eq:decomp1}
\end{multline}
In the above equation, the first line includes the contributions from discrete series $D_{1+t}^\pm$ states (and we have combined the highest and lowest weight UIRs into a single sum), and the term out front is the normalization factor (see \cref{App:Repthy}). The second line contains the contributions coming from principal series states $\pi_\nu$.\footnote{One can incorporate the complementary series $\gamma_\Delta$ by appropriately shifting the principal series contour. }

\paragraph{Principal series contribution:} We now proceed to write down the principal series contribution to \eqref{eq:decomp1}. We have already done most of the work in \cref{sec:freefock}. This argument is similar to one that appeared in \cite{Hogervorst:2021uvp}. Given that our Fock space is built atop the Bunch-Davies state $\ket{\Omega}$ the symmetries of the problem require: 
\begin{align}
    \left\langle\Omega\right\vert\phi(X)\left\lvert\tfrac{1}{2}\left(1 + i \nu\right), n\right\rangle &= c(\nu) \phi_n^{E,\Delta=\tfrac{1}{2}(1+i\nu)}(X)~,\\ \left\langle\tfrac{1}{2}\left(1 + i \nu\right), n\right\vert\phi(Y)\left\vert \Omega\right\rangle&=c(\nu)^* \phi_n^{*E,\Delta=\tfrac{1}{2}(1+i\nu)}(Y)~,
\end{align}
where the mode functions are given in \eqref{eq:eucmode} and subsequent equations. The undetermined coefficient $c(\nu)$ contains information about the field $\phi$ and its interactions. Now, we can use \eqref{eq:2pteuclideandecom} to express: 
\begin{equation}
\sum_{n\in \mathbb{Z}}\left\langle\Omega\right\vert\phi(X)\left\lvert\tfrac{1}{2}\left(1 + i \nu\right), n\right\rangle\left\langle\tfrac{1}{2}\left(1 + i \nu\right), n\right\vert\phi(Y)\left\vert \Omega\right\rangle = \rho_\phi(\nu)\, G_{\rm f}^{\Delta=\tfrac{1}{2}(1+i\nu)}\left(u(X,Y)\right)
\end{equation}
where $G_{\rm f}^\Delta(u)$ is given in \eqref{eq:massivefreeg} and $\rho_\phi(\nu)\equiv |c(\nu)|^2\geq 0$.

\paragraph{Discrete series contribution:} Identifying the discrete series mode functions is a subtle problem. As described in \cref{sec:discreterevesit}, this is because, on this representation and in stark contrast to the principal series, the corresponding Euclidean Laplacian operator for the discrete series is non-invertible. The non-invertibility of the Euclidean differential operator indicates a gauge redundant structure in the discrete series sector, and suggests \cite{Bros:2010wa} we must modify the Klein-Gordon representation of the Casimir operator. A potential modification is described at length in \cite{Bros:2010wa}, but we will not review it here. We  simply quote the following: 
\begin{equation}
    \sum_{ n\neq\{-t,..,t\}}\frac{\Gamma(|n|+1+t)}{\Gamma(|n|-t)}\bra{\Omega}\phi(X)\ket{1+t,n}\bra{1+t,n}\phi(Y)\ket{\Omega}=\sigma_\phi(t;\alpha_t)\, H_{\rm f}^{\Delta=1+t}\left(u(X,Y);\alpha_t\right)~.
\end{equation}
The left hand side of this equation suggests that we may be able to prove that $\sigma_\phi(t;\alpha_t)\ge 0$, but since the functions $H_{\rm f}^{\Delta=1+t}(u;\alpha)$ are not positive on the Euclidean section, at this stage we can not impose any positive conditions on the density $\sigma_\phi$. Moreover, we have left in the inherent ambiguity with respect to the choice of $\alpha_t$. 

\paragraph{Final answer:} We are now ready to write down the full K\"all\'en-Lehmann representation for an interacting scalar field $\phi$ in two dimensions: 
\begin{equation}
    G_{\rm i}(X,Y)=\sum_{t=0}^\infty\sigma_\phi(t;\alpha_t)\, H_{\rm f}^{\Delta=1+t}\left(u(X,Y);\alpha_t\right)+\int_{-\infty}^\infty d\nu\,\rho_\phi(\nu)\, G_{\rm f}^{\Delta=\tfrac{1}{2}(1+i\nu)}\left(u(X,Y)\right)~.
\end{equation}
A remark is in order. While we were not able to constrain the sign of $\sigma_\phi$, nor its dependence on the ambiguous parameters $\alpha_t$, it must be that the right hand side leads to a unitary two point function for an interacting scalar theory in de Sitter. This means that the total sum of discrete and principal series contributions must result in a two point function with appropriate properties: positivity on the Euclidean section as well as a positive coincident point limit that grows at most logarithmically in $u$. While the density on the discrete series may or may not be sign definite, this certainly constrains how they may contribute to the final answer.

Given the discussion around \eqref{eq:latetimediscreteseries2pt}, it is interesting to note that, if present, the discrete series contributions will dominate over the principal series contributions at the late-time boundary. While this seems bizarre, it tells us that we need to either search for a general principle that would exclude the contribution of the discrete series in the two-point function of scalars in dS$_2$, or understand how and why they may arise in interacting scalar field theories. We have no way to discount the discrete series representations based on group theoretic arguments alone, unless the de Sitter group is gauged---as is the case when gravity is turned on. In quantum field theory on a rigid dS$_2$ background, the discrete series states are, without a doubt, present in the tensor product Hilbert space of two species of particles, even in the absence of interactions. We take this as an invitation to try and understand the physical nature of these vexing representations. 

\section{Discrete series operators in BF theory on \texorpdfstring{dS$_2$}{ds2}}\label{discretesec}

This section is concerned with BF gauge theories on dS$_2$. Our interest in these theories stems from the fact that, in a suitable gauge, the equation of motion
\begin{equation}
    \square_{\rm dS}\phi=-t(t+1)\phi~,\qquad\qquad  t\in \mathbb{Z}
\end{equation}
arises quite naturally. One may use this to conclude that the discrete series UIRs have a role to play in BF theories on dS$_2$, but this is too quick. After all, BF theories are topological, meaning they are insensitive to the background on which they live, and there should be no imprint of the dS$_2$ background once we have appropriately quantized the theory. However, there is a sense in which the discrete series UIRs are realized at the level of the pre-Hilbert space of the field operators, in the sense of \eqref{confOps}. That is, the discrete series UIRs are realized by the linearized field equations of the $SL(N,\mathbb{R})$ BF-theory, with $N\ge2$, as we will demonstrate below. These field operators do not survive the imposition of the gauge constraints, but if, for example,  we add a boundary to dS$_2$, for example along a worldline in the static patch, we can imagine breathing life into these modes.  

Moreover, semiclassically,  the $SL(N,\mathbb{R})$ gauge theory is a two-dimensional version of higher-spin theory  \cite{Alkalaev:2013fsa,Alkalaev:2019xuv,Alkalaev:2020kut,Grumiller:2013swa} whose gravitational subsector is governed by the $SL(2,\mathbb{R})$ embedding inside of $ SL(N,\mathbb{R})$. The field operators play a crucial role in formulating the gauge theory, but they are subject to the gauge constraints. Imposing these gauge constraints, for the theory quantized on a spatial circle, cuts down the size of the pre-Hilbert space and leads to a physical (gauge-invariant) Hilbert space absent of any non-trivial $SL(2,\mathbb{R})$ representations. Nonetheless, the presence of the discrete series UIR, at the level of the pre-gauged operator algebra, plays an important role in characterizing the entanglement structure of the theory \cite{Kitaev:2005dm,Levin:2006zz}, as well as providing a convenient basis for expressing the wavefunctionals  \cite{Henneaux:1985nw,Isler:1989hq,Chamseddine:1989yz}. These pre-Hilbert space states may offer a guiding principle for a microphysical completion of the theory, as in \cite{Anninos:2011ui,Anninos:2017eib}. 

The setup in this section may be compared with an analogous one in dS$_4$. At the free level, the discrete series UIR of $SO(1,4)$ may be realized as the single-particle Hilbert space of (partially) massless gauge fields of spin $s$, with $s=1,2,\ldots$ (along with their fermionic counterparts \cite{Letsios:2020twa})  (see \cref{sec:ds4discrete}). Turning on interactions among these fields, which include the  linearized graviton for $s = 2$, requires that we gauge any residual symmetries, {\it including} the $SO(1,4)$ isometry group of dS$_4$, which is a subgroup of the diffeomorphism group \cite{Higuchi:1991tk,Higuchi:1991tm}. As such, non-trivial $SO(1,4)$ UIRs are projected out of the physical Hilbert space of the interacting theory. On its own, the  free theory leads to a somewhat misleading picture. For example: A model that encapsulates these ideas is the Vasiliev theory with $\Lambda>0$ \cite{Vasiliev:1990en,Vasiliev:1999ba}, which has an infinite tower of interacting higher-spin gauge fields in dS$_4$.  The  gauge-invariant Hilbert space of the Vasiliev theory was argued in  \cite{Anninos:2017eib} to be dramatically reduced at the microscopic level. Related remarks for the theories at hand will be given in  section \ref{sec:outlook}.

\subsection{Abelian BF-theory}
The Abelian BF-theory with compact $U(1)$ gauge group is governed by the action
\begin{equation}\label{BFaction}
S_{\text{BF}} =    \frac{k}{4\pi}\int d^2x\sqrt{-g} \, B \, \epsilon^{\mu\nu} F_{\mu\nu}~, \quad\quad F_{\mu\nu} = \partial_\mu A_\nu - \partial_\nu A_\mu~.
\end{equation}
Here, $B$ is a real compact scalar field $B\cong B+2\pi$, and $\epsilon^{\mu\nu}$ is the antisymmetric tensor with $\epsilon^{T\vartheta}=+{1}/{\sqrt{-g}}$. The theory is invariant under gauge transformations $A_\mu \to A_\mu + \partial_\mu \omega$ with $\omega$ a compact scalar of radius $2\pi$. 
Although we will consider the theory on the spacetime \eqref{confmetric}, the action \eqref{BFaction} is independent of the metric. Also note that similar to the case of Chern-Simons theory, the parameter $k$ in front of the action is quantized. 

The classical equations of motion are given by
\begin{equation}\label{BFeoms}
\partial_\mu B = 0~, \quad\quad  \square_{\text{dS}} \phi = 0~,
\end{equation}
where we have picked the Lorenz gauge: $A^\mu \equiv \epsilon^{\mu\nu} \partial_\nu \phi$.  We note that the constant part of $\phi$ is absent from the field configuration space, as it does not affect the physical field $A_\mu$. The second equation in (\ref{BFeoms}) is equivalent to that of a free massless scalar field in dS$_2$, i.e. \eqref{eq:massiveeom} with $m^2 =0$, but now with the zero-mode removed by construction. The solutions are
\begin{equation}\label{oscphi}
{\phi}(T,\vartheta) = \chi \frac{T}{2\pi}+  \sum_{n\in\mathbb{Z}/\{0\}} \frac{e^{i n \vartheta}}{2\pi} \left( {\alpha}_n \frac{\sin n T}{n}  + {\beta}_n \cos n T  \right) \underset{\lim_{T\rightarrow0^-}}{\approx}  \chi \frac{T}{2\pi}+   \sum_{n\in\mathbb{Z}/\{0\}} \frac{e^{i n \vartheta}}{2\pi} \left( T {\alpha}_n    +{\beta}_n  \right)~,
\end{equation}
subject to reality conditions ${\alpha}_n^* = {\alpha}_{-n}$ and ${\beta}_n^* = {\beta}_{-n}$, and we have singled out $\alpha_0\equiv\chi$. Recall that $\beta_0$ is not included in the above sum because the gauge field $A_\mu$ is insensitive to the constant mode of $\phi$.  Upon quantization, $B$ and $\phi$ are promoted to quantum operators, and (\ref{BFeoms}) become operator equations. The operator ${\chi}$ is associated to the Wilson loop and forms a canonical pair with the constant mode of ${B}$. Under the decomposition (\ref{oscphi}), one might conclude that the modes $\alpha_n$ and $\beta_n$  can be organised in terms of the $D^\pm_{\Delta}$ UIRs with $\Delta=1$. $\chi$ and  the constant mode of ${B}$, which are the only operators that survive the gauge constraints, furnish a singlet representation of $SL(2,\mathbb{R})$. 

Having discussed the transformation properties of the operators $B$ and $\phi$, we note that it is {\it not} the case that the physical state space furnishes the $\Delta=1$ UIRs. Starting from the above-mentioned pre-Hilbert space,  properly imposing the gauge constraints will result in an enormous reduction of the physical Hilbert space. 

\paragraph{Hilbert space of Abelian BF-theory.} There are many ways to quantize this theory, but we will begin with a way that quickly identifies the Hilbert space \cite{Seiberg:Fun}. This will be done in temporal gauge, where we set $A_T=0$. This means we must also impose the constraint generated by $A_T$ at the level of the action, namely: 
\begin{equation}\label{constraintB}
    \partial_\vartheta B=0~. 
\end{equation}
Let us define $\chi\equiv\oint  d\vartheta A_\vartheta$, which is the piece of $A_\vartheta$ invariant under the residual gauge freedom $A_\vartheta \to A_\vartheta+ \partial_\vartheta \omega(\vartheta)$, and $b= \frac{1}{2\pi}\oint d\vartheta B$. 
Thus our gauge-fixed action is a simple quantum mechanical model:
\begin{equation}
    S_{\rm BF}^{\rm g.f.}=\frac{k}{2\pi}\int dT \,b\,\dot{\chi}~.
\end{equation}
 Recall that by flux quantization $b\cong b+2\pi$ and moreover, by compactness of the gauge group $\chi\cong \chi+2\pi$. Thus the fields $b$ and $\chi$ are canonically conjugate pairs, with commutation relation: 
\begin{equation}\label{eq:commrelationchib}
    [\chi,b]=\frac{2\pi i}{k}~. 
\end{equation}
Because of the compactness of the fields, the well-defined observables are $e^{ib}$ and $e^{i\chi}$. We can work in an eigenspace of the operator $e^{ib}$, which is spanned by square-integrable wavefunctions $\Psi_n(\chi) = e^{i n \chi}/\sqrt{2\pi}$ with $n \in \mathbb{Z}$. Recalling that the operator $b = -\tfrac{2\pi i}{k} \partial_\chi$, conjugate to $\chi$, is also compact, such that $e^{i b} = e^{i (b+ 2\pi q)}$ with $q \in \mathbb{Z}$. We require that
\begin{equation}
-\tfrac{2\pi i}{k} \partial_\chi \Psi_n(\chi) \cong  \left( -\tfrac{2\pi i}{k} \partial_\chi + 2\pi  \right) \Psi_n(\chi)~.
\end{equation}
The above expression informs us that $n$ should in fact be defined modulo $k$ such that the correct range is $n = 0,1,\ldots,k-1$. Thus the gauge-invariance significantly reduces the Hilbert space, rendering a naively infinite Hilbert space into one that is $k$-dimensional, and therefore incapable of carrying the $\Delta = 1$ UIR. Going back to our original gauge-theoretic variables, the Hilbert space is spanned by the wavefunctionals
\begin{equation}\label{schroU1}
\Psi_n \left[ A_\mu \right] = \frac{1}{\sqrt{2\pi}} \, e^{i n \oint d\vartheta A_\vartheta }~, \quad\quad n = 0,1,\ldots,k-1~,
\end{equation}
where $B = -i \delta / \delta A_\vartheta$.

\subsection{\texorpdfstring{$SL(2,\mathbb{R})$}{SL(2,R)} BF-theory as JT gravity} \label{sec:BFsl2r}
We now proceed to $SL(2, \mathbb{R})$ BF-theory. The theory is semi-classically equivalent to a JT gravity \cite{Henneaux:1985nw,Isler:1989hq,Chamseddine:1989yz}
built from gravitational degrees of freedom including a two-dimensional metric $g_{\mu\nu}$ and a scalar $\phi$, and it is in these variables that we will analyze the model. The theory is governed by the following action:
 \begin{equation}
     S_{\text{JT}} = \frac{1}{16 \pi G}\int_{\mathcal{M}} d^2 x\sqrt{-g} \phi \left(R- \frac{2}{\ell^2} \right) 
     \label{def:SJT}
 \end{equation}
 where we take $\ell^2$ to be positive such that the theory admits dS$_2$ solutions. 
 As described in  \cite{Isler:1989hq,Chamseddine:1989yz,Iliesiu:2019xuh}, we may recast the above action as an $SL(2,\mathbb{R})$ gauge theory (in the first order formalism) by combining the zweibein $e^a_\mu$ and the spin-connection $\omega_{01}$ into an $SL(2,\mathbb{R})$ gauge field, while collecting the field $\phi$ and additional auxiliary fields (imposing the torsionless condition) into the adjoint-valued scalar $B$.  Though related, neither the $SL(2,\mathbb{R})$ gauge symmetry, nor any subgroup thereof,  of the BF-theory  (\ref{def:SJT}) is to be identified with the $SL(2,\mathbb{R})$ isometry of the dS$_2$ vacuum which is in fact slightly broken by the general classical solution.
 
 The equations of motion stemming from (\ref{def:SJT}) are
 \begin{align}
 	\delta\phi:&\hspace{1cm}R - \frac{2}{\ell^2}= 0  \label{eqn:geom}\, ,\\
     \delta g_{\mu\nu}:& \hspace{1cm} \left(\ell^2 \nabla_\mu \nabla_\nu +  g_{\mu\nu}\right) \phi = 0  \, . \label{eqn:phieom}
 \end{align}
In global coordinates, the solution is
\begin{equation}\label{JTsolns}
\frac{ds^2}{\ell^2} =     \frac{ -dT^2 + d\vartheta^2 }{\sin^2 T}~, \quad\quad  \phi =  \gamma_0 \cot T +  \left(\gamma_{1} e^{i \vartheta } + \gamma_{-1} e^{-i \vartheta } \right) \csc T~.
\end{equation}
The breaking of de Sitter invariance is evidenced by a non-trivial profile for the dilaton $\phi$.\footnote{Note that taking the trace of \eqref{eqn:phieom} leads to
 \begin{equation}
     \ell^2 \square_{\text{dS}} \phi(\vartheta^+, \, \vartheta^-)  = -2  \phi(\vartheta^+, \, \vartheta^-)  \, , \label{eqn:phieomfinal}
 \end{equation}
 which is a tachyonic free scalar equation with $m^2\ell^2 = -2$. However, since $\phi$ is subject to a symmetric tensor's worth of equations, there are only three linearly independent solutions. We will make further comments on this observation below, but take note that this is precisely the number of Euclidean zero-modes admitted by the Euclidean continuation of \eqref{eqn:phieomfinal}.} 

We will parallel the previous section's discussion on the Abelian BF theory. First we will pick a gauge where the $\Delta=2$ discrete series equation of motion appears, allowing for a description of this UIR, at least at the level of the pre-Hilbert space. Afterwards we will explain how imposing the gauge constraints significantly cuts down the size of the Hilbert space. To this end, we parameterize the global dS$_2$ geometry in Weyl gauge as:
\begin{equation}\label{weyl}
\frac{ds^2}{\ell^2} =  e^{2\omega(\vartheta^+, \, \vartheta^-)} \,  \frac{4 \, d\vartheta^+ d\vartheta^-}{\sin^2 (\vartheta^+- \vartheta^-)}~,
\end{equation}
where the solution in  (\ref{confmetric}) corresponds to $\omega(\vartheta^+, \, \vartheta^-)=0$ with $T = \vartheta^+ - \vartheta^-$ and $\vartheta = \vartheta^+ + \vartheta^-$. Equation (\ref{eqn:geom}) governing the Weyl factor is then 
\begin{equation}\label{eqn:r2}
\ell^2 \square_{\text{dS}} \omega(\vartheta^+, \, \vartheta^-) = 1 - e^{2\omega(\vartheta^+, \, \vartheta^-)} \approx - 2 \omega(\vartheta^+, \, \vartheta^-)  ~,
\end{equation}
where in the second expression we have expanded for small $\omega(\vartheta^+, \, \vartheta^-)$ and $\square_{\text{dS}}$ is the Laplacian with respect to the metric \eqref{JTsolns}, which is the same as \eqref{weyl} with $\omega(\vartheta^+, \, \vartheta^-)=0$. As noted in \cite{Anninos:2018svg}, this linearized equation governing $\omega(\vartheta^+, \, \vartheta^-)$ is the equation of a free scalar on a fixed dS$_2$ background with a {tachyonic mass} $m^2\ell^2 = -2$, corresponding to the discrete series equation for $\Delta=2$. The non-linear nature of equation \eqref{eqn:r2} can be thought of as self-interactions for the field $\omega$. Therefore, the Weyl gauge here should be thought of as paralleling the Lorenz gauge in the Abelian BF theory. 

In two-dimensions, a local region of the geometry is determined entirely by the Ricci scalar, so up to global effects, the general solution space to  (\ref{eqn:r2}) must be the set of diffeomorphisms preserving the Weyl gauge (\ref{weyl})
\begin{equation}\label{eq:genweylsoln}
e^{2\omega(\vartheta^+, \, \vartheta^-)} = \sin^2 (\vartheta^+ - \vartheta^-) \frac{\partial_+ f(\vartheta^+) \partial_- g(\vartheta^-)}{\sin^2 \left(f(\vartheta^+) - g(\vartheta^-)\right)}~, 
\end{equation}
and 
\begin{equation}\label{eq:genphisol}
    \phi =  \gamma_0 \cot\left(f(\vartheta^+)-g(\vartheta^-) \right) +  \left\lbrace\gamma_{1} e^{i \left[ f(\vartheta^+)+g(\vartheta^-) \right] } + \gamma_{-1} e^{-i \left[f(\vartheta^+)+g(\vartheta^-) \right] } \right\rbrace \csc \left(f(\vartheta^+)-g(\vartheta^-) \right)~.
\end{equation}
with $f(\vartheta^+)$ and $g(\vartheta^-)$  smooth functions. The classical solution space is therefore labeled by the modes of the functions $f$ and $g$, as well as the three parameters $\gamma_i$. In the Abelian BF theory, the zero-mode of the scalar $\phi$ in Lorenz gauge was projected out, allowing us to identify candidate operators that might furnish a $\Delta=1$ UIR---at least at the level of the pre-Hilbert space. Our task now is to identify a similar mechanism in this setting, but for $\Delta=2$, which requires projecting out three zero-modes. 

Which are the zero-modes that must be projected out? The natural candidates are either the three Killing symmetries of the spacetime, or the  three non-isometric conformal-Killing transformations. Of these the former are physical, since the de Sitter invariance of the background is broken by the dilaton profile, whereas the latter correspond to field redefinitions of Weyl factor $\omega$, and therefore do not describe physically inequivalent data. We will now show precisely how this plays out. Starting from the linearized equation \eqref{eqn:r2} and using the variables $T = \vartheta^+ - \vartheta^-$ and $\vartheta = \vartheta^+ + \vartheta^-$, we must solve:
\begin{equation}\label{linomega}
\sin^2 T \left( -\partial_T^2 + \partial_\vartheta^2 \right) \omega(T, \vartheta) = - 2 \omega(T,\vartheta)~.
\end{equation}
The solutions may be expressed in terms of angular modes around the spatial $S^1$, labelled by $n \in \mathbb{Z}$. For $n \in \mathbb{Z}/\{-1,0,1\}$, the solutions are
\begin{equation}\label{delta2}
\omega(T, \vartheta) =  \sqrt{|\sin T|}  \, \sum_{n \in \mathbb{Z}/\{-1,0,1\}} e^{i n \vartheta} \left( \alpha_n \sqrt{\frac{2}{\pi}}   Q_{  |n|-1/2}^{{3}/{2}}(\cos T) +\beta_n  \sqrt{\frac{\pi}{2}} P_{  |n|-1/2}^{{3}/{2}}(\cos T) \right)~,
\end{equation}
with $\alpha^*_n = \alpha_{-n}$ and $\beta^*_n = \beta_{-n}$. Near the future boundary $T \to 0^-$, we have two characteristic behaviors given by $\omega(T, \vartheta) \sim T^{-1}$ and $\omega(T, \vartheta) \sim T^2$. Concretely,
\begin{equation}
\omega(T, \vartheta) \approx    \sum_{n \in \mathbb{Z}/\{-1,0,1\}} e^{i n \vartheta} \left( \alpha_n \, \frac{|n|(n^2-1)}{3} \,T^2+ \frac{\beta_n}{ T} \right)~.
\end{equation}
We must treat the $n \in \{-1,0,1\}$ modes separately. These are given by
\begin{multline}\label{omega3}
   \omega_{\{-1,0,1\}}(T, \vartheta)  = \left\lbrace  \alpha_0 \left(1-T\cot T\right) + \tfrac{1}{2}\left( \alpha_{-1} e^{-i \vartheta } + \alpha_1 e^{i \vartheta }  \right) \left(T\csc T-\cos T\right)\right. \\\left.+\beta_0 \cot T  +\left( \beta_{-1} e^{-i \vartheta } + \beta_1 e^{i \vartheta }  \right) \csc T   \right\rbrace \, .
\end{multline} 
Near $T\sim 0^-$, these modes behave as
\begin{equation}
   \omega_{\{-1,0,1\}}(T, \vartheta)  \underset{T\rightarrow0^-}{\approx}  \sum_{n \in \{-1,0,1\}} e^{i n \vartheta} \left( \alpha_n \, \frac{T^2}{3}+ \frac{\beta_n}{ T} \right)\, .
\end{equation} 
Due to the reality conditions on $\omega(T, \vartheta)$, $\beta_0$ and $\alpha_0$ are real valued. Which of these modes correspond to Killing symmetries, and which correspond to conformal Killing transformations? 

To determine this, let us expand the nonlinear solution \eqref{eq:genweylsoln} near the future boundary, where $T\rightarrow 0^-$, and we will furthermore choose the slice $g(x) = f(x)\equiv h(2x)$. On this slice one finds : 
\begin{equation}\label{schw}
  e^{2\omega(\vartheta^+, \, \vartheta^-)}|_{bdy}= 1 + \frac{T^2}{3}\left(2\,\text{Sch}\left(h\left(\vartheta\right),\vartheta\right)+4h'\left(\vartheta\right)^2-1\right) \ldots~.
\end{equation}
This equation is invariant under the transformation 
\begin{equation}
    \tan h(\vartheta)\rightarrow\frac{a \tan h(\vartheta)+b}{c\tan h(\vartheta)+d}~, \qquad\qquad ad-bc=1~,
\end{equation}
therefore, we see that there are three modes that leave the late time form of the metric invariant.  Moreover, these modes are proportional to $T^2$ at late times, suggesting that the $\alpha_n$ for $n=\{-1,0,1\}$ are associated to the (broken) Killing symmetries, and therefore should not be discarded.

A clearer exposition, making contact with the discussion around the tachyonic equation of motion for the $\Delta=2$ discrete series UIR, can be achieved by repeating the above analysis at the linearized level. To linear order, the solutions \eqref{eq:genweylsoln} are 
\begin{equation}\label{eq:linweyl}
    2\omega(\vartheta^+,\vartheta^-) = \epsilon\left\lbrace\delta g'(\vartheta^-)+\delta f'(\vartheta^+)-2 \delta f(\vartheta^+) \cot
   (\vartheta^+-\vartheta^-)+2 \delta g(\vartheta^-) \cot
   (\vartheta^+-\vartheta^-)\right\rbrace~,
\end{equation}
where we have taken $f(\vartheta^+) = \vartheta^+ + \epsilon \delta f(\vartheta^+)$ and $g(\vartheta^-) = \vartheta^- + \epsilon\delta g(\vartheta^-)$. Taking the late-time limit $T\rightarrow 0$ along the slice $ \delta g(x)=\delta f(x) \equiv\delta h(2x)$, we gain an equation for the Weyl mode in terms of the linearized Schwarzian derivative: 
\begin{equation}
    \omega \approx \frac{2}{3} T^2\epsilon\left(\delta h'\left(\vartheta\right) + \delta h''' \left({\vartheta}\right)\right)~.
\end{equation}
Expanding $\delta h$ in modes $\delta h(\vartheta)=\sum_n c_n e^{i n\vartheta}$, we see that the modes with $n=\{-1,0,1\}$ do not change the Weyl factor at late times.

\paragraph{Conformal Killing vectors.} We now expect the $\beta_n$ modes for $n=\{-1,0,1\}$ to be associated with conformal Killing transformations. These are field redefinitions of $\omega$, and therefore are not in the physical phase space, leading us to discard them. To see how this works out, we study the conformal Killing equation:
\begin{equation}\label{cke}
\nabla_\mu \xi_\nu +\nabla_\nu  \xi_\mu = g_{\mu\nu}\nabla_\rho \xi^\rho ~.
\end{equation}
For the case of dS$_2$ in global coordinates (\ref{confmetric}) there are six globally well-defined conformal Killing vectors, smooth with respect to the Euclidean continuation to the two-sphere. Of these, three are divergence-free Killing vector fields that do not alter the original metric, and are essentially the three $\alpha_{-1,0,1}$ zero-modes identified earlier. The remaining three are {conformal} Killing vector fields that non-trivially transform the metric Weyl factor.
Contracting the conformal Killing equation (\ref{cke}) with a covariant derivative yields
\begin{equation}
\ell^2 \square_{\text{dS}}\xi_\nu  = -\xi_\nu ~,
\end{equation}
 from which it follows that
 \begin{equation}
\ell^2 \square_{\text{dS}}\nabla_\mu\xi^\mu = - 2\nabla_\mu \xi^\mu~.
 \end{equation}
Upon identifying $\omega = \nabla^\mu \xi_\mu$, this is precisely the linearized form of \eqref{linomega}. The three globally well-defined conformal Killing vector fields are given by (labelling vector fields as $V\equiv V^\mu\partial_{\mu}$)
 \begin{equation}
 \xi_1  = \partial_T~, \quad   \xi_2  = \cos T \sin \vartheta\partial_T + \cos\vartheta \sin{T} \partial_\vartheta~, \quad
 \xi_3 = \cos T \cos\vartheta \partial_T - \sin \vartheta \sin{T} \partial_\vartheta~. 
 \end{equation}
 Their corresponding, non-vanishing, divergence is
 \begin{equation}
 \omega=\nabla_\mu \left(a_1 \, \xi^\mu_1 +a_2 \, \xi^\mu_2 + a_3 \, \xi^\mu_3 \right) = \beta_0 \cot T+  \left( \beta_{-1} e^{-i \vartheta } + \beta_1 e^{i \vartheta } \right) \csc T\, ,
 \end{equation}
where $\beta_0 = -2 a_1$, $\beta_{\pm1} =  -a_3 \mp i a_2$. These are the $\beta_{\pm1}$ and $\beta_0$ modes in (\ref{omega3}), which we can now interpret as the subset of globally well-defined conformal Killing vector fields.
 We thus interpret the three $\beta_n$ modes of \eqref{omega3} in terms of a residual redundancy in the Weyl parameterisation (\ref{weyl}). 
 
 Thus, just as removing the constant part of $\phi$ ($\beta_0$ in \eqref{oscphi}) when working in the Lorenz gauge $A^\mu = \epsilon^{\mu\nu} \partial_\nu \phi$ in the Abelian $U(1)$ BF-theory was necessary so as to not overcount field configurations;  failure to remove the $\beta_{-1,0,1}$ modes from the configuration space of Weyl factors, would lead to an overcounting of field configurations.\footnote{It follows that the configuration space of the boundary values of $\omega(T,\vartheta)$ should be understood in terms of a quotient space of the set of all boundary metrics modulo the $SL(2,\mathbb{R})$ redundancy, a feature that has prominently appeared in the AdS$_2$ considerations of the model \cite{Maldacena:2016upp,Kitaev:2017awl}. This echoes naturally with the group theoretic necessity of removing the three modes $n=\{-1,0,1\}$ from the $\Delta=2$ discrete series UIR.}
 
 The three remaining  $\alpha_n$ modes are then naturally paired to the three modes $\gamma_n$ of the dilaton solution space (\ref{JTsolns}). This again mirrors the story in the $U(1)$ BF example. There, the canonical pairing was between the constant part of $B$ and the Wilson Loop operator $\alpha_0\equiv\chi$. Indeed, a canonical analysis reveals that $-4\pi\partial_T \omega$ is momentum conjugate to $\phi$ \cite{Maldacena:2019cbz}, and the $\alpha_n$ modes have an appropriate Poisson bracket algebra with the $\gamma_n$ modes of \eqref{JTsolns}. Under the action of the Killing isometries, the triplets furnish a three-dimensional, and hence non-unitary, irreducible representation of $SL(2,\mathbb{R})$.
$\,$\newline\newline
{\textbf{Hilbert space of dS$_2$ JT gravity.}} Upon quantizing the JT gravity theory (\ref{def:SJT}), the fields $\phi(T,\vartheta)$ and $g_{\mu\nu}(T,\vartheta)$ are promoted to operators. The wave equation (\ref{linomega}) becomes an operator equation governing ${\omega}(T,\vartheta)$. As we have just shown, the operator modes in the decomposition (\ref{delta2}) furnish a $\Delta=2$ irreducible representation of $SL(2,\mathbb{R})$. The operators ${\alpha}_n$ and ${\beta}_n$, for $n \in \mathbb{Z}/\{-1,0,1\}$, appearing in (\ref{delta2}) constitute a basis for the representation. The highest- and lowest-weight towers are labelled by quantum numbers $n = 2,3,\ldots$ and $n = -2,-3,\ldots$ as expected. The operators ${\alpha}_n$ with $n \in \{-1,0,1\}$ are naturally paired up with the three modes $\gamma_n$ associated to ${\phi}$ in (\ref{JTsolns}). As mentioned, these furnish a three-dimensional non-unitary irreducible representation of $SL(2,\mathbb{R})$. 

But as in the Abelian BF-theory, although the operator algebra furnishes the $\Delta=2$ representation, the physical Hilbert space is much smaller. This reduction arises when the gauge constraints of the BF theory are properly imposed. Instead, the Hilbert space is spanned by a family of states parameterised by a single real parameter, indicating a Hilbert space of a quantum mechanical, rather than quantum field theoretic, nature. Concretely, one can consider the problem in the Schrödinger picture, as  we did for the $U(1)$ BF-theory in (\ref{schroU1}). Here, one studies wavefunctionals $\Psi_\Sigma[h(u),\phi(u)]$ of the induced metric $h(u)$ and dilaton field $\phi(u)$ on a Cauchy surface whose points are labelled by $u$. The wavefunctions are subject to the constraints of the JT theory. As shown in \cite{Henneaux:1985nw,Isler:1989hq,Iliesiu:2020zld,Chamseddine:1989yz}, one can solve the constraints exactly. The spatial diffeomorphism redundancy allows us to pick a gauge of constant $\phi(u) = \phi_0$. In addition one must impose the associated momentum constraint on the wavefunctions. Prior to doing so, the wavefunctions are functionals of the boundary metric $h(u)$. (In the AdS$_2$ context, this would be the functional of the Schwarzian mode \cite{Maldacena:2016upp}.) The momentum constraint further enforces that the $\Psi_\Sigma[h(u),\phi_0]$ are  functions of the constant mode of $h(u)$ only (see appendix C of \cite{Maldacena:2019cbz} for example). Thus, the physical state-space of the $SL(2,\mathbb{R})$ BF-theory on a global spatial slice is insensitive to the $\Delta=2$ conformal operators stemming from (\ref{delta2}).

\subsection{\texorpdfstring{$SL(N,\mathbb{R})$}{SL(N,R)} BF-theory, briefly} \label{SLNBF}

Our discussion permits a direct generalization to the $SL(N,\mathbb{R})$ BF-theory. We comment on this case briefly here, and leave a general analysis to future work. The $SL(N,\mathbb{R})$ BF-theory  has been studied in \cite{Alkalaev:2013fsa,Alkalaev:2019xuv,Alkalaev:2020kut,Grumiller:2013swa} as a higher-spin extension of JT gravity. Provided the $SL(2,\mathbb{R})$ is principally embedded in the $SL(N,\mathbb{R})$ gauge group, the theory contains a spectrum of fields of spin $s = 2,3,\ldots,N$. These can be viewed as decomposing the adjoint representation of $\mathfrak{sl}(N,\mathbb{R})$ into traceless and symmetric $\mathfrak{sl}(2,\mathbb{R})$ tensors as follows:
\begin{equation}
t^a = \oplus_{k=1}^{N-1} t^{A_1\ldots A_k}~, \quad\quad a = 1,\ldots, N^2-1~,
\end{equation}
where $t^a$ is a generator of $\mathfrak{sl}(N,\mathbb{R})$ in the adjoint, and the $t^{A_1\ldots A_k}$ are rank-$k$ traceless symmetric tensors of $\mathfrak{sl}(2,\mathbb{R})$, each with $(2k+1)$ components, such that the sum gives a total of $N^2-1$ components. This is indeed the dimension of the adjoint representation of $\mathfrak{sl}(N,\mathbb{R})$. 

The gravitational subsector of the theory is captured by the $SL(2,\mathbb{R}) \subset SL(N,\mathbb{R})$ embedding, and the models admit a (near) dS$_2$ vacuum solution. The equation governing fluctuations (in a suitably chosen gauge) about the dS$_2$ vacuum generalizes (\ref{eqn:r2}) to the following collection of fluctuation equations  \cite{Alkalaev:2020kut}
\begin{equation}\label{alkalaev}
\ell^2 \square_{\text{dS}} \omega_s(T, \vartheta) = s(1-s) \omega_s(T, \vartheta)  ~, \quad s = 2,3,\ldots,N~.
\end{equation}
The Schwarzian boundary mode is extended to an $SL(N,\mathbb{R})$ version. The three-redundant modes observed in the $SL(2,\mathbb{R})$ case are replaced by $(2s+1)$ redundant modes for each $s$. As for the previous cases, although there are operators furnishing the discrete series UIRs for $\Delta=2,3,\ldots,N$ the state-space of the theory does not. The details of this will be presented in future work. At $N\to \infty$ the gauge-symmetry is generated by an infinite dimensional higher-spin algebra \cite{Alkalaev:2019xuv}, which is an extension of the $N\to\infty$ limit of $SL(N,\mathbb{R})$, reminiscent of the algebra governing four-dimensional Vasiliev theory \cite{Vasiliev:1990en,Vasiliev:1999ba}.
%

\section{Structures at \texorpdfstring{$\mathcal{I}^+$}{iplus}, contact terms, and gravitational constraints}\label{sec:contactterms}

The purpose of this section is to present certain important structures that arise when considering matter-field observables coupled to gravity. Specifically, we will show how to implement the gauge constraints on the late-time matter-field observables, and the implications that arise as a result of imposing these constraints. Throughout the section, we work in the semiclassical limit, where the fluctuations of the metric field are suppressed. Nevertheless, quantum gravity will play a role inasmuch as it requires us to impose the gauge constraints on the matter.  These constraints are of particular importance for the theories discussed in the previous section \ref{discretesec}, which realise the discrete series UIR in the pre-Hilbert space of a gauge theory. The importance of gravitational constraints were pointed out in early work  \cite{Moncrief:1978te,Moncrief:1979bg,Higuchi:1991tk,Higuchi:1991tm,Marolf:2008it}, and explored more recently in \cite{Anninos:2017eib,Chandrasekaran:2022cip,Chakraborty:2023yed}. 

To construct gauge-invariant observables at the late-time boundary, we will consider expectation values of conformal operators (made out of matter fields) integrated over the late-time spatial slice. Thus, the implementation of the diffeomorphism constraints is performed at the late-time conformal boundary, in a similar way to how vertex operators are integrated over the string worldsheet in order to construct operators invariant under the Virasoro constraints. Because we advocate integrating over the spatial slice, it will be crucial to keep track of any and all contact terms that arise in late-time (pre-integrated) dS$_2$ correlation functions. We thus start with a discussion on the allowed structures present in late-time correlation functions. 
\subsection{Allowed structures for correlators on \texorpdfstring{$\mathcal{I}^+$}{iplus}}
In dS$_2$, the future boundary is conformal to an $S^1$, and conformal operators on the conformal circle transform as described in \eqref{confOps}:
\begin{equation}
\left[ \widehat{L}_0, \mathcal{O}_{\Delta,n}\right]= n \mathcal{O}_{\Delta,n}~, \quad\quad \left[\widehat{L}_\pm, \mathcal{O}_{\Delta,n}\right] = \left( n \pm \Delta \right) \mathcal{O}_{\Delta,n\pm1}~.
\end{equation}
This is the type of setup imagined in  dS/CFT  where it is posited that interacting quantum fields on a  de Sitter background reorganize themselves into a Euclidean CFT on the future boundary at $\mathcal{I}^+$. We can combine this collection of operators into a single object called a local quasi-primary field 
\begin{equation}
\mathcal{O}_\Delta(\vartheta)\equiv\sum_{n}\mathcal{O}_{\Delta,n}\frac{e^{-in\vartheta}}{\sqrt{2\pi}}~,
\end{equation}
which transforms covariantly under $SL(2,\mathbb{R})$ coordinate transformations
\begin{equation}
\mathcal{O}_\Delta(\vartheta)\rightarrow \left(\frac{\partial\vartheta'}{\partial\vartheta}\right)^\Delta\mathcal{O}_\Delta(\vartheta')~,\qquad\qquad \tan\frac{\vartheta'}{2}=\frac{a\tan\frac{\vartheta}{2}+b}{c\tan\frac{\vartheta}{2}+d}~,
\end{equation}
with $(a,b,c,d)\in \mathbb{R}$ and $\text{det}\begin{pmatrix} a & b \\ c & d\end{pmatrix}=1$. The three generators that exponentiate to form the group elements are: 
\begin{equation}
e^{i\lambda H} \longleftrightarrow \begin{pmatrix} 1 & \lambda\\ 0 & 1\end{pmatrix}~,\qquad e^{i\lambda K}  \longleftrightarrow  \begin{pmatrix} 1 & 0\\ -\lambda & 1\end{pmatrix}~,\qquad 
e^{i\lambda D}  \longleftrightarrow  \begin{pmatrix} 1+\frac{\lambda}{2} & 0\\ 0 & \frac{1}{1+\frac{\lambda}{2}}\end{pmatrix}~,
\end{equation}
and the action of these generators on a quasi-primary field of dimension $\Delta$ is
\begin{align}
H&\equiv~~2i\cos\frac{\vartheta}{2} \left[\Delta\,\sin \frac{\vartheta}{2} -\cos \frac{\vartheta}{2} \partial_\vartheta\right]~,\\
K&\equiv-2i\sin \frac{\vartheta}{2} \left[\Delta\,\cos \frac{\vartheta}{2} +\sin\ \frac{\vartheta}{2} \partial_\vartheta\right]~,\\
D&\equiv -i\left[\Delta\,\cos\vartheta+\sin\vartheta\,\partial_\vartheta\right]~.
\end{align}
These Hermitian generators are related to the complexified ones described previously via the relationship
\begin{equation}
H = L_0 - \frac{1}{2}\left(L_++L_-\right) ~, \quad K = L_0+ \frac{1}{2}\left(L_++L_-\right)~, \quad D=-\frac{i}{2}\left(L_+-L_-\right)~,
\end{equation}
where the operators 
\begin{equation}
L_0=-i\partial_\vartheta~,\qquad L_{\pm}=e^{\mp i\vartheta}\left(\mp\Delta-i\partial_\vartheta\right)~,
\end{equation}
generate the algebra of $SL(2,\mathbb{R})$ as given in \eqref{eq:sl2ralgebra}. 

Under this general structure correlation functions of quasi-primary operators transform as expected, namely
\begin{equation}\label{eq:conftransnpoint}
    \langle\mathcal{O}_{\Delta_1}(\vartheta_1)\mathcal{O}_{\Delta_2}(\vartheta_2)\dots \mathcal{O}_{\Delta_n}(\vartheta_n)\rangle=\left[\prod_{i=1}^n\left.\left(\frac{\partial\vartheta'}{\partial \vartheta}\right)^{\Delta_i}\right\rvert_{\vartheta=\vartheta_i}\right] \langle\mathcal{O}_{\Delta_1}(\vartheta_1')\mathcal{O}_{\Delta_2}(\vartheta_2')\dots \mathcal{O}_{\Delta_n}(\vartheta_n')\rangle~.
\end{equation}
and if the underlying theory is invariant under $SL(2,\mathbb{R})$ transformations, then the correlation functions must be invariant as well. We will now raise some points which, to our knowledge, have not previously been highlighted in the literature. 

\paragraph{Two-point function:} In general treatments on conformal field theory, one determines that invariance under \eqref{eq:conftransnpoint} entirely fixes the two point function of quasi-primary operators: 
\begin{equation}
    \langle\mathcal{O}_{\Delta_1}(\vartheta_1)\mathcal{O}_{\Delta_2}(\vartheta_2)\rangle=\begin{cases} \frac{c_{12}}{\left[\sin^2\left(\frac{\vartheta_1-\vartheta_2}{2}\right)\right]^{\frac{\Delta_1+\Delta_2}{2}}}~, & \Delta_1=\Delta_2\\
    0~, & \Delta_1\neq\Delta_2\end{cases}~\qquad\qquad (\text{typical})~.
\end{equation}
This above statement presupposes that no local contact terms contribute to the correlation function of two operators. The logic behind this reasoning stems from the formulation of CFT in Euclidean signature. In this setting, contact terms in correlation functions correspond to ultraviolet ambiguities in the definition of the local operator $\mathcal{O}_{\Delta}(\vartheta)$. 

The setting for dS/CFT is different. Correlation functions at the future boundary encode the entire bulk history. As we will soon demonstrate in an example, local contact terms in correlation functions can now arise naturally from the bulk Heisenberg algebra of quantum fields. Therefore they are not UV ambiguities, but rather, are fixed by the canonical structure of the bulk Hilbert space. With this in mind, we present the most general structure allowed for a two-point function which is invariant under \eqref{eq:conftransnpoint}:  
\begin{equation}\label{eq:twopointfunctinccontact}
    \langle\mathcal{O}_{\Delta_1}(\vartheta_1)\mathcal{O}_{\Delta_2}(\vartheta_2)\rangle=\begin{cases} a_{12}\,\delta(\vartheta_1-\vartheta_2)~, & \Delta_1=1-\Delta_2\\ b_{12}\, g_{\Delta_1+\Delta_2}(\vartheta_1-\vartheta_2)+\frac{c_{12}}{\left[\sin^2\left(\frac{\vartheta_1-\vartheta_2}{2}\right)\right]^{\frac{\Delta_1+\Delta_2}{2}}}~, & \Delta_1=\Delta_2\\
    0~, & \text{otherwise}\end{cases}~.
\end{equation}
The function $g_\Delta$ only contains contact terms
\begin{equation}
    g_\Delta(\vartheta)=\begin{cases}\delta(\vartheta)~, & \Delta=1\\ 
    \delta'(\vartheta)~, & \Delta=2\\
    \delta''(\vartheta)+\frac{\delta(\vartheta)}{4}~, & \Delta=3\\
    \delta'''(\vartheta)+\delta'(\vartheta)~, & \Delta=4\\
    \delta^{(4)}(\vartheta)+\frac{5}{2}\delta''(\vartheta)+\frac{9}{16}\delta(\vartheta)~, & \Delta=5\\
    \delta^{(5)}(\vartheta)+5\delta'''(\vartheta)+4\delta'(\vartheta)~, & \Delta=6\\
    \vdots\end{cases}~.\label{eq:contacterms}
\end{equation}
Crucially, the contact term proportional to $a_{12}$ in \eqref{eq:twopointfunctinccontact} occurs when an operator is paired with an operator in a shadow representation. For principal series operators, this means we have paired an operator with its complex conjugate. It is worth noting that the late-time behaviour of the discrete series two-point function \eqref{eq:latetimediscreteseries2pt} is in tension with the structures present in \eqref{eq:twopointfunctinccontact}. 

\paragraph{Three-point function:} In a similar vein to the discussion above, standard CFT treatments state that invariance under \eqref{eq:conftransnpoint} also completely fixes the three point function: 
\begin{multline}
    \langle\mathcal{O}_{\Delta_1}(\vartheta_1)\mathcal{O}_{\Delta_2}(\vartheta_2)\mathcal{O}_{\Delta_3}(\vartheta_2)\rangle=\\\frac{c_{123}}{\left[\sin^2\left(\frac{\vartheta_1-\vartheta_2}{2}\right)\right]^{\frac{\Delta_1+\Delta_2-\Delta_3}{2}}\left[\sin^2\left(\frac{\vartheta_1-\vartheta_3}{2}\right)\right]^{\frac{\Delta_1+\Delta_3-\Delta_2}{2}}\left[\sin^2\left(\frac{\vartheta_2-\vartheta_3}{2}\right)\right]^{\frac{\Delta_2+\Delta_3-\Delta_1}{2}}} \qquad (\text{typical})~.
\end{multline}
Again, this presupposes that none of the points are coincident. Allowing for contact terms, we find
\begin{align}
    \langle\mathcal{O}_{\Delta_1}(\vartheta_1)\mathcal{O}_{\Delta_2}(\vartheta_2)&\mathcal{O}_{\Delta_3}(\vartheta_2)\rangle\nonumber = \\
    &c_{123}\left(b_{12}\,g_{\Delta_1+\Delta_2-\Delta_3}(\vartheta_1-\vartheta_2)+\left[\sin^2\left(\frac{\vartheta_1-\vartheta_2}{2}\right)\right]^{-\frac{\Delta_1+\Delta_2-\Delta_3}{2}}\right)\nonumber\\
    &~~\times\left(b_{13}\,g_{\Delta_1+\Delta_3-\Delta_2}(\vartheta_1-\vartheta_3)+\left[\sin^2\left(\frac{\vartheta_1-\vartheta_3}{2}\right)\right]^{-\frac{\Delta_1+\Delta_3-\Delta_2}{2}}\right)\nonumber\\
    &~~\times\left(b_{23}\,g_{\Delta_2+\Delta_3-\Delta_1}(\vartheta_2-\vartheta_3)+\left[\sin^2\left(\frac{\vartheta_2-\vartheta_3}{2}\right)\right]^{-\frac{\Delta_2+\Delta_3-\Delta_1}{2}}\right)~,
\end{align}
and the function $g_\Delta$ is given in \eqref{eq:contacterms}. One must exercise care when using this expression, as we must tune the coefficients such that the correlation function make sense as in a distributional sense. 

As we will now demonstrate, these contact terms can unambigulously arise at the late time boundary of de Sitter. We will demonstrate this by considering the correlation functions of a  free field in the principal series. 

\subsection{Contact terms in the two-point function}
Let us now return to the setting of a free-principal series field, with $\Delta=\frac{1}{2}(1+i\nu)$ as in \cref{sec:freefock}. There, we introduced the bulk scalar field $\phi(\tau,\vartheta)$ and its canonical conjugate
\begin{equation}
\pi(\tau,\vartheta)=\cosh\tau\,\partial_\tau\phi~,
\end{equation}
subject to the quantization condition \eqref{eq:canonicalcomm}
\begin{equation}\label{eq:canoinicalpiphi}
[\phi(\tau,\vartheta),\pi(\tau,\vartheta')]=i\delta(\vartheta-\vartheta')~.
\end{equation}
The late-time behavior of this free field operator is
\begin{eqnarray}
\lim_{\tau \to \infty} {\phi}(\tau,\vartheta) &\approx& e^{-\Delta\tau} \mathcal{O}_{\Delta}(\vartheta) +  e^{-(1-\Delta)\tau} \mathcal{O}^\dag_{\Delta}(\vartheta)~, \\
\lim_{\tau \to \infty} {\pi}(\tau,\vartheta) &\approx& -\frac{1}{2}\left[{\Delta}\,e^{(1-\Delta)\tau} \mathcal{O}_{\Delta}(\vartheta) +{(1-\Delta)}\,e^{\Delta\tau} \mathcal{O}^\dag_{\Delta}(\vartheta) \right]~,
\end{eqnarray}
where the operator $\mathcal{O}_{\Delta}(\vartheta)$ and its complex conjugate $\mathcal{O}^\dag_{\Delta}(\vartheta)$ transform as conformal quasi-primary operators in the principal series with weight $\frac{1}{2}(1+i\nu)$ and $\frac{1}{2}(1-i\nu)$ respectively \cite{Sengor:2019mbz,Sengor:2021zlc}.
Recalling the definitions \eqref{eq:phiprincipaldef}-\eqref{eq:phaseeuclideanmodes} we write these operators as follows: 
\begin{align}\label{psop}
    \mathcal{O}_\Delta(\vartheta)&=\frac{\Gamma\left(\frac{1}{2}-\Delta\right)}{2\pi}\sum_{n=-\infty}^\infty\left[a_n^\Delta\,  e^{i(\alpha_n+\beta_n)-i\frac{\pi}{2}(\Delta-|n|)-in\vartheta}+a_n^{\Delta\dag}\,  e^{-i(\alpha_n-\beta_n)+i\frac{\pi}{2}(\Delta-|n|)+in\vartheta}\right]~,\\
    \mathcal{O}_\Delta^\dag(\vartheta)&=\frac{\Gamma\left(\Delta-\frac{1}{2}\right)}{2\pi}\sum_{n=-\infty}^\infty\left[a_n^{\Delta}\,  e^{i(\alpha_n-\beta_n)-i\frac{\pi}{2}(1-\Delta-|n|)-in\vartheta}+a_n^{\Delta\dag}\,  e^{-i(\alpha_n+\beta_n)+i\frac{\pi}{2}(1-\Delta-|n|)+in\vartheta}\right]~,
\end{align}
where we have defined the following phase: 
\begin{equation}
    e^{2i\beta_n}=\frac{\Gamma(\Delta-|n|)}{\Gamma(1-\Delta-|n|)}~.
\end{equation}
Given \eqref{eq:canoinicalpiphi}, we must have that 
\begin{equation}\label{contact}
    [\mathcal{O}_\Delta(\vartheta),\mathcal{O}^\dag_\Delta(\vartheta')]=\frac{i}{\Delta-\frac{1}{2}}\delta(\vartheta-\vartheta')=\frac{2}{\nu}\delta(\vartheta-\vartheta')~,
\end{equation}
as one can check using the algebra of creation and annihilation operators. As a result of this rigid structure, one readily finds: 
\begin{align}
    \bra{\Omega}\mathcal{O}_\Delta(\vartheta)\mathcal{O}_\Delta^\dag(\vartheta')\ket{\Omega}&=\frac{1+\coth\left(\frac{\pi\nu}{2}\right)}{\nu}\delta(\vartheta-\vartheta')~,\\
    \bra{\Omega}\mathcal{O}_\Delta(\vartheta)\mathcal{O}_\Delta(\vartheta')\ket{\Omega}&=\frac{\Gamma(\Delta)\Gamma\left(\frac{1}{2}-\Delta\right)}{4\pi^{3/2}}\left[\sin^2\left(\frac{\vartheta-\vartheta'}{2}\right)\right]^{-\Delta}~.\label{contact2}
\end{align}
Notably, the contact term is rigid, stemming from the canonical quantization condition imposed on the bulk scalar field $\phi$, reproducing the structure in \eqref{eq:twopointfunctinccontact}. 

There is a related discontinuity that one can extract when bulk operators are null separated. This can be seen in the Wightman two-point function, which for a free theory reads
\begin{equation}\label{2pt}
\langle \Omega | \phi(\tau,\vartheta) \phi(\tau',\vartheta')  | \Omega \rangle = \frac{\Gamma(\Delta)\Gamma(1-\Delta)}{4\pi} \, {_2}F_1\left(\Delta,1-\Delta,1,1-\frac{u}{2}\right)~, 
\end{equation}
where 
\begin{equation}
u =1+\sinh \tau\sinh \tau'-\cos(\vartheta-\vartheta')\cosh \tau \cosh \tau'~.
\end{equation}
The  two-point function (\ref{2pt}) exhibits a branch cut along $u=0$, that is, for null-separated points, which results in a discontinity across the cut as we approach from a spacelike direction ($u \rightarrow 0^+$), or a timelike direction ($u \rightarrow 0^-$):
\begin{equation} 
\left( \lim_{u\to 0^-}  - \lim_{u\to 0^+} \right) \langle \Omega | \phi(\tau,\vartheta) \phi(\tau',\vartheta') | \Omega\rangle  = - \frac{i}{4}~.
\end{equation}
One can further compute
\begin{equation}
\langle \Omega |  \phi(\tau,\vartheta) \pi(\tau',\vartheta') | \Omega\rangle  = \frac{\Gamma(\Delta)\Gamma(1-\Delta)}{4\pi} \cosh\tau'\partial_{\tau'}\,  {_2}F_1\left(\Delta,1-\Delta,1,1-\frac{u}{2}\right) ~.
\label{eq:2pointplanar}
\end{equation}
For the above correlator, the singular behavior extends along the light-cone from the $\delta$-function singularity on the equal time slice \eqref{eq:canoinicalpiphi}. This singular structure will be present in the operator algebra of any interacting quantum field theory on a rigid de Sitter background.

\subsection{Contact terms in higher-point functions}

The operator algebra (\ref{contact}), or somewhat more concretely \eqref{eq:canoinicalpiphi}, persists in an interacting theory. Consequently, higher point functions must obey these operator algebras. For perturbatively small interactions, the operators $\mathcal{O}_{\Delta}(x)$ and $\mathcal{O}^\dag_{\Delta}(x)$ remain good conformal operators up to small corrections. Take, for example, an equal-time $2n$-point function
\begin{equation}
G^{(2n)}(\vartheta_1,\ldots,\vartheta_n; \vartheta_{n+1},\ldots,\vartheta_{2n}) \equiv \langle \Omega |  \prod_{i=1}^n   \mathcal{O}_{\Delta}(\vartheta_i)   \prod_{i=1}^n \mathcal{O}^\dag_{\Delta}(\vartheta_{n+i})  | \Omega\rangle~.
\end{equation}
The above correlation function will be invariant under permutations of $\mathcal{X}_1 = \{ \vartheta_1,\ldots, \vartheta_n\}$ and $\mathcal{X}_2 = \{ \vartheta_{n+1},\ldots, \vartheta_{2n}\}$. However, exchanging elements between $\mathcal{X}_1$ and $\mathcal{X}_2$ leads to non-trivial structure. For example, at tree level order we have
\begin{multline}
 \langle \Omega |  \left(\prod_{i=1}^{n-1}   \mathcal{O}_{\Delta}(\vartheta_i) \right) [\mathcal{O}_\Delta(\vartheta_n),\mathcal{O}^\dag_\Delta(\vartheta_{n+1})] \left( \prod_{i=1}^n \mathcal{O}^\dag_{\Delta}(\vartheta_{n+i})\right)  | \Omega\rangle =\\  \frac{2}{\nu}  \delta(\vartheta_{n}-\vartheta_{n+1})\times G^{(2n-2)}(\vartheta_1,\ldots,\vartheta_{n-1}; \vartheta_{n+2},\ldots,\vartheta_{2n})~,
\end{multline}
Relations such as the above, which follow from the \emph{Lorentzian} canonical nature of the bulk de Sitter theory, are an important structural feature of the space of correlation functions at the late-time surface.

\subsection{Integrated operators and gravity}
We now consider what happens when we couple the quantum field theory in question to gravity, bearing in mind the aforementioned contact terms. Let us assume the existence of a semiclassical gravitational theory, which permits a de Sitter solution with small fluctuations. This might be, for example, a de Sitter version of JT gravity coupled to matter fields \cite{Anninos:2017hhn,Maldacena:2019cbz,Cotler:2019nbi} or two-dimensional gravity with $\Lambda>0$ coupled to a CFT with a large positive central charge as in \cite{Anninos:2021ene,Muhlmann:2022duj}. 

For the sake of simplicity, we take the matter theory to have a pair of free massive scalars $\phi$ and $\tilde\phi$, each with $\Delta=\frac{1}{2}(1+i\nu)$. Observables must be diffeomorphism invariant, and consequently also de Sitter invariant since the $SL(2,\mathbb{R})$ de Sitter isometries are a subgroup of the diffeomorphism group. On the late-time surface at $\mathcal{I}^+$, we require that the invariant operators are $SL(2,\mathbb{R})$ invariant with respect to the conformal transformations of the boundary $S^1$. Recalling that the one-form $d\vartheta$ transforms with weight minus one, this can be achieved by integrating an operator of weight $\Delta=1$ over the boundary direction. One such example  of a \textbf{non-Hermitian} boundary operator invariant under the residual $SL(2,\mathbb{R})$, we take
\begin{equation}
\mathcal{O}_{\Delta}^{\text{grav}} = \int_0^{2\pi} d\vartheta : \mathcal{O}_{\Delta}(\vartheta) \tilde{\mathcal{O}}^\dag_{\Delta}(\vartheta) :~,
\end{equation}
where $\mathcal{O}_{\Delta}(\vartheta)$ and $\tilde{\mathcal{O}}_{\Delta}(\vartheta)$ are two distinct principal series conformal operators as in (\ref{psop}), transforming with $\Delta=\frac{1}{2}(1+i\nu)$. One can also build de Sitter invariant states. One of them is the Bunch-Davies vacuum $|\Omega\rangle$. Acting with $\mathcal{O}_{\Delta}^{\text{grav}}$ on $|\Omega\rangle$  yields a de Sitter invariant state, but one that is not normalizable. A more systematic way of constructing de Sitter invariant states is discussed in \cite{Higuchi:1991tk,Higuchi:1991tm,Marolf.2009yal,Chakraborty:2023yed} and specifically for dS$_2$ in \cite{Marolf:2008it}. We can  consider expectation values of de Sitter invariant operators. For instance, 
\begin{equation}
\langle \Omega| \mathcal{O}_{\Delta}^{\text{grav}}   \mathcal{O}_{\Delta}^{\text{grav}} |\Omega\rangle = \frac{\csch(\pi\nu)}{4\pi \nu} \int_{S^1\times S^1}   \frac{d\vartheta d\vartheta'}{\sin^2\left(\frac{\vartheta-\vartheta'}{2}\right)}~.
\end{equation}
 The above expression is divergent and sensitive to the coincident point limit but avoids the appearance of any contact terms. Additionally, this expression, and corresponding higher-point correlation functions, are of the type  that appears when considering open string amplitudes on the disk. To regularize the integral, we can consider a point-splitting cutoff as in \cite{Friedan:2012hi,Liu:1987nz}, namely whenever the points collide we split them by a small amount $\varepsilon$. Specifically, we define 
 \begin{equation}
    x\equiv\frac{\vartheta+\vartheta'}{2}\,\qquad\qquad y\equiv \vartheta-\vartheta'~, 
\end{equation}
and write
\begin{align}\label{grav2pt}
\langle \Omega | \mathcal{O}_{\Delta}^{\text{grav}}  \mathcal{O}_{\Delta}^{\text{grav}} |\Omega\rangle_{\rm reg.}  &=  \frac{\csch(\pi\nu)}{4\pi \nu}\int_{\frac{\varepsilon}{2}}^{2\pi-\frac{\varepsilon}{2}}dx\left(\int_{-x}^{-\varepsilon}\frac{dy}{\sin^{2}\frac{y}{2}}+\int_{\varepsilon}^{x}\frac{dy}{\sin^{2}\frac{y}{2}}\right)~,\nonumber\\
&\underset{\varepsilon\rightarrow0}{\approx}\frac{4\csch(\pi\nu)}{\nu}\left(\frac{1}{\varepsilon}-\frac{1}{2\pi}+O(\varepsilon)\right)~.
\end{align}
which diverges linearly in $\varepsilon$. Higher-point functions will exhibit the same type of divergence. 

We can compare the structure of the $SL(2,\mathbb{R})$ invariant gravitational correlators in dS$_2$ to the invariant volume of $SL(2,\mathbb{R})$, as computed in \cite{Liu:1987nz}:
\begin{equation}
\text{vol} \, SL(2,\mathbb{R}) = \frac{1}{32} \, \int_{S^1 \times S^1 \times S^1} \frac{d\varphi_1 d\varphi_2 d\varphi_3}{| \sin \frac{\varphi_1-\varphi_2}{2}\sin \frac{\varphi_1-\varphi_3}{2}\sin \frac{\varphi_2-\varphi_3}{2}|}~.
\end{equation}
 Upon regularization we note the same linear divergence as for \eqref{grav2pt}. The work of \cite{Liu:1987nz} goes a step further (at least in the context of open string theory) and argues, based on Weyl invariance, that one can meaningfully extract a constant term from $\text{vol} \, SL(2,\mathbb{R})$, which turns out to be $-\pi^2/2$. 

If one were concerned with the non-Hermiticity of the operators $\mathcal{O}_\Delta^{\rm grav}$ defined above, we may also build the analogous diffeomorphism invariant operators out of the late-time operators of a single field in order to analyze the contribution of the contact term (\ref{contact2}). In this case we take
\begin{equation}
\mathcal{Q}_\Delta^{\text{grav}} =\int_0^{2\pi} d\vartheta : \mathcal{O}_{\Delta}(\vartheta) {\mathcal{O}}^\dag_{\Delta}(\vartheta) :~,
\label{def:opcontact}
\end{equation}
leading to: 
\begin{equation}
   \langle \Omega | \mathcal{Q}^{\text{grav}}   \mathcal{Q}^{\text{grav}} |\Omega\rangle =\frac{\csch(\pi\nu)}{4\pi \nu} \int_{S^1\times S^1}   \frac{d\vartheta d\vartheta'}{\sin^2\left(\frac{\vartheta-\vartheta'}{2}\right)}+\frac{\csch^2\left(\frac{\pi\nu}{2}\right)}{\nu^2} \int_{S^1\times S^1}  d\vartheta d\vartheta'\,\delta^2(\vartheta-\vartheta')~.
\end{equation}
 A very similar calculation to \eqref{grav2pt} yields\footnote{We  make the substitution $\int_{S^1\times S^1}  d\vartheta d\vartheta'\,\delta^2(\vartheta-\vartheta')=2\pi\delta(0)$, for reasons we hope are not too obscure.}
\begin{equation}
\langle \Omega | \mathcal{Q}^{\text{grav}}   \mathcal{Q}^{\text{grav}} |\Omega\rangle \underset{\varepsilon\rightarrow0}{\approx}\frac{4\csch(\pi\nu)}{\nu}\left(\frac{1}{\varepsilon}+\frac{\pi}{\nu}\coth\left(\frac{\pi\nu}{2}\right)\delta(0)-\frac{1}{2\pi}\right)+O(\varepsilon) ~. 
\end{equation}
The contact term contribution, unlike the contribution of separated points, has not been previously discussed in the context of the volume of $SL(2, \mathbb{R})$. Nonetheless, upon replacing the $\delta$ function with a limiting Gaussian, it has a similar linearly divergent behaviour. The suggestion that this too could be an appearance of the volume of the group, and therefore regularizable in the same way is intriguing. 

It seems any such gravitational correlators built out of Hermitian combinations of these late-time operators (\eqref{def:opcontact} is one example) must include contractions between $\mathcal{O}$ and $\mathcal{O}^\dag$ and so cannot avoid contributions from these types of contact terms. The situation is reversed in the case of correlators built from late-time operators for complementary series fields, as detailed in \cite{Sengor:2021zlc}, the operators  for $\Delta \in \left(0, \tfrac{1}{2}\right)$ are
\begin{equation} 
\lim_{\tau\rightarrow\infty}\phi_\Delta(\tau, \vartheta) = e^{-\tau\Delta} \alpha_\Delta(\vartheta) + e^{-(1-\Delta)\tau}\beta_{1-\Delta}(\vartheta)~.
\end{equation}
The operators $\alpha$ and $\beta$ are Hermitian, but contact terms, such as the ones described above, can only arise in correlators that mix $\alpha$ and $\beta$. This implies that integrated gravitational correlators can be constructed of Hermitian observables that do not encounter contact terms, for example, for $\Delta=1/2$ the two-point function of 
\begin{equation}
\mathcal{A}^{\text{grav}} = \int_{S^1} d\vartheta :\alpha_{\frac{1}{2}}(\vartheta) \alpha_{\frac{1}{2}}(\vartheta):~ ,   
\end{equation}
will be free of contact term singularities.

To recap: The implementation of gravitational constraints in this way renders observables diffeomorphism invariant at the expense of locality. This is not to say that quasi-local physics is doomed. Since recent observations suggest that an exponentially expanding universe is a good model for the cosmological era we are currently entering, we must reflect on how one is to describe the quasi-local physics of everyday experience. Perhaps the quasilocal description emerges in a relational sense, much like the relation between an observer's physical orientation relative to the chair they are reading this paper in.

\section{Microphysical outlook: a holographic proposal}\label{sec:outlook}

We would like to end our discussion by pointing out a microphysical model whose operator content furnishes an infinite tower of discrete series UIRs, namely $D^
\pm_{s}$ with $s=2,4,\ldots$ and moreover has an infinitely large symmetry: The $q=2$ SYK model, whose Hamiltonian governs $\mathbf{N}$ quantum mechanical Majorana fermions $\psi_i$, with $i=1,\ldots,\mathbf{N}$ subject a random two-body interaction \cite{Maldacena:2016hyu,Kitaev:2017awl,Anninos:2016szt}. Although free, the model exhibits an emergent conformal symmetry at low energies and the operator spectrum can be organised in terms of their properties under $SL(2,\mathbb{R})$. The discrete series operators take the form $\mathcal{O}_\Delta(\vartheta) = \psi_i \, \partial_\vartheta^{\Delta-1} \psi_i$ with $\Delta=2,4,\ldots$ We discuss the potential role of $\mathcal{O}_\Delta$ in terms of a microphysical completion of a dS$_2$ theory endowed with an infinite tower of higher-spin fields --- a de Sitter version of the theories discussed in \cite{Alkalaev:2013fsa,Alkalaev:2019xuv,Alkalaev:2020kut,Grumiller:2013swa}.

\subsection{The \texorpdfstring{$q=2$}{} SYK model}

The general $q$ SYK model describes $\mathbf{N}$ Majorana fermions $\psi_i$ interacting via a $q$-body interaction. Following  \cite{Maldacena:2016hyu,Kitaev:2017awl}, the Euclidean action is given by
\begin{equation}\label{eq:sykham}
	S_{\rm UV}=\int_{S^1} d\vartheta\left[\frac{1}{2} \psi_i\partial_\vartheta{\psi}_i-i^{q/2} \sum_{1\leq i_1<i_2<\dots< i_q\leq \mathbf{N}}J_{i_1\dots i_q} \psi_{i_1}\dots \psi_{i_q}\right]~,
\end{equation}
where the factor of $i^{q/2}$ in \eqref{eq:sykham} is required by Hermiticity, 
and the couplings $J_{i_1\dots i_q}$ are sampled from a Gaussian with zero mean and variance
\begin{equation}
	\left\langle J_{i_1\dots i_q}^2\right\rangle=\frac{J^2(q-1)!}{\mathbf{N}^{q-1}}~,
\end{equation}
{where $J$ has units of energy and characterizes the variance of the distribution on $J_{i_1\dots i_q}$.} The coordinate $\vartheta \sim \vartheta + 2\pi$ is a coordinate on a Euclidean $S^1$ which is meant to evoke the boundary coordinate of global dS$_2$. We can obtain an effective, disorder-averaged theory by integrating in a bilocal field $G(\vartheta,\vartheta')$ given by the fermionic two-point function
\begin{equation}
	G(\vartheta,\vartheta')= \frac{1}{\mathbf{N}}\sum_{i=1}^{\mathbf{N}} \left\langle  T \psi_i(\vartheta)\psi_i(\vartheta')\right\rangle~.
\end{equation}
Here $T$ signifies Euclidean time ordering. Integrating over both the couplings $J_{i_1\dots i_q}$ and the fermions leads to the effective action
\begin{equation}
	{S_{\rm UV}^{\rm eff}}=-\frac{\mathbf{N}}{2}\log\det\left(\partial_\vartheta-\Sigma\right)+\frac{\mathbf{N}}{2}\int_{S^1\times S^1} d\vartheta d\vartheta'\, \left[\Sigma(\vartheta,\vartheta')G(\vartheta,\vartheta')-\frac{J^2}{q}G(\vartheta,\vartheta')^q\right]~.
\end{equation}
Because of the anticommuting fermions, the bilocal field must obey $G(\vartheta,\vartheta')=-G(\vartheta',\vartheta)$. 
The saddle point equations are given by
\begin{equation}
	\partial_\vartheta G(\vartheta,\vartheta')-\int_{S^1} d\upsilon\, \Sigma(\vartheta,v)G(\upsilon,\vartheta')=\delta(\vartheta-\vartheta')~,\quad \quad\Sigma(\vartheta,\vartheta')= J^2 G(\tau,\tau')^{q-1}~.
\end{equation}
At low energies, we can drop the local derivative term and are left, focusing on $q=2$, with the following equation of motion: 
\begin{equation}\label{eq:lowenergyeomq2}
	J^2\int_{S^1} d\upsilon\, G(\vartheta,\upsilon)G(\upsilon,\vartheta')=-\delta(\vartheta-\vartheta')~.
\end{equation}
As does the general $q$-body SYK model, the $q=2$ theory exhibits a reparameterization invariance
\begin{equation}\label{eq:diffsq2}
	\vartheta\rightarrow f(\vartheta)~, \quad \quad G(\vartheta_1,\vartheta_2)\rightarrow [f'(\vartheta_1)]^{1/2} G(f_1,f_2) [f'(\vartheta_2)]^{1/2}~,
\end{equation}
where $f(\vartheta)$ is a monotonic map. The fermion operators $\psi_i$ transform as primaries of scaling dimension $\Delta=1/2$. The  saddle point solution is given by
\begin{equation}
G^{(cl)}(\vartheta_1 , \vartheta_2) = \frac{1}{2\pi J} \, \sin^{-1}   \frac{\vartheta_1 - \vartheta_2}{2}~,
\end{equation}
and at low energies all reparameterisations (\ref{eq:diffsq2}) of the above are also solutions. The low energy sector of the model contains a tower of conformal primary operators $\mathcal{O}_\Delta(\vartheta) = \mathbf{N}^{-1} \sum_i \psi_i \partial_\vartheta^{\Delta-1} \psi_i$, for $\Delta \in 2\mathbb{Z}^+$, where  $\Delta=2,4,\ldots$ are the respective conformal dimensions. {Had we considered complex fermions, we would have $\Delta_p = p$ for all positive integers.}

The presence of an infinite tower of conformal operators, each of integer weight, is suggestive of an integrable structure with an infinite enhancement of symmetries. We now show this is indeed the case. 

\subsection{Infinite symmetries of the \texorpdfstring{$q=2$}{} SYK model}

The saddle point equation (\ref{eq:lowenergyeomq2})  is reminiscent of matrix multiplication $G\cdot G^T= -J^{-2} \, \mathds{1}$ which is invariant under 
\begin{equation}
	G\rightarrow O\cdot G\cdot O^T
\end{equation}
with $O$ an orthogonal matrix. The Lie algebra of the orthogonal matrices is spanned by the skew-symmetric  matrices. 
In the continuum case, $O$ is replaced by a function of two times, and each time coordinate serves the purpose of a continuous matrix index. We thus observe that (\ref{eq:lowenergyeomq2}) is invariant under the  transformation
\begin{equation}\label{OGO}
	G(\vartheta,\vartheta')\rightarrow \int_{S^1 \times S^1} d \upsilon d\zeta\, O(\vartheta,\upsilon)G(\upsilon,\zeta)O(\vartheta',\zeta)~,
\end{equation}
where $O(\vartheta,\vartheta')$ satisfies
\begin{equation}\label{ortho}
	\int_{S^1} d \upsilon \, O(\vartheta,\upsilon) O(\vartheta',\upsilon) =\delta(\vartheta - \vartheta') \, .
\end{equation}
At the infinitesimal level, we can expand $O(\vartheta,\upsilon) = \delta(\vartheta,\upsilon) + \xi(\vartheta,\upsilon)$, where it follows from (\ref{ortho}) that $\xi(\vartheta,\upsilon)$ is an anti-symmetric function $\xi(\vartheta,\upsilon)=-\xi(\upsilon,\vartheta)$. To first order in $\xi(\vartheta,\upsilon)$, we find that the transformation of $G$ is
\begin{equation}
	\delta_\xi G(\vartheta,\vartheta')=\int_{S^1} d \upsilon \,\left( \xi(\vartheta,\upsilon)G(\upsilon,\vartheta')- G(\vartheta,\upsilon)\xi(\upsilon,\vartheta')\right)~.
\end{equation}
We can determine the commutator of two such transformations by computing $[\delta_{\xi'},\delta_{\xi}]\equiv\delta_{\xi'}\delta_\xi-\delta_\xi\delta_{\xi'}$~. A little algebra reveals
\begin{equation}
	[\delta_{\xi'},\delta_{\xi}]=\delta_{\xi'\circ \, \xi}
\end{equation}
where
\begin{equation}
	\left[\xi'\circ\xi\right](\vartheta,\vartheta') \equiv \int d\upsilon\,\left[\xi'(\vartheta,\upsilon)\xi(\upsilon,\vartheta')-\xi(\vartheta,\upsilon)\xi'(\upsilon,\vartheta')\right]~.
\end{equation}
The commutator of an infinitesimal reparameterisation or $\vartheta$, which is (\ref{eq:diffsq2}) with  $f(\vartheta)=\vartheta + \varepsilon(\vartheta)$,  with an orthogonal generator $[\delta_\varepsilon,\delta_\xi]$ yields
\begin{multline}
	[\delta_\varepsilon,\delta_\xi]G(\vartheta,\vartheta')=\int_{S^1} d\upsilon\left[\left( \varepsilon(\vartheta)\partial_\vartheta+\varepsilon(\upsilon)\partial_\upsilon\right)\xi(\vartheta,\upsilon)+\frac{\varepsilon'(\vartheta)+\varepsilon'(\upsilon)}{2}\xi(\vartheta,\upsilon)\right] G(\upsilon,\vartheta')\\-\int_{S^1} d\upsilon \,G(\vartheta,\upsilon)\left[\left( \varepsilon(\upsilon)\partial_\upsilon+\varepsilon(\vartheta')\partial_{\vartheta'}\right)\xi(\upsilon,\vartheta')+\frac{\varepsilon'(\upsilon)+\varepsilon'(\vartheta')}{2}\xi(\upsilon,\vartheta')\right]~,
\end{multline}
which is the action of a reparameterised $\xi$. As for the reparameterisation symmetries, although (\ref{OGO}) is a symmetry of the strict low energy action, it is broken by the leading irrelevant contribution to the effective action 
\begin{align}
	{S_{\rm UV}^{\rm eff}}&= \frac{\mathbf{N}}{2}\int_{S^1\times S^1} d\vartheta d\vartheta'\, \delta(\vartheta-\vartheta') \partial_\vartheta G^{(cl)}(\vartheta,\vartheta')~,\\
 &=\frac{\mathbf{N}}{4\pi J}\int_{S^1\times S^1}d\vartheta d\vartheta'\, \delta(\vartheta-\vartheta')\partial_\vartheta\int_{S^1 \times S^1} d \upsilon d\zeta\, \frac{O(\vartheta,\upsilon)O(\vartheta',\zeta)}{\sin\frac{\upsilon-\zeta}{2}}~,
\end{align}
which diverges at coincident points. Extracting the soft mode action from this divergence can be done by a heuristic point-splitting analysis (see section 3 of \cite{Anninos:2017cnw}) where we split the coincident points by a small amount $\delta\varepsilon$. We find the following contribution
\begin{equation}
S^{\rm eff}_{\rm breaking}=\frac{\mathbf{N}}{2\pi J } \int_{S^1\times S^1} d\vartheta   d\nu\,  \left[\text{p.v.}\int \frac{d\xi}{\sin \xi}  \, O(\vartheta +\delta\varepsilon,\nu+\xi)\partial_\vartheta O(\vartheta,\nu-\xi)\right]~, 
\end{equation}
where p.v. denotes the Cauchy principal value. The above is reminiscent of the contribution to the soft-mode sector in SYK theories with global symmetries \cite{Yoon:2017nig,Anninos:2017cnw}. In the case at hand, these will give rise to an infinite enhancement of the soft sector. 

Although many of the transformations (\ref{OGO}) lead to a non-vanishing soft-mode action and are consequently softly broken, the SYK model with $q=2$ has an infinite number of physical symmetries due to the fact that the underlying theory is free. This is most easily seen from the perspective of the Euclidean fermionic action
\begin{equation}
S_{\rm UV} = \frac{1}{2}\int_{S^1 \times S^1} d\vartheta d\vartheta'  \psi_i(\vartheta) \delta(\vartheta-\vartheta') \left( {\delta_{ij}}    \partial_{\vartheta'} - i   J_{ij}   \right) \psi_j(\vartheta')~, 
\end{equation} 
where the couplings $J_{ij} = -J_{ji}$ are sampled from a Gaussian with zero mean and variance $\left\langle J_{ij}^2\right\rangle={J^2}/{\mathbf{N}}$.
For any given realisation of the couplings, the following non-local transformation
\begin{equation}
\psi_i(\vartheta) \to \int_{S^1} d{\vartheta'}  Q_{ij} (\vartheta,\vartheta') \psi_j(\vartheta')~,
\end{equation}
 transforms the action as follows
\begin{equation}
S_{\rm UV} =  \frac{1}{2}\int_{(S^1)^4} d\vartheta d\vartheta'd\vartheta''d\vartheta'''  Q_{il}(\vartheta,\vartheta'')\psi_l(\vartheta'') \delta(\vartheta-\vartheta') \left( \delta_{ij}    \partial_{\vartheta'} - i   J_{ij}   \right) Q_{jk}(\vartheta',\vartheta''')\psi_k(\vartheta''')~.
\end{equation} 
Thus for $Q_{ij} (\vartheta,\vartheta')$'s satisfying
\begin{equation}
 \int_{S^1 \times S^1}d{\vartheta} d{\vartheta'}  \delta(\vartheta-\vartheta')Q_{il} (\vartheta,\vartheta'') \left(\delta_{ij}  \partial_{\vartheta'} - i J_{ij} \right)  Q_{jk} (\vartheta',\vartheta''') =  \delta(\vartheta''-\vartheta''')\left( \delta_{lk} \partial_{\vartheta'''} - i  J_{lk} \right) ~,
\end{equation}
we have a symmetry of the action.

Ordinarily, non-local field transformations are not permitted but here we are viewing $\vartheta$ as a coordinate on the future spacelike boundary of dS$_2$. As such, the locality properties of fields can be relaxed.  At low energies, and upon averaging over the couplings the above symmetry becomes  (\ref{OGO}) and (\ref{ortho}). What we see here is that the infinite low energy symmetry is deformed into an infinite symmetry of the ultraviolet fermionic theory.

\subsection{Higher-spin \texorpdfstring{dS$_2$}{dS2} dual?}

We have seen that the $q=2$ model displays a highly symmetric low energy sector encoding an infinite tower of conformal primaries $\mathcal{O}_\Delta$ with $\Delta \in 2\mathbb{Z}^+$. It is tempting to suggest that the $\mathcal{O}_\Delta$ are captured by an underlying dS$_2$ theory with an infinite tower of operators in the discrete series UIR, echoing ideas expressed for the AdS$_2$ case in \cite{Gonzalez:2018enk,Alkalaev:2019xuv}. A potentially relevant class of models exhibiting such properties are the $N\to\infty$ extensions of the $SL(N,\mathbb{R})$ BF-theories \cite{Alkalaev:2020kut,Grumiller:2013swa} discussed in section \ref{SLNBF}. Guided by the linearized equations (\ref{alkalaev}), the higher-spin bulk operators at $\mathcal{I}^+$ extend the $SL(2,\mathbb{R})$ $\Delta=2$ operator (\ref{delta2}) to an infinite tower of operators with $\Delta=2,3,\ldots,\infty$. 

We propose that the bulk late-time conformal operators are  microscopically constructed from two towers of $q=2$ SYK conformal operators
\begin{equation}
\mathcal{O}_\Delta(\vartheta)  = \frac{1}{\mathbf{N}} \sum_{i=1}^\mathbf{N} \psi_i \partial_\vartheta^{\Delta-1} \psi_i~, \quad   \tilde{\mathcal{O}}_\Delta(\vartheta) =  \frac{1}{\mathbf{N}} \sum_{i=1}^\mathbf{N} \chi_i \partial_{\vartheta}^{\Delta-1} \chi_i~, \qquad \text{with} \qquad \Delta \in 2\mathbb{Z}^+~.
\end{equation}
Here $\psi_i$ and $\chi_i$, with $i=1,\ldots,\mathbf{N}$, are two collections of  $q=2$ SYK fermions. The reason we have two towers of operators in the bulk dS$_2$ is the higher spin extension of the observation that there are two collections of operator modes, $\alpha_n$ and $\beta_n$, associated with the mode expansion bulk operator $\omega(T,\vartheta)$ in (\ref{delta2}). 
In addition, there should be operators associated to the adjoint valued $B(T,\vartheta)$ field, which has non-trivial commutation relations with the $SL(N,\mathbb{R})$ gauge field. It is natural to construct these out of fermionic operators also, which have non-trivial commutation relations with the $\psi_i$ and $\chi_i$. To this end, we note that for a sufficiently large number $\mathbf{N}$ of fermionic operators satisfying the standard anti-commutation relations, one can approximate the bosonic creation/annihilation operator algebra with arbitrary precision \cite{Holstein:1940zp}. 

Due to the presence of a timelike boundary, the AdS$_2$ version of the $SL(N,\mathbb{R})$ BF-theory has a slightly broken infinite dimensional higher-spin algebra governed by a soft-sector \cite{Gonzalez:2018enk}, as well as an infinite set of conserved physical boundary charges extending the conservation of energy associated to ordinart JT gravity. In the dS$_2$ case, all physical symmetries must be further gauged. Given that the microphysical operators $\psi_i$ are built from quantum mechanical fermions subject to an infinite dimensional symmetry, such a gauging might result in a finite-dimensional Hilbert space \cite{Fischler:2000,Banks:2006rx,Bousso:2000nf,Parikh:2004wh}.

\appendix

\section*{Acknowledgements}

It is a great pleasure to thank Teresa Bautista,   Max Downing, Eleanor Harris, Kurt Hinterbichler, Dami\'an Galante, Diego Hofman, Austin Joyce, Vasileios Letsios, Manuel Loparco, Beatrix M\"uhlmann, Guilherme Pimentel, Kamran Salehi Vaziri, Vladimir Schaub, and Zimo Sun. D.A. is funded by the Royal Society under the grant
“The Atoms of a deSitter Universe”. G.\c{S}. acknowledges funds from Europe Union's Horizon 2020 MSCA-IF grant agreement No 840709 SymAcc, European Structural and Investment Funds and the Czech Ministry of Education, Youth and Sports (MSMT) Project CoGraDS with
grant number-CZ.02.1.01/0.0/0.0/15003/0000437) and T\"{U}BITAK (The Scientific and
Technological Research Council of Turkey) 2232 - B International Fellowship for Early Stage Researchers
programme with project number 121C138 at different stages of this work. B.P. is funded by the STFC under grant number ST/V506771/1. 

\section{Representation theory of \texorpdfstring{$SL(2,\mathbb{R})$}{SL(2,R)}}\label{App:Repthy}

We review the Unitary Irreducible Representations (UIRs) of $SL(2,\mathbb{R})$. For more detailed reviews of this subject we refer the reader to \cite{Kitaev:2017hnr, Sun:2021thf}. We provide a short summary here for convenience and to fix our conventions, making the link to the global coordinate parameterisation explicit. 

\paragraph{Global decomposition.} $SL(2,\mathbb{R})$ is a non-compact connected simple real Lie group. As such, it does not permit non-trivial finite dimensional UIRs. We can derive the UIRS of the group  $SL(2,\mathbb{R})$ by induction on $K= SO(2)$, the maximal compact subgroup. This approach is reviewed in \cite{Dobrev:1977qv,Repka1978TensorR,Sengor:2019mbz,Sun:2021thf}. We consider the Lie algebra $\mathfrak{sl}(2,\mathbb{R})$. The complexified generators defined in equation \eqref{def:sl2r} obey the commutation relations 
\begin{equation}
[L_0,L_\pm] = \mp L_\pm~, \hspace{1cm} [L_+,L_-] = 2L_0~.
\end{equation}
 $L_0$ generates K and is associated to the spatial translations in the global coordinate parameterisation \eqref{globalds2}. We derive UIRS by insisting on the following reality conditions, derived from the hermiticity of the de Sitter generators in \eqref{def:dsgens}
 \begin{equation}
    L_0^\dag = L_0 ~, \hspace{1cm} L_\pm^\dag = L_\mp \, .
 \end{equation}
 The quadratic Casimir is given, as in the text, by 
\begin{equation}
\mathcal{C} \equiv L_0^2-\frac{1}{2}\left(L_{-}L_{+}+L_{+}L_{-}\right)\, . 
\end{equation}
 We consider states which simultaneously diagonalise $L_0$ and $\mathcal{C}$,
\begin{equation}
    L_0\ket{n,\Delta} = -n\ket{n,\Delta}~, \quad \mathcal{C}\ket{n,\Delta} = \Delta(\Delta-1)\ket{n,\Delta}~ . \label{def:estates}
\end{equation}
Including half integer spin fields leads one to consider even and odd UIRs of $SL(2,\mathbb{R})$. These are defined respectively by the action of the exponentiation of $L_0$
\begin{equation}
    e^{2\pi i L_0} \ket{n, \Delta} = \pm\ket{n,\Delta}\, .
\end{equation}
It is clear that $n\in \mathbb{Z}$ or $n \in \mathbb{Z}+ \frac{1}{2}$ for even and odd representations respectively. We can immediately see that the reality of the Casimir eigenvalue implies one of either $\Delta = 
    \frac{1}{2}+ i\nu$, $\nu \in \mathbb{R}$ or $\Delta \in \mathbb{R}$ holds. We further seek to normalise the states such that $\braket{n,\Delta|n,\Delta} \geq 0$ and the action of the ladder operators satisfies
\begin{equation}
   L_\pm \ket{n,\Delta} = -(n\pm\Delta)\ket{n \pm 1,\Delta}\,.
\end{equation}
Unitarity demands $\bra{n,\Delta}L_-\ket{n+1,\Delta}=\bra{n+1,\Delta}L_+\ket{n,\Delta}^*$ and $\braket{n,\Delta|n,\Delta}>0$, (see \cite{Sun:2021thf}) which leads to the following condition
\begin{equation}
    \frac{\braket{n+1,\Delta|n+1,\Delta}}{\braket{n,\Delta|n,\Delta}} = \frac{n+1-\Delta}{n+\Delta^*}\equiv\lambda_n~,\qquad\implies \qquad\lambda_n > 0 \, ,
    \label{eqn:poscon}
\end{equation}
for all $n$. For $\Delta\in \mathbb{R}$, it will be useful to rewrite
\begin{equation}\label{eq:speciallambda}
    \lambda_n=\frac{\left(n+\tfrac{1}{2}\right)^2-\left(\Delta-\tfrac{1}{2}\right)^2}{(n+\Delta)^2}>0~,\qquad\qquad \Delta\in\mathbb{R}~.
\end{equation}
Thus, even UIRs are  permitted for the principal series when $\Delta =\frac{1}{2}(1+ i \nu)$, as well as for the complementary series when  $\Delta \in (0,1)$, and for the discrete series  $\Delta\in \mathbb{Z}^+$ . The same follows for odd UIRs in the principal and discrete series. However the positivity condition \eqref{eq:speciallambda} for $n= -\frac{1}{2}$ in the complementary series can not be satisfied
 and thus there are no odd complementary series UIRs. To summarise, the UIRs of $SL(2,\mathbb{R})$ are given by:
 \begin{itemize}
    \item {\textbf{Even and odd principal series}}, $\mathcal{\pi^\pm_\nu}$: We have $\Delta =\frac{1}{2}(1+ i \nu)$ and $n\in\mathbb{Z}$ or $n \in \mathbb{Z}+\frac{1}{2}$ for the even, resp., odd representations. The states can be consistently normalised as follows:
    \begin{equation}\braket{n,\Delta| m, \Delta}_{\pi_\nu} = \delta_{n,m}\, .\end{equation}
    \item {\textbf{Complementary series}}, $\gamma_\Delta$: In this case we are confined to the range $\Delta \in (0,1)$. The complementary series must be even, and thus $n\in \mathbb{Z}$. For this representation, we can take: 
    \begin{equation}\braket{n,\Delta|m,\Delta}_{\gamma_\Delta} = \frac{\Gamma(n+1-\Delta)}{\Gamma(n+\Delta)}\delta_{n,m}\, .\end{equation}
    \item  {\textbf{Even and odd discrete series}}, $D^\pm_{\Delta, \pm}$: For the even and odd cases, where $\Delta \in \mathbb{Z}_+$ or $\Delta \in \mathbb{N} +\frac{1}{2}$, respectively, we have a pair of UIRs for which  $L_{\pm}\ket{n=\mp\Delta,\Delta}=0,$ corresponding to  $D^\pm_\Delta$: the highest- and lowest-weight representations. In this case we must normalize the states as follows:
    \begin{equation}\braket{n,\Delta|m,\Delta}_{D^\pm_\Delta} = \frac{\Gamma(\mp n+1-\Delta)}{\Gamma(\mp n+\Delta)}\delta_{n,m}\, .\label{eq:discretesernorm}\end{equation}
    \item  {\textbf{Trivial Representation}}:  $\Delta = 0$ and $ n = 0$.  
\end{itemize}

\section{Two conformal particles on the circle\label{app:circle}}

In this appendix, we consider the two-particle Hilbert space content, for particles in the principal series UIR. For a related analysis see section 3.2 of \cite{Penedones:2023uqc}. We will be using the conventions of \cite{Anous:2020nxu}. We start with the {Hermitian} generators acting on wavefunctions of a single degree of freedom $\theta\in[0,2\pi)$
\begin{align}
H_\Delta^\theta&\equiv~~2i\cos\frac{\theta}{2} \left[\Delta\,\sin \frac{\theta}{2} -\cos \frac{\theta}{2} \partial_\theta\right]~,\label{eq:hdff}\\
K_\Delta^\theta&\equiv-2i\sin \frac{\theta}{2} \left[\Delta\,\cos \frac{\theta}{2} +\sin\ \frac{\theta}{2} \partial_\theta\right]~,\label{eq:kdff}\\
D_\Delta^\theta&\equiv -i\left[\Delta\,\cos\theta+\sin\theta\,\partial_\theta\right]~.\label{eq:ddff}
\end{align}
These hermitian generators are related to the complexified ones of \eqref{def:sl2r} by 
\begin{equation}
H^\theta_\Delta = L_0^{\Delta,\theta} - \frac{1}{2}\left(L_+^{\Delta,\theta}+L_-^{\Delta,\theta}\right) ~, \quad K^\theta_\Delta = L_0^{\Delta,\theta}+ \frac{1}{2}\left(L_+^{\Delta,\theta}+L_-^{\Delta,\theta}\right)~, \quad D^\theta_\Delta=\frac{-i}{2}\left(L_+^{\Delta,\theta}-L_-^{\Delta,\theta}\right)~.
\end{equation}
The operators (\ref{eq:hdff}-\ref{eq:ddff}) obey the algebra
\begin{equation}
	\left[D^\theta_\Delta,H^\theta_\Delta\right]=iH^\theta_\Delta~,\qquad\left[D^\theta_\Delta,K^\theta_\Delta\right]=-iK^\theta_\Delta~,\qquad \left[K^\theta_\Delta,H^\theta_\Delta\right]=2iD^\theta_\Delta~.
\end{equation}
These square to the trivial quadratic Casimir
\begin{equation}
	\mathcal{C}^\theta =\frac{1}{2}(H^\theta_\Delta K^\theta_\Delta+K^\theta_\Delta H^\theta_\Delta)-\left(D^\theta_\Delta\right)^2=\Delta(\Delta-1)
\end{equation}
If we pick the standard inner product on this Hilbert space:
\begin{equation}
	(f,g)=\int_0^{2\pi}d\theta\, f^*(\theta)\,g(\theta)~,
\end{equation}
then the operators (\ref{eq:hdff}-\ref{eq:ddff}) are self-adjoint with respect to this inner product if and only if $\Delta=\frac{1}{2}(1+i\nu)$ with $\nu\in\mathbb{R}$, also known as the principal series.

\subsection*{Single particle Hilbert space}
To build the single particle Hilbert space, we construct the compact operator $L_0^{\Delta,\theta}$ and raising/lowering operators $L_\pm^{\Delta,\theta}$ 
 \begin{equation}\label{eq:dffopsqm}
 L_0^{\Delta,\theta}=\frac{1}{2}\left(H_\Delta^\theta+K_\Delta^\theta\right)=-i\partial_\theta~,\,\qquad L_\pm^{\Delta,\theta}=\frac{1}{2}\left(H_\Delta^\theta-K_\Delta^\theta\right)\mp iD_\Delta^\theta=e^{\mp i\theta}\left(\mp\Delta-i\partial_\theta\right)~,
 \end{equation}
 The Hilbert space is spanned by states of definite $\mathcal{C}^\theta$ and $L_0^{\Delta,\theta}$. These are states  $\psi_n(\theta)$ satisfying: 
\begin{equation}
	L_0^{\Delta,\theta}\,\psi_n(\theta)=-n\,\psi_n(\theta)~, \qquad \mathcal{C}^\theta \,\psi_n(\theta)=\Delta(\Delta-1)\,\psi_n(\theta)
\end{equation}
with $n\in\mathbb{Z}$. These imply the action of the raising and lowering operators will be: 
\begin{equation}\label{eq:Lpmaction}
L^{\Delta,\theta}_\pm\psi_n(\theta)=-(n\pm\Delta)\psi_{n\pm1}(\theta)
\end{equation}
Since the Casimir is trivial, the wavefunctions are easy to compute
\begin{equation}\label{eq:l0eigs}
	\psi_n(\theta)=\frac{1}{\sqrt{2\pi}}e^{-in\theta}
\end{equation}
and are orthonormal with respect to the standard inner product
\begin{equation}
	(\psi_k,\psi_n)=\int_0^{2\pi}d\theta\, \psi^*_k(\theta)\psi_n(\theta)=\delta_{kn}~.
\end{equation}

\subsection*{Two-particle Hilbert space}
Let us now construct the two-particle Hilbert space, built out of the tensor product of two single-particle Hilbert spaces on the conformal circle. We would like to thus consider the diagonal $SL(2,\mathbb{R})$ algebra constructed as
\begin{equation}
H\equiv H_{\Delta_1}^{\theta_1}+H_{\Delta_2}^{\theta_2}~,\qquad\qquad
K\equiv  H_{\Delta_1}^{\theta_1}+H_{\Delta_2}^{\theta_2}~, \qquad\qquad
D\equiv  D_{\Delta_1}^{\theta_1}+D_{\Delta_2}^{\theta_2}
\end{equation}
which obey the same algebra as before. Now, the quadratic Casimir operator
\begin{equation}
    	\mathcal{C} =\frac{1}{2}(H K+K H)-D^2
\end{equation}
is nontrivial and given explicitly by 
\begin{multline}
\mathcal{C}  =-4 \sin^2\left(\frac{\theta_1-\theta_2}{2}\right)\partial_{\theta_1}\partial_{\theta_2}+2\sin\left({\theta_1-\theta_2}\right)\left[\Delta_2\partial_{\theta_1}-\Delta_1\partial_{\theta_2}\right]\\+\left[\Delta_1(\Delta_1-1)+\Delta_2(\Delta_2-1)+2\Delta_1\Delta_2\cos(\theta_1-\theta_2)\right]~.
\end{multline}
To build the two-particle Hilbert space, we proceed exactly as before. First we construct the compact $L_0$ generator, as well as the raising/lowering operators $L_\pm$, defined as
 \begin{equation}
 L_0=\frac{1}{2}\left(H+K\right)=-i\partial_{\theta_1}-i\partial_{\theta_2}~,\,\qquad\qquad L_\pm=\frac{1}{2}\left(H-K\right)\mp iD~. 
\end{equation}
The two-particle Hilbert space is spanned by  wavefunctions that satisfy
\begin{equation}
	L_0\,\psi_n^\Delta(\theta_1,\theta_2)=-n\,\psi_n^\Delta(\theta_1,\theta_2)~, \qquad \mathcal{C} \,\psi_n^\Delta(\theta_1,\theta_2)=\Delta(\Delta-1)\,\psi_n^\Delta(\theta_1,\theta_2)~. \label{eq:twoparticlerep}
\end{equation}
What is interesting in the two-particle case, is that, while $\Delta_1$ and $\Delta_2$ are required to take values in the principal series by Hermiticity, the eigenvalue $\Delta$ is allowed to take values in any of the $SL(2,\mathbb{R})$ representations. It is possible to solve \eqref{eq:twoparticlerep} explicitly. The solution is: 
\begin{align}
\psi_n^\Delta(\theta_1,\theta_2)=&\frac{e^{-in\left(\frac{\theta_1+\theta_2}{2}\right)+i\Delta\left(\frac{\theta_1-\theta_2}{2}\right)}}{\left[\sin^2\left(\frac{\theta_1-\theta_2}{2}\right)\right]^{\frac{\Delta_1+\Delta_2-\Delta}{2}}}\times\nonumber\\
&\Bigg[~~c_1 e^{+i (n+\Delta_1-\Delta_2)\left(\frac{\theta_1-\theta_2}{2}\right)}\, {}_2F_1\left(\Delta+n,\,\Delta+\Delta_1-\Delta_2,\,1+n+\Delta_1-\Delta_2,\,e^{i(\theta_1-\theta_2)}\right)\nonumber\\
&+c_2 e^{-i (n+\Delta_1-\Delta_2)\left(\frac{\theta_1-\theta_2}{2}\right)}\, {}_2F_1\left(\Delta-n,\,\Delta-\Delta_1+\Delta_2,\,1-n-\Delta_1+\Delta_2,\,e^{i(\theta_1-\theta_2)}\right)\bigg]~.\label{eq:2partwavefunction}
\end{align}
The choices of $c_1$ and $c_2$ are predicated by the normaliseability of these wavefunctions using the inner product
\begin{equation}\label{eq:twoparticlecirclenorm}
	(\psi_n^\Delta,\psi_m^{\Delta'})=\int_0^{2\pi}\int_0^{2\pi}d\theta_1 d\theta_2\, (\psi_n^\Delta(\theta_1,\theta_2))^*\psi_m^{\Delta'}(\theta_1,\theta_2)=c_{n}\,\delta_{nm}\delta_{\Delta\Delta'}~.
\end{equation}
where $\delta_{\Delta\Delta'}$ is a stand-in for the appropriate Kronecker or Dirac delta function, depending upon  which representation we are dealing with, and $c_n$ is the appropriate coefficient that ensures that the generators act faithfully on the representation, see \cref{App:Repthy}. To actually check normalizability, it is easier to change coordinates to 
\begin{equation}
    x\equiv\frac{\theta_1+\theta_2}{2}\,\qquad\qquad y\equiv \theta_1-\theta_2~, 
\end{equation}
and the integration measure becomes
\begin{equation}
    \int_0^{2\pi}\int_0^{2\pi}d\theta_1 d\theta_2\rightarrow \int_0^{2\pi}dx\int_{-x}^{x}dy~.
\end{equation}

\subsection{Discrete highest weight \texorpdfstring{$\Delta\in D_\Delta^+$}{dplus}}\label{app:dplus}
To check that the $D_\Delta^+$ UIR appears in the tensor product Hilbert space of two Principal Series quantum mechanical degrees of freedom, we simply need to check that the highest weight wavefunction $\psi_{n=-\Delta}^\Delta(\theta_1,\theta_2)$ is normalizable. This wavefunction satisfies
\begin{equation}
  L_0 \psi_{n=-\Delta}^\Delta(\theta_1,\theta_2)=\Delta\,\psi_{n=-\Delta}^\Delta(\theta_1,\theta_2)~,\qquad\qquad  L_+ \psi_{n=-\Delta}^\Delta(\theta_1,\theta_2)=0~, \label{eq:highestweightcircle}
\end{equation}
for $\Delta=1+t$ and $t=0,1,2,\dots$. The rest of the highest weight module can be generated by acting successively with $L_-$. It is easy to find the wavefunction satisfying \eqref{eq:highestweightcircle}:
\begin{equation}
    \psi^{1+t}_{-(1+t)}(\theta_1,\theta_2)=\frac{1}{N(t,\mu,\nu)}\, e^{i(1+t)\frac{\theta_1+\theta_2}{2}+\frac{\mu-\nu}{4}(\theta_1-\theta_2)}\left[\sin \left(\frac{\theta_1-\theta_2}{2}\right)\right]^{t-i\left(\frac{\mu-\nu}{2}\right)}~,
\end{equation}
where we have taken $\Delta_1\equiv\frac{1}{2}(1+i\nu)$ and $\Delta_2\equiv\frac{1}{2}(1+i\mu)$, and $N(t,\mu,\nu)$ is an overall factor that ensures that this wavefunction is properly normalized (see \eqref{eq:discretesernorm}). Normalizability of the wavefunction requires $t\geq0$ and single-valuedness of the center-of-mass wavefunction fixes $t$ to be an integer. It is straightforward to verify that these wavefunctions are square-integrable with respect to the inner product \eqref{eq:twoparticlecirclenorm}, and a simple calculation yields
\begin{equation}\label{eq:normalizationhighestweight}
    N(t,\mu,\nu)=2^{1-t}\sinh\left[\frac{\pi}{2}(\mu-\nu)\right]\sqrt{(2t+1)!\sum_{s=0}^{2t}\frac{(-1)^{1+s-t}\left[(s-t)^2-\left(\frac{\mu-\nu}{2}\right)^2\right]}{\left[(s-t)^2+\left(\frac{\mu-\nu}{2}\right)^2\right]^2}\binom{2t}{s}}~.
\end{equation}
With this normalization, we have verified that the action of $L_-$ on this state follows \eqref{eq:Lpmaction}. Thus we have shown that, in this quantum mechanics of two decoupled principal series degrees of freedom, the tensor product Hilbert space contains every possible discrete highest weight module. Let us now show that this is also true for the discrete lowest weight modules. 

\subsection{Discrete lowest weight \texorpdfstring{$\Delta\in D_\Delta^-$}{dminus}}\label{app:dminus}
This exercise is exactly the same as before. Now we search for a wavefunction that satisfies 
\begin{equation}
  L_0 \psi_{n=\Delta}^\Delta(\theta_1,\theta_2)=-\Delta\,\psi_{n=\Delta}^\Delta(\theta_1,\theta_2)~,\qquad\qquad  L_- \psi_{n=\Delta}^\Delta(\theta_1,\theta_2)=0~, \label{eq:lowesttweightcircle}
\end{equation}
for $\Delta=1+t$ and $t=0,1,2,\dots$. The remainder of the lowest weight module can be generated by acting successively with $L_+$. It is again easy to find the wavefunction satisfying \eqref{eq:lowesttweightcircle}:
\begin{equation}
    \psi^{1+t}_{1+t}(\theta_1,\theta_2)=\frac{1}{N(t,\mu,\nu)}\, e^{-i(1+t)\frac{\theta_1+\theta_2}{2}-\frac{\mu-\nu}{4}(\theta_1-\theta_2)}\left[\sin \left(\frac{\theta_1-\theta_2}{2}\right)\right]^{t-i\left(\frac{\mu-\nu}{2}\right)}~,
\end{equation}
where we have again taken $\Delta_1\equiv\frac{1}{2}(1+i\nu)$ and $\Delta_2\equiv\frac{1}{2}(1+i\mu)$, and $N(t,\mu,\nu)$ is given in \eqref{eq:normalizationhighestweight} and ensures appropriate normalization according to \eqref{eq:discretesernorm}.

\subsection{Principal and complementary series}

For normalizability of the Principal or complementary series, we will re-solve \eqref{eq:twoparticlerep}, this time for $n=0$, specifically. The general solution can be written as
\begin{equation}
    \psi_0^\Delta=\frac{1}{\left[\sin^2\left(\frac{\theta_1-\theta_2}{2}\right)\right]^{\frac{\Delta_1+\Delta_2-\frac{1}{2}}{2}}}\left\lbrace b_1\, P_{\Delta_1-\Delta_2-\frac{1}{2}}^{\Delta-\frac{1}{2}}\left[\cos\left(\frac{\theta_1-\theta_2}{2}\right)\right]+ b_2\, Q_{\Delta_1-\Delta_2-\frac{1}{2}}^{\Delta-\frac{1}{2}}\left[\cos\left(\frac{\theta_1-\theta_2}{2}\right)\right]\right\rbrace~, 
    \end{equation}
 where $P_\mu^\nu(x)$ and $Q_\mu^\nu(x)$ are associated Legendre functions. 
 It suffices to determine if $\psi_0^\Delta$ can be made normalizable for some choice of $b_{1,2}$. Specifically, we need, for the complementary series:
 \begin{equation}
     \int_0^{2\pi}\int_0^{2\pi}d\theta_1 d\theta_2\, (\psi_0^\Delta(\theta_1,\theta_2))^*\psi_0^{\Delta'}(\theta_1,\theta_2)=\delta(\Delta-\Delta')~.
 \end{equation}
 However, for $\Delta,\Delta'\in[0,1]$, it is easy to verify that the above wavefunctions are not oscillatory over the domain of $\theta_{1,2}\in [0,2\pi]$, making it impossible for the wavefunctions to be plane-wave normalizable. This precludes the complementary series from appearing in the tensor product of two principal series Hilbert spaces.  
 
 For the principal series, we need:
 \begin{equation}
     \int_0^{2\pi}\int_0^{2\pi}d\theta_1 d\theta_2\, \left(\psi_0^{\frac{1}{2}+i\gamma}(\theta_1,\theta_2)\right)^*\psi_0^{\frac{1}{2}+i\gamma'}(\theta_1,\theta_2)=\delta(\gamma-\gamma')~. 
 \end{equation}
 For these parameters, the above wavefunctions are appropriately oscillatory, and if we were sufficiently patient, we could use the results of \cite{bielski2013orthogonality} to find which choice of $b_{1,2}$ gives the desired normalization, but we leave this as an exercise.  
 
\subsection{Tensor products from Harish-Chandra characters\label{sec:tensorchar} }

One way to demonstrate the presence of discrete series representations in the decomposition of the tensor product of principle series representations is a simple calculation using the Harish-Chandra characters. Starting from $H_\Delta^\theta$ in \eqref{eq:hdff}, the principle series character for the representation $\Delta=\frac{1}{2}(1+i\nu)$ can be derived as follows (see section in A.2 of \cite{Anninos:2020hfj} or section 3 of \cite{Anous:2020nxu}):
\begin{equation}\label{eq:principalcharacter}
    \chi_{\pi_\nu}(t)\equiv\int d\theta\bra{\theta}e^{-itH_\Delta^\theta}\ket{\theta} = \frac{e^{-\frac{1}{2}(1+i \nu)t}+ e^{-\frac{1}{2}(1-i \nu)t}}{|1-e^{-t}|} = \left\lvert\csch \frac{t}{2}\right\rvert \cos \frac{\nu t}{2}~.
\end{equation}
On the other hand, for $\Delta \in \mathbb{Z}_+$ the discrete series Harish-Chandra character is (see equation (5.18) of \cite{Sun:2021thf}):
\begin{equation}
 \chi_{D^+_\Delta\oplus D^-_\Delta}(t) = \frac{2e^{-\Delta t}}{1-e^{-t}}  ~, 
\end{equation}
where the factor of $2$ comes from the fact that $D^+_\Delta$ and $D^-_\Delta$ have the same character.

We will now show that the discrete series appears in the tensor product Hilbert space of two principal series representations. To do this, let us multiply the Harish-Chandra characters of two different principal series representations. 
\begin{align}
    \chi_{\pi_\nu}(t) \times \chi_{\pi_\mu} (t)&= \csch^2 \frac{t}{2} \cos\frac{\nu t}{2}\cos \frac{\mu t}{2} \, ,\label{eq:prestep} \\
     &= \int_{-\infty}^\infty  d\lambda\,\rho( \lambda)\, \chi_{\pi_\lambda}(t)  \, . 
\end{align}
We can formally recover the density $\rho(\lambda)$ by dividing \eqref{eq:prestep} by $\left\lvert\csch \frac{t}{2}\right\rvert$ and performing inverse cosine transform (see \cite{Penedones:2023uqc} for a similar discussion). The result is
\begin{align}
    \rho(\lambda ) &= -\frac{1}{8\pi}\left\lbrace 8\gamma + \sum_{\sigma_1,\sigma_2,\sigma_3= 0}^1 \psi\left( \frac{1}{2}+ \frac{i}{2}\left[(-1)^{\sigma_1} \nu + (-1)^{\sigma_2} \mu + (-1)^{\sigma_3} \lambda)\right]\right)\right\rbrace \,.
\end{align}
Here $\psi(x)\equiv\frac{\Gamma'(x)}{\Gamma(x)}$ is the digamma function, and $\gamma$ The Euler-Mascheroni constant. The digamma function $\psi(x)$ has poles whenever $x$ is a non-positive integer, with residue $-1$. Note, given the form or $\chi_{\pi_\lambda}(t)$ in \eqref{eq:principalcharacter}, that $\rho(\lambda)$ and $\rho(\lambda)+c$ will give the same result when integrated against a character. Thus $\rho(\lambda)$ is not a well-defined concept in its own right, but makes sense when integrated against a certain class of  distributions. 

Let us consider the case $\mu = \nu $, where the appearance of the discrete series manifests itself most clearly since a piece of the density is independent of $\nu$. In this case we can write
\begin{equation}
    \rho_{\mu=\nu}(\lambda)\equiv \rho^{\rm indep.}(\lambda)+\rho^{\rm dep.}(\lambda)
\end{equation}
with
\begin{equation}
    \rho^{\rm indep.}(\lambda)=-\frac{1}{4\pi}\left\lbrace\psi\left( \frac{1}{2}(1+i\lambda)\right)+\psi\left( \frac{1}{2}(1-i\lambda)\right)\right\rbrace
\end{equation}
and
\begin{equation}
    \rho^{\rm dep.}(\lambda)=-\frac{1}{8\pi}\left\lbrace8\gamma+\sum_{\sigma_1,\sigma_2=0}^1\psi\left( \frac{1}{2}+\frac{i}{2}\left[(-1)^{\sigma_1}2\nu+(-1)^{\sigma_2}\lambda\right]\right)\right\rbrace~.
\end{equation}
Now we proceed to evaluate
\begin{equation}
    \int_{-\infty}^\infty  d\lambda\,\rho^{\rm indep.}( \lambda)\, \chi_{\pi_\lambda}(t) ~.
\end{equation}
The functions $\psi\left( \frac{1}{2}(1\pm i\lambda)\right)$ have poles for $\lambda=\pm 2i\left(n+\tfrac{1}{2}\right)$, respectively, with $n=0,1,2,\dots$. Proceeding carefully, for terms multiplying  $e^{-\tfrac{t}{2}(1+i\lambda)}$ we must close the contour in the lower-half $\lambda$-plane, and for terms multiplying  $e^{-\tfrac{t}{2}(1-i\lambda)}$ we must close the contour in the upper-half $\lambda$-plane. Carefully noting the orientation of each of the contours, we obtain
\begin{equation}
    \int_{-\infty}^\infty  d\lambda\,\rho^{\rm indep.}( \lambda)\, \chi_{\pi_\lambda}(t)=2\sum_{n=0}^{\infty}\frac{ e^{-(1+n)t}}{1-e^{-t}}=\sum_{\Delta=1}^\infty \chi_{D^+_\Delta\oplus D^-_\Delta}(t) ~.
\end{equation}
Thus
\begin{equation}
     \chi^2_{\pi_\nu}(t)=\sum_{\Delta=1}^\infty \chi_{D^+_\Delta\oplus D^-_\Delta}(t)+\int_{-\infty}^\infty  d\lambda\,\rho^{\rm dep.}( \lambda)\, \chi_{\pi_\lambda}(t)~,
\end{equation}
and we clearly see the discrete series Hilbert space emerging from an analysis of the character of the tensor product of two principal series representations. This nicely complements the analysis in appendices \ref{app:dplus} and \ref{app:dminus}.

\bibliographystyle{utphys.bst}
\bibliography{references}

\providecommand{\href}[2]{#2}\begingroup\raggedright\begin{thebibliography}{100}

\bibitem{Wigner:1939cj}
E.~P. Wigner, ``{On Unitary Representations of the Inhomogeneous Lorentz
  Group},'' \href{http://dx.doi.org/10.2307/1968551}{{\em Annals Math.}
  {\bfseries 40} (1939) 149--204}.

\bibitem{Weinberg:1995mt}
S.~Weinberg, {\em {The Quantum theory of fields. Vol. 1: Foundations}}.
\newblock Cambridge University Press, 6, 2005.

\bibitem{Dirac:1935zz}
P.~A.~M. Dirac, ``{The Electron Wave Equation in De-Sitter Space},''
  \href{http://dx.doi.org/10.2307/1968649}{{\em Annals Math.} {\bfseries 36}
  (1935) 657--669}.

\bibitem{nachtmann1967quantum}
O.~Nachtmann, ``Quantum theory in de-sitter space,'' {\em Communications in
  Mathematical Physics} {\bfseries 6} (1967) 1--16.

\bibitem{borner1969classical}
G.~B{\"o}rner and H.~P. D{\"u}rr, ``Classical and quantum fields in de sitter
  space.,'' tech. rep., Max-Planck-Institut fuer Physik und Astrophysik,
  Munich, 1969.

\bibitem{Tagirov:1972vv}
E.~A. Tagirov, ``{Consequences of field quantization in de Sitter type
  cosmological models},''
  \href{http://dx.doi.org/10.1016/0003-4916(73)90047-X}{{\em Annals Phys.}
  {\bfseries 76} (1973) 561--579}.

\bibitem{Deser:1983mm}
S.~Deser and R.~I. Nepomechie, ``{Gauge Invariance Versus Masslessness in De
  Sitter Space},'' \href{http://dx.doi.org/10.1016/0003-4916(84)90156-8}{{\em
  Annals Phys.} {\bfseries 154} (1984) 396}.

\bibitem{Deser:2001us}
S.~Deser and A.~Waldron, ``{Partial masslessness of higher spins in (A)dS},''
  \href{http://dx.doi.org/10.1016/S0550-3213(01)00212-7}{{\em Nucl. Phys. B}
  {\bfseries 607} (2001) 577--604},
  \href{http://arxiv.org/abs/hep-th/0103198}{{\ttfamily arXiv:hep-th/0103198}}.

\bibitem{Deser:2001pe}
S.~Deser and A.~Waldron, ``{Gauge invariances and phases of massive higher
  spins in (A)dS},''
  \href{http://dx.doi.org/10.1103/PhysRevLett.87.031601}{{\em Phys. Rev. Lett.}
  {\bfseries 87} (2001) 031601},
  \href{http://arxiv.org/abs/hep-th/0102166}{{\ttfamily arXiv:hep-th/0102166}}.

\bibitem{Deser:2003gw}
S.~Deser and A.~Waldron, ``{Arbitrary spin representations in de Sitter from dS
  / CFT with applications to dS supergravity},''
  \href{http://dx.doi.org/10.1016/S0550-3213(03)00348-1}{{\em Nucl. Phys. B}
  {\bfseries 662} (2003) 379--392},
  \href{http://arxiv.org/abs/hep-th/0301068}{{\ttfamily arXiv:hep-th/0301068}}.

\bibitem{Higuchi:2010xt}
A.~Higuchi, D.~Marolf, and I.~A. Morrison, ``{On the Equivalence between
  Euclidean and In-In Formalisms in de Sitter QFT},''
  \href{http://dx.doi.org/10.1103/PhysRevD.83.084029}{{\em Phys. Rev. D}
  {\bfseries 83} (2011) 084029},
  \href{http://arxiv.org/abs/1012.3415}{{\ttfamily arXiv:1012.3415 [gr-qc]}}.

\bibitem{Vasiliev:1990en}
M.~A. Vasiliev, ``{Consistent equation for interacting gauge fields of all
  spins in (3+1)-dimensions},''
  \href{http://dx.doi.org/10.1016/0370-2693(90)91400-6}{{\em Phys. Lett. B}
  {\bfseries 243} (1990) 378--382}.

\bibitem{Bros:1990cu}
J.~Bros, ``{Complexified de Sitter space: Analytic causal kernels and
  Kallen-Lehmann type representation},''
  \href{http://dx.doi.org/10.1016/0920-5632(91)90119-Y}{{\em Nucl. Phys. B
  Proc. Suppl.} {\bfseries 18} (1991) 22--28}.

\bibitem{Bros:1995js}
J.~Bros and U.~Moschella, ``{Two point functions and quantum fields in de
  Sitter universe},'' \href{http://dx.doi.org/10.1142/S0129055X96000123}{{\em
  Rev. Math. Phys.} {\bfseries 8} (1996) 327--392},
  \href{http://arxiv.org/abs/gr-qc/9511019}{{\ttfamily arXiv:gr-qc/9511019}}.

\bibitem{Bros:2009bz}
J.~Bros, H.~Epstein, M.~Gaudin, U.~Moschella, and V.~Pasquier, ``{Triangular
  invariants, three-point functions and particle stability on the de Sitter
  universe},'' \href{http://dx.doi.org/10.1007/s00220-009-0875-4}{{\em Commun.
  Math. Phys.} {\bfseries 295} (2010) 261--288},
  \href{http://arxiv.org/abs/0901.4223}{{\ttfamily arXiv:0901.4223 [hep-th]}}.

\bibitem{Bros:2010wa}
J.~Bros, H.~Epstein, and U.~Moschella, ``{Scalar tachyons in the de Sitter
  universe},'' \href{http://dx.doi.org/10.1007/s11005-010-0406-4}{{\em Lett.
  Math. Phys.} {\bfseries 93} (2010) 203--211},
  \href{http://arxiv.org/abs/1003.1396}{{\ttfamily arXiv:1003.1396 [hep-th]}}.

\bibitem{Bros:2010rku}
J.~Bros, H.~Epstein, and U.~Moschella, ``{Particle decays and stability on the
  de Sitter universe},''
  \href{http://dx.doi.org/10.1007/s00023-010-0042-7}{{\em Annales Henri
  Poincare} {\bfseries 11} (2010) 611--658},
  \href{http://arxiv.org/abs/0812.3513}{{\ttfamily arXiv:0812.3513 [hep-th]}}.

\bibitem{Epstein:2012zz}
H.~Epstein, ``{Remarks on quantum field theory on de Sitter and anti-de Sitter
  space-times},'' \href{http://dx.doi.org/10.1007/s12043-012-0312-7}{{\em
  Pramana} {\bfseries 78} (2012) 853--864}.

\bibitem{Epstein:2014jaa}
H.~Epstein and U.~Moschella, ``{de Sitter tachyons and related topics},''
  \href{http://dx.doi.org/10.1007/s00220-015-2308-x}{{\em Commun. Math. Phys.}
  {\bfseries 336} no.~1, (2015) 381--430},
  \href{http://arxiv.org/abs/1403.3319}{{\ttfamily arXiv:1403.3319 [hep-th]}}.

\bibitem{Epstein:2018wfh}
H.~Epstein and U.~Moschella, ``{Topological surprises in de Sitter QFT in
  two-dimensions},'' \href{http://dx.doi.org/10.1142/S0217751X18450094}{{\em
  Int. J. Mod. Phys. A} {\bfseries 33} no.~34, (2018) 1845009},
  \href{http://arxiv.org/abs/1901.10874}{{\ttfamily arXiv:1901.10874
  [hep-th]}}.

\bibitem{Joung:2006gj}
E.~Joung, J.~Mourad, and R.~Parentani, ``{Group theoretical approach to quantum
  fields in de Sitter space. I. The Principle series},''
  \href{http://dx.doi.org/10.1088/1126-6708/2006/08/082}{{\em JHEP} {\bfseries
  08} (2006) 082}, \href{http://arxiv.org/abs/hep-th/0606119}{{\ttfamily
  arXiv:hep-th/0606119}}.

\bibitem{Joung:2007je}
E.~Joung, J.~Mourad, and R.~Parentani, ``{Group theoretical approach to quantum
  fields in de Sitter space. II. The complementary and discrete series},''
  \href{http://dx.doi.org/10.1088/1126-6708/2007/09/030}{{\em JHEP} {\bfseries
  09} (2007) 030}, \href{http://arxiv.org/abs/0707.2907}{{\ttfamily
  arXiv:0707.2907 [hep-th]}}.

\bibitem{Marolf:2008it}
D.~Marolf and I.~Morrison, ``{Group Averaging of massless scalar fields in 1+1
  de Sitter},'' \href{http://dx.doi.org/10.1088/0264-9381/26/3/035001}{{\em
  Class. Quant. Grav.} {\bfseries 26} (2009) 035001},
  \href{http://arxiv.org/abs/0808.2174}{{\ttfamily arXiv:0808.2174 [gr-qc]}}.

\bibitem{Marolf:2010nz}
D.~Marolf and I.~A. Morrison, ``{The IR stability of de Sitter QFT: results at
  all orders},'' \href{http://dx.doi.org/10.1103/PhysRevD.84.044040}{{\em Phys.
  Rev. D} {\bfseries 84} (2011) 044040},
  \href{http://arxiv.org/abs/1010.5327}{{\ttfamily arXiv:1010.5327 [gr-qc]}}.

\bibitem{Basile:2016aen}
T.~Basile, X.~Bekaert, and N.~Boulanger, ``{Mixed-symmetry fields in de Sitter
  space: a group theoretical glance},''
  \href{http://dx.doi.org/10.1007/JHEP05(2017)081}{{\em JHEP} {\bfseries 05}
  (2017) 081}, \href{http://arxiv.org/abs/1612.08166}{{\ttfamily
  arXiv:1612.08166 [hep-th]}}.

\bibitem{Marolf:2010zp}
D.~Marolf and I.~A. Morrison, ``{The IR stability of de Sitter: Loop
  corrections to scalar propagators},''
  \href{http://dx.doi.org/10.1103/PhysRevD.82.105032}{{\em Phys. Rev. D}
  {\bfseries 82} (2010) 105032},
  \href{http://arxiv.org/abs/1006.0035}{{\ttfamily arXiv:1006.0035 [gr-qc]}}.

\bibitem{Marolf:2011sh}
D.~Marolf and I.~A. Morrison, ``{The IR stability of de Sitter QFT: Physical
  initial conditions},''
  \href{http://dx.doi.org/10.1007/s10714-011-1233-3}{{\em Gen. Rel. Grav.}
  {\bfseries 43} (2011) 3497--3530},
  \href{http://arxiv.org/abs/1104.4343}{{\ttfamily arXiv:1104.4343 [gr-qc]}}.

\bibitem{Anninos:2019oka}
D.~Anninos, D.~M. Hofman, and J.~Kruthoff, ``{Charged Quantum Fields in
  AdS$_2$},'' \href{http://dx.doi.org/10.21468/SciPostPhys.7.4.054}{{\em
  SciPost Phys.} {\bfseries 7} no.~4, (2019) 054},
  \href{http://arxiv.org/abs/1906.00924}{{\ttfamily arXiv:1906.00924
  [hep-th]}}.

\bibitem{Anous:2020nxu}
T.~Anous and J.~Skulte, ``{An invitation to the principal series},''
  \href{http://dx.doi.org/10.21468/SciPostPhys.9.3.028}{{\em SciPost Phys.}
  {\bfseries 9} no.~3, (2020) 028},
  \href{http://arxiv.org/abs/2007.04975}{{\ttfamily arXiv:2007.04975
  [hep-th]}}.

\bibitem{Sengor:2019mbz}
G.~Seng\"or and C.~Skordis, ``{Unitarity at the Late time Boundary of de
  Sitter},'' \href{http://dx.doi.org/10.1007/JHEP06(2020)041}{{\em JHEP}
  {\bfseries 06} (2020) 041}, \href{http://arxiv.org/abs/1912.09885}{{\ttfamily
  arXiv:1912.09885 [hep-th]}}.

\bibitem{Sengor:2021zlc}
G.~Sengor and C.~Skordis, ``{Scalar two-point functions at the late-time
  boundary of de Sitter},'' \href{http://arxiv.org/abs/2110.01635}{{\ttfamily
  arXiv:2110.01635 [hep-th]}}.

\bibitem{Sun:2021thf}
Z.~Sun, ``{A note on the representations of $\text{SO}(1,d+1)$},''
  \href{http://arxiv.org/abs/2111.04591}{{\ttfamily arXiv:2111.04591
  [hep-th]}}.

\bibitem{Letsios:2023qzq}
V.~A. Letsios, ``{(Non-)unitarity of strictly and partially massless fermions
  on de Sitter space},'' \href{http://arxiv.org/abs/2303.00420}{{\ttfamily
  arXiv:2303.00420 [hep-th]}}.

\bibitem{Enayati:2022hed}
M.~Enayati, J.-P. Gazeau, H.~Pejhan, and A.~Wang, ``{The de Sitter group and
  its representations: a window on the notion of de Sitterian elementary
  systems},'' \href{http://arxiv.org/abs/2201.11457}{{\ttfamily
  arXiv:2201.11457 [math-ph]}}.

\bibitem{Takook:2023yum}
M.~V. Takook, J.~P. Gazeau, and E.~Huguet, ``{Asymptotic states and $S$-matrix
  operator in de Sitter ambient space formalism},''
  \href{http://arxiv.org/abs/2304.04756}{{\ttfamily arXiv:2304.04756
  [hep-th]}}.

\bibitem{Loparco:2023rug}
M.~Loparco, J.~Penedones, K.~Salehi~Vaziri, and Z.~Sun, ``{The
  K\"all\'en-Lehmann representation in de Sitter spacetime},''
  \href{http://arxiv.org/abs/2306.00090}{{\ttfamily arXiv:2306.00090
  [hep-th]}}.

\bibitem{Higuchi:1986wu}
A.~Higuchi, ``{Symmetric Tensor Spherical Harmonics on the $N$ Sphere and Their
  Application to the De Sitter Group SO($N$,1)},''
  \href{http://dx.doi.org/10.1063/1.527513}{{\em J. Math. Phys.} {\bfseries 28}
  (1987) 1553}. [Erratum: J.Math.Phys. 43, 6385 (2002)].

\bibitem{barutfronsdal}
A.~O. Barut and C.~Fronsdal, ``On non-compact groups. ii. representations of
  the 2 + 1 lorentz group,'' {\em Proceedings of the Royal Society of London.
  Series A, Mathematical and Physical Sciences} {\bfseries 287} no.~1411,
  (1965) 532--548. \url{http://www.jstor.org/stable/2415039}.

\bibitem{mukunda}
J.~G. Kuriyan, N.~Mukunda, and E.~C.~G. Sudarshan, ``Master analytic
  representation: Reduction of o(2, 1) in an o(1, 1) basis,''
  \href{http://dx.doi.org/10.1063/1.1664551}{{\em Journal of Mathematical
  Physics} {\bfseries 9} no.~12, (1968) 2100--2108},
  \href{http://arxiv.org/abs/https://doi.org/10.1063/1.1664551}{{\ttfamily
  https://doi.org/10.1063/1.1664551}}. \url{https://doi.org/10.1063/1.1664551}.

\bibitem{Witten:2001kn}
E.~Witten, ``{Quantum gravity in de Sitter space},'' in {\em {Strings 2001:
  International Conference}}.
\newblock 6, 2001.
\newblock \href{http://arxiv.org/abs/hep-th/0106109}{{\ttfamily
  arXiv:hep-th/0106109}}.

\bibitem{Maldacena:2002vr}
J.~M. Maldacena, ``{Non-Gaussian features of primordial fluctuations in single
  field inflationary models},''
  \href{http://dx.doi.org/10.1088/1126-6708/2003/05/013}{{\em JHEP} {\bfseries
  05} (2003) 013}, \href{http://arxiv.org/abs/astro-ph/0210603}{{\ttfamily
  arXiv:astro-ph/0210603}}.

\bibitem{Strominger:2001pn}
A.~Strominger, ``{The dS / CFT correspondence},''
  \href{http://dx.doi.org/10.1088/1126-6708/2001/10/034}{{\em JHEP} {\bfseries
  10} (2001) 034}, \href{http://arxiv.org/abs/hep-th/0106113}{{\ttfamily
  arXiv:hep-th/0106113}}.

\bibitem{Anninos:2011ui}
D.~Anninos, T.~Hartman, and A.~Strominger, ``{Higher Spin Realization of the
  dS/CFT Correspondence},''
  \href{http://dx.doi.org/10.1088/1361-6382/34/1/015009}{{\em Class. Quant.
  Grav.} {\bfseries 34} no.~1, (2017) 015009},
  \href{http://arxiv.org/abs/1108.5735}{{\ttfamily arXiv:1108.5735 [hep-th]}}.

\bibitem{Anninos:2017eib}
D.~Anninos, F.~Denef, R.~Monten, and Z.~Sun, ``{Higher Spin de Sitter Hilbert
  Space},'' \href{http://dx.doi.org/10.1007/JHEP10(2019)071}{{\em JHEP}
  {\bfseries 10} (2019) 071}, \href{http://arxiv.org/abs/1711.10037}{{\ttfamily
  arXiv:1711.10037 [hep-th]}}.

\bibitem{Isono:2020qew}
H.~Isono, H.~M. Liu, and T.~Noumi, ``{Wavefunctions in dS/CFT revisited:
  principal series and double-trace deformations},''
  \href{http://dx.doi.org/10.1007/JHEP04(2021)166}{{\em JHEP} {\bfseries 04}
  (2021) 166}, \href{http://arxiv.org/abs/2011.09479}{{\ttfamily
  arXiv:2011.09479 [hep-th]}}.

\bibitem{Anninos:2014lwa}
D.~Anninos, T.~Anous, D.~Z. Freedman, and G.~Konstantinidis, ``{Late-time
  Structure of the Bunch-Davies De Sitter Wavefunction},''
  \href{http://dx.doi.org/10.1088/1475-7516/2015/11/048}{{\em JCAP} {\bfseries
  11} (2015) 048}, \href{http://arxiv.org/abs/1406.5490}{{\ttfamily
  arXiv:1406.5490 [hep-th]}}.

\bibitem{Anninos:2014hia}
D.~Anninos, R.~Mahajan, D.~Radi\v{c}evi\'c, and E.~Shaghoulian,
  ``{Chern-Simons-Ghost Theories and de Sitter Space},''
  \href{http://dx.doi.org/10.1007/JHEP01(2015)074}{{\em JHEP} {\bfseries 01}
  (2015) 074}, \href{http://arxiv.org/abs/1405.1424}{{\ttfamily arXiv:1405.1424
  [hep-th]}}.

\bibitem{Arkani-Hamed:2015bza}
N.~Arkani-Hamed and J.~Maldacena, ``{Cosmological Collider Physics},''
  \href{http://arxiv.org/abs/1503.08043}{{\ttfamily arXiv:1503.08043
  [hep-th]}}.

\bibitem{Sleight:2019hfp}
C.~Sleight and M.~Taronna, ``{Bootstrapping Inflationary Correlators in Mellin
  Space},'' \href{http://dx.doi.org/10.1007/JHEP02(2020)098}{{\em JHEP}
  {\bfseries 02} (2020) 098}, \href{http://arxiv.org/abs/1907.01143}{{\ttfamily
  arXiv:1907.01143 [hep-th]}}.

\bibitem{Sleight:2020obc}
C.~Sleight and M.~Taronna, ``{From AdS to dS Exchanges: Spectral
  Representation, Mellin Amplitudes and Crossing},''
  \href{http://arxiv.org/abs/2007.09993}{{\ttfamily arXiv:2007.09993
  [hep-th]}}.

\bibitem{Goodhew:2020hob}
H.~Goodhew, S.~Jazayeri, and E.~Pajer, ``{The Cosmological Optical Theorem},''
  \href{http://dx.doi.org/10.1088/1475-7516/2021/04/021}{{\em JCAP} {\bfseries
  04} (2021) 021}, \href{http://arxiv.org/abs/2009.02898}{{\ttfamily
  arXiv:2009.02898 [hep-th]}}.

\bibitem{Melville:2021lst}
S.~Melville and E.~Pajer, ``{Cosmological Cutting Rules},''
  \href{http://dx.doi.org/10.1007/JHEP05(2021)249}{{\em JHEP} {\bfseries 05}
  (2021) 249}, \href{http://arxiv.org/abs/2103.09832}{{\ttfamily
  arXiv:2103.09832 [hep-th]}}.

\bibitem{Arkani-Hamed:2018kmz}
N.~Arkani-Hamed, D.~Baumann, H.~Lee, and G.~L. Pimentel, ``{The Cosmological
  Bootstrap: Inflationary Correlators from Symmetries and Singularities},''
  \href{http://dx.doi.org/10.1007/JHEP04(2020)105}{{\em JHEP} {\bfseries 04}
  (2020) 105}, \href{http://arxiv.org/abs/1811.00024}{{\ttfamily
  arXiv:1811.00024 [hep-th]}}.

\bibitem{Baumann:2019oyu}
D.~Baumann, C.~Duaso~Pueyo, A.~Joyce, H.~Lee, and G.~L. Pimentel, ``{The
  cosmological bootstrap: weight-shifting operators and scalar seeds},''
  \href{http://dx.doi.org/10.1007/JHEP12(2020)204}{{\em JHEP} {\bfseries 12}
  (2020) 204}, \href{http://arxiv.org/abs/1910.14051}{{\ttfamily
  arXiv:1910.14051 [hep-th]}}.

\bibitem{Baumann:2020dch}
D.~Baumann, C.~Duaso~Pueyo, A.~Joyce, H.~Lee, and G.~L. Pimentel, ``{The
  Cosmological Bootstrap: Spinning Correlators from Symmetries and
  Factorization},'' \href{http://arxiv.org/abs/2005.04234}{{\ttfamily
  arXiv:2005.04234 [hep-th]}}.

\bibitem{Hogervorst:2021uvp}
M.~Hogervorst, J.~a. Penedones, and K.~S. Vaziri, ``{Towards the
  non-perturbative cosmological bootstrap},''
  \href{http://arxiv.org/abs/2107.13871}{{\ttfamily arXiv:2107.13871
  [hep-th]}}.

\bibitem{Penedones:2023uqc}
J.~Penedones, K.~Salehi~Vaziri, and Z.~Sun, ``{Hilbert space of Quantum Field
  Theory in de Sitter spacetime},''
  \href{http://arxiv.org/abs/2301.04146}{{\ttfamily arXiv:2301.04146
  [hep-th]}}.

\bibitem{Arkani-Hamed:2017fdk}
N.~Arkani-Hamed, P.~Benincasa, and A.~Postnikov, ``{Cosmological Polytopes and
  the Wavefunction of the Universe},''
  \href{http://arxiv.org/abs/1709.02813}{{\ttfamily arXiv:1709.02813
  [hep-th]}}.

\bibitem{Benincasa:2022omn}
P.~Benincasa, ``{Wavefunctionals/S-matrix techniques in de Sitter},''
  \href{http://dx.doi.org/10.22323/1.406.0358}{{\em PoS} {\bfseries CORFU2021}
  (2022) 358}, \href{http://arxiv.org/abs/2203.16378}{{\ttfamily
  arXiv:2203.16378 [hep-th]}}.

\bibitem{Albayrak:2023hie}
S.~Albayrak, P.~Benincasa, and C.~D. Pueyo, ``{Perturbative Unitarity and the
  Wavefunction of the Universe},''
  \href{http://arxiv.org/abs/2305.19686}{{\ttfamily arXiv:2305.19686
  [hep-th]}}.

\bibitem{Benincasa:2018ssx}
P.~Benincasa, ``{From the flat-space S-matrix to the Wavefunction of the
  Universe},'' \href{http://arxiv.org/abs/1811.02515}{{\ttfamily
  arXiv:1811.02515 [hep-th]}}.

\bibitem{Galante:2023uyf}
D.~A. Galante, ``{Modave lectures on de Sitter space \& holography},''
  \href{http://dx.doi.org/10.22323/1.435.0003}{{\em PoS} {\bfseries Modave2022}
  (2023) 003}, \href{http://arxiv.org/abs/2306.10141}{{\ttfamily
  arXiv:2306.10141 [hep-th]}}.

\bibitem{Spradlin:2001pw}
M.~Spradlin, A.~Strominger, and A.~Volovich, ``{Les Houches lectures on de
  Sitter space},'' in {\em {Les Houches Summer School: Session 76: Euro Summer
  School on Unity of Fundamental Physics: Gravity, Gauge Theory and Strings}},
  pp.~423--453.
\newblock 10, 2001.
\newblock \href{http://arxiv.org/abs/hep-th/0110007}{{\ttfamily
  arXiv:hep-th/0110007}}.

\bibitem{Anninos:2012qw}
D.~Anninos, ``{De Sitter Musings},''
  \href{http://dx.doi.org/10.1142/S0217751X1230013X}{{\em Int. J. Mod. Phys. A}
  {\bfseries 27} (2012) 1230013},
  \href{http://arxiv.org/abs/1205.3855}{{\ttfamily arXiv:1205.3855 [hep-th]}}.

\bibitem{Bousso:2002fq}
R.~Bousso, ``{Adventures in de Sitter space},'' in {\em {Workshop on Conference
  on the Future of Theoretical Physics and Cosmology in Honor of Steven
  Hawking's 60th Birthday}}, pp.~539--569.
\newblock 5, 2002.
\newblock \href{http://arxiv.org/abs/hep-th/0205177}{{\ttfamily
  arXiv:hep-th/0205177}}.

\bibitem{Bargmann:1946me}
V.~Bargmann, ``{Irreducible unitary representations of the Lorentz group},''
  \href{http://dx.doi.org/10.2307/1969129}{{\em Annals Math.} {\bfseries 48}
  (1947) 568--640}.

\bibitem{Harish-Chandra.1952}
Harish-Chandra, ``{Plancherel Formula for the 2 × 2 Real Unimodular Group},''
  \href{http://dx.doi.org/10.1073/pnas.38.4.337}{{\em Proceedings of the
  National Academy of Sciences} {\bfseries 38} no.~4, (1952) 337--342}.

\bibitem{Thomas:1941}
L.~H. Thomas, ``On unitary representations of the group of de sitter space,''
  {\em Annals of Mathematics} {\bfseries 42} no.~1, (1941) 113--126.
  \url{http://www.jstor.org/stable/1968990}.

\bibitem{Higuchi:1986py}
A.~Higuchi, ``{Forbidden Mass Range for Spin-2 Field Theory in De Sitter
  Space-time},'' \href{http://dx.doi.org/10.1016/0550-3213(87)90691-2}{{\em
  Nucl. Phys. B} {\bfseries 282} (1987) 397--436}.

\bibitem{Dobrev:1977qv}
V.~K. Dobrev, G.~Mack, V.~B. Petkova, S.~G. Petrova, and I.~T. Todorov,
  \href{http://dx.doi.org/10.1007/BFb0009678}{{\em {Harmonic Analysis on the
  n-Dimensional Lorentz Group and Its Application to Conformal Quantum Field
  Theory}}}, vol.~63.
\newblock Springer, 1977.

\bibitem{Repka1978TensorR}
J.~Repka, ``{Tensor Products of Unitary Representations of SL 2 (R)},''
  \href{http://dx.doi.org/10.2307/2373909}{{\em American Journal of
  Mathematics} {\bfseries 100} no.~4, (8, 1978) 747}.

\bibitem{nachtmann1968dynamische}
O.~Nachtmann, ``Dynamische stabilit{\"a}t im de-sitter-raum.,'' {\em
  Oesterreichische Akademie Wissenschaften Mathematisch naturwissenschaftliche
  Klasse Sitzungsberichte Abteilung} {\bfseries 176} (1968) 363--379.

\bibitem{Kallen:1952zz}
G.~Kallen, ``{On the definition of the Renormalization Constants in Quantum
  Electrodynamics},''
  \href{http://dx.doi.org/10.1007/978-3-319-00627-7_90}{{\em Helv. Phys. Acta}
  {\bfseries 25} no.~4, (1952) 417}.

\bibitem{Lehmann:1954xi}
H.~Lehmann, ``{On the Properties of propagation functions and renormalization
  contants of quantized fields},''
  \href{http://dx.doi.org/10.1007/BF02783624}{{\em Nuovo Cim.} {\bfseries 11}
  (1954) 342--357}.

\bibitem{Hollands:2011we}
S.~Hollands, ``{Massless interacting quantum fields in deSitter spacetime},''
  \href{http://dx.doi.org/10.1007/s00023-011-0140-1}{{\em Annales Henri
  Poincare} {\bfseries 13} (2012) 1039--1081},
  \href{http://arxiv.org/abs/1105.1996}{{\ttfamily arXiv:1105.1996 [gr-qc]}}.

\bibitem{DiPietro:2021sjt}
L.~Di~Pietro, V.~Gorbenko, and S.~Komatsu, ``{Analyticity and Unitarity for
  Cosmological Correlators},''
  \href{http://arxiv.org/abs/2108.01695}{{\ttfamily arXiv:2108.01695
  [hep-th]}}.

\bibitem{Alkalaev:2013fsa}
K.~B. Alkalaev, ``{On higher spin extension of the Jackiw-Teitelboim gravity
  model},'' \href{http://dx.doi.org/10.1088/1751-8113/47/36/365401}{{\em J.
  Phys. A} {\bfseries 47} (2014) 365401},
  \href{http://arxiv.org/abs/1311.5119}{{\ttfamily arXiv:1311.5119 [hep-th]}}.

\bibitem{Alkalaev:2019xuv}
K.~Alkalaev and X.~Bekaert, ``{Towards higher-spin AdS$_2$/CFT$_1$
  holography},'' \href{http://dx.doi.org/10.1007/JHEP04(2020)206}{{\em JHEP}
  {\bfseries 04} (2020) 206}, \href{http://arxiv.org/abs/1911.13212}{{\ttfamily
  arXiv:1911.13212 [hep-th]}}.

\bibitem{Isler:1989hq}
K.~Isler and C.~A. Trugenberger, ``{A Gauge Theory of Two-dimensional Quantum
  Gravity},'' \href{http://dx.doi.org/10.1103/PhysRevLett.63.834}{{\em Phys.
  Rev. Lett.} {\bfseries 63} (1989) 834}.

\bibitem{Chamseddine:1989yz}
A.~H. Chamseddine and D.~Wyler, ``{Gauge Theory of Topological Gravity in
  (1+1)-Dimensions},''
  \href{http://dx.doi.org/10.1016/0370-2693(89)90528-5}{{\em Phys. Lett. B}
  {\bfseries 228} (1989) 75--78}.

\bibitem{Higuchi:1991tk}
A.~Higuchi, ``{Quantum linearization instabilities of de Sitter space-time.
  1},'' \href{http://dx.doi.org/10.1088/0264-9381/8/11/009}{{\em Class. Quant.
  Grav.} {\bfseries 8} (1991) 1961--1981}.

\bibitem{Higuchi:1991tm}
A.~Higuchi, ``{Quantum linearization instabilities of de Sitter space-time.
  2},'' \href{http://dx.doi.org/10.1088/0264-9381/8/11/010}{{\em Class. Quant.
  Grav.} {\bfseries 8} (1991) 1983--2004}.

\bibitem{Marolf.2009yal}
D.~Marolf and I.~A. Morrison, ``{Group averaging for de Sitter free fields},''
  \href{http://dx.doi.org/10.1088/0264-9381/26/23/235003}{{\em Classical and
  Quantum Gravity} {\bfseries 26} no.~23, (2009) 235003},
  \href{http://arxiv.org/abs/0810.5163}{{\ttfamily 0810.5163}}.

\bibitem{Grumiller:2013swa}
D.~Grumiller, M.~Leston, and D.~Vassilevich, ``{Anti-de Sitter holography for
  gravity and higher spin theories in two dimensions},''
  \href{http://dx.doi.org/10.1103/PhysRevD.89.044001}{{\em Phys. Rev. D}
  {\bfseries 89} no.~4, (2014) 044001},
  \href{http://arxiv.org/abs/1311.7413}{{\ttfamily arXiv:1311.7413 [hep-th]}}.

\bibitem{Alkalaev:2020kut}
K.~Alkalaev and X.~Bekaert, ``{On BF-type higher-spin actions in two
  dimensions},'' \href{http://dx.doi.org/10.1007/JHEP05(2020)158}{{\em JHEP}
  {\bfseries 05} (2020) 158}, \href{http://arxiv.org/abs/2002.02387}{{\ttfamily
  arXiv:2002.02387 [hep-th]}}.

\bibitem{Martinec:2014uva}
E.~J. Martinec and W.~E. Moore, ``{Modeling Quantum Gravity Effects in
  Inflation},'' \href{http://dx.doi.org/10.1007/JHEP07(2014)053}{{\em JHEP}
  {\bfseries 07} (2014) 053}, \href{http://arxiv.org/abs/1401.7681}{{\ttfamily
  arXiv:1401.7681 [hep-th]}}.

\bibitem{Bautista:2015wqy}
T.~Bautista and A.~Dabholkar, ``{Quantum Cosmology Near Two Dimensions},''
  \href{http://dx.doi.org/10.1103/PhysRevD.94.044017}{{\em Phys. Rev. D}
  {\bfseries 94} no.~4, (2016) 044017},
  \href{http://arxiv.org/abs/1511.07450}{{\ttfamily arXiv:1511.07450
  [hep-th]}}.

\bibitem{Bautista:2019jau}
T.~Bautista, A.~Dabholkar, and H.~Erbin, ``{Quantum Gravity from Timelike
  Liouville theory},'' \href{http://dx.doi.org/10.1007/JHEP10(2019)284}{{\em
  JHEP} {\bfseries 10} (2019) 284},
  \href{http://arxiv.org/abs/1905.12689}{{\ttfamily arXiv:1905.12689
  [hep-th]}}.

\bibitem{Anninos:2021ene}
D.~Anninos, T.~Bautista, and B.~M\"uhlmann, ``{The two-sphere partition
  function in two-dimensional quantum gravity},''
  \href{http://dx.doi.org/10.1007/JHEP09(2021)116}{{\em JHEP} {\bfseries 09}
  (2021) 116}, \href{http://arxiv.org/abs/2106.01665}{{\ttfamily
  arXiv:2106.01665 [hep-th]}}.

\bibitem{Muhlmann:2021clm}
B.~M\"uhlmann, ``{The two-sphere partition function in two-dimensional quantum
  gravity at fixed area},''
  \href{http://dx.doi.org/10.1007/JHEP09(2021)189}{{\em JHEP} {\bfseries 09}
  no.~189, (2021) 189}, \href{http://arxiv.org/abs/2106.04532}{{\ttfamily
  arXiv:2106.04532 [hep-th]}}.

\bibitem{Muhlmann:2022duj}
B.~M\"uhlmann, ``{The two-sphere partition function from timelike Liouville
  theory at three-loop order},''
  \href{http://dx.doi.org/10.1007/JHEP05(2022)057}{{\em JHEP} {\bfseries 05}
  (2022) 057}, \href{http://arxiv.org/abs/2202.04549}{{\ttfamily
  arXiv:2202.04549 [hep-th]}}.

\bibitem{Maldacena:2019cbz}
J.~Maldacena, G.~J. Turiaci, and Z.~Yang, ``{Two dimensional Nearly de Sitter
  gravity},'' \href{http://dx.doi.org/10.1007/JHEP01(2021)139}{{\em JHEP}
  {\bfseries 01} (2021) 139}, \href{http://arxiv.org/abs/1904.01911}{{\ttfamily
  arXiv:1904.01911 [hep-th]}}.

\bibitem{Cotler:2019nbi}
J.~Cotler, K.~Jensen, and A.~Maloney, ``{Low-dimensional de Sitter quantum
  gravity},'' \href{http://dx.doi.org/10.1007/JHEP06(2020)048}{{\em JHEP}
  {\bfseries 06} (2020) 048}, \href{http://arxiv.org/abs/1905.03780}{{\ttfamily
  arXiv:1905.03780 [hep-th]}}.

\bibitem{Anninos:2017hhn}
D.~Anninos and D.~M. Hofman, ``{Infrared Realization of dS$_2$ in AdS$_2$},''
  \href{http://dx.doi.org/10.1088/1361-6382/aab143}{{\em Class. Quant. Grav.}
  {\bfseries 35} no.~8, (2018) 085003},
  \href{http://arxiv.org/abs/1703.04622}{{\ttfamily arXiv:1703.04622
  [hep-th]}}.

\bibitem{Anninos:2018svg}
D.~Anninos, D.~A. Galante, and D.~M. Hofman, ``{De Sitter horizons \&
  holographic liquids},'' \href{http://dx.doi.org/10.1007/JHEP07(2019)038}{{\em
  JHEP} {\bfseries 07} (2019) 038},
  \href{http://arxiv.org/abs/1811.08153}{{\ttfamily arXiv:1811.08153
  [hep-th]}}.

\bibitem{Schlingemann:1999mk}
D.~Schlingemann, ``{Euclidean field theory on a sphere},''
  \href{http://arxiv.org/abs/hep-th/9912235}{{\ttfamily arXiv:hep-th/9912235}}.

\bibitem{Anninos:2020hfj}
D.~Anninos, F.~Denef, Y.~T.~A. Law, and Z.~Sun, ``{Quantum de Sitter horizon
  entropy from quasicanonical bulk, edge, sphere and topological string
  partition functions},'' \href{http://dx.doi.org/10.1007/JHEP01(2022)088}{{\em
  JHEP} {\bfseries 01} (2022) 088},
  \href{http://arxiv.org/abs/2009.12464}{{\ttfamily arXiv:2009.12464
  [hep-th]}}.

\bibitem{Mottola:1984ar}
E.~Mottola, ``{Particle Creation in de Sitter Space},''
  \href{http://dx.doi.org/10.1103/PhysRevD.31.754}{{\em Phys. Rev. D}
  {\bfseries 31} (1985) 754}.

\bibitem{Bousso:2001mw}
R.~Bousso, A.~Maloney, and A.~Strominger, ``{Conformal vacua and entropy in de
  Sitter space},'' \href{http://dx.doi.org/10.1103/PhysRevD.65.104039}{{\em
  Phys. Rev. D} {\bfseries 65} (2002) 104039},
  \href{http://arxiv.org/abs/hep-th/0112218}{{\ttfamily arXiv:hep-th/0112218}}.

\bibitem{Guijosa:2005qi}
A.~Guijosa, D.~A. Lowe, and J.~Murugan, ``{A Prototype for dS/CFT},''
  \href{http://dx.doi.org/10.1103/PhysRevD.72.046001}{{\em Phys. Rev. D}
  {\bfseries 72} (2005) 046001},
  \href{http://arxiv.org/abs/hep-th/0505145}{{\ttfamily arXiv:hep-th/0505145}}.

\bibitem{Folacci:1992xc}
A.~Folacci, ``{BRST quantization of the massless minimally coupled scalar field
  in de Sitter space: Zero modes, euclideanization and quantization},''
  \href{http://dx.doi.org/10.1103/PhysRevD.46.2553}{{\em Phys. Rev. D}
  {\bfseries 46} (1992) 2553--2559},
  \href{http://arxiv.org/abs/0911.2064}{{\ttfamily arXiv:0911.2064 [gr-qc]}}.

\bibitem{Allen:1987tz}
B.~Allen and A.~Folacci, ``{The Massless Minimally Coupled Scalar Field in De
  Sitter Space},'' \href{http://dx.doi.org/10.1103/PhysRevD.35.3771}{{\em Phys.
  Rev. D} {\bfseries 35} (1987) 3771}.

\bibitem{Bonifacio:2018zex}
J.~Bonifacio, K.~Hinterbichler, A.~Joyce, and R.~A. Rosen, ``{Shift Symmetries
  in (Anti) de Sitter Space},''
  \href{http://dx.doi.org/10.1007/JHEP02(2019)178}{{\em JHEP} {\bfseries 02}
  (2019) 178}, \href{http://arxiv.org/abs/1812.08167}{{\ttfamily
  arXiv:1812.08167 [hep-th]}}.

\bibitem{Ford:1984hs}
L.~H. Ford, ``{Quantum Instability of De Sitter Space-time},''
  \href{http://dx.doi.org/10.1103/PhysRevD.31.710}{{\em Phys. Rev. D}
  {\bfseries 31} (1985) 710}.

\bibitem{Allen:1985ux}
B.~Allen, ``{Vacuum States in de Sitter Space},''
  \href{http://dx.doi.org/10.1103/PhysRevD.32.3136}{{\em Phys. Rev. D}
  {\bfseries 32} (1985) 3136}.

\bibitem{DiFrancesco:1997nk}
P.~Di~Francesco, P.~Mathieu, and D.~Senechal,
  \href{http://dx.doi.org/10.1007/978-1-4612-2256-9}{{\em {Conformal Field
  Theory}}}.
\newblock Graduate Texts in Contemporary Physics. Springer-Verlag, New York,
  1997.

\bibitem{Folacci:1996dv}
A.~Folacci, ``{Toy model for the zero mode problem in the conformal sector of
  de Sitter quantum gravity},''
  \href{http://dx.doi.org/10.1103/PhysRevD.53.3108}{{\em Phys. Rev. D}
  {\bfseries 53} (1996) 3108--3117}.

\bibitem{Hinterbichler:2016fgl}
K.~Hinterbichler and A.~Joyce, ``{Manifest Duality for Partially Massless
  Higher Spins},'' \href{http://dx.doi.org/10.1007/JHEP09(2016)141}{{\em JHEP}
  {\bfseries 09} (2016) 141}, \href{http://arxiv.org/abs/1608.04385}{{\ttfamily
  arXiv:1608.04385 [hep-th]}}.

\bibitem{Brust:2016zns}
C.~Brust and K.~Hinterbichler, ``{Partially Massless Higher-Spin Theory},''
  \href{http://dx.doi.org/10.1007/JHEP02(2017)086}{{\em JHEP} {\bfseries 02}
  (2017) 086}, \href{http://arxiv.org/abs/1610.08510}{{\ttfamily
  arXiv:1610.08510 [hep-th]}}.

\bibitem{Pethybridge:2021rwf}
B.~Pethybridge and V.~Schaub, ``{Tensors and Spinors in de Sitter Space},''
  \href{http://arxiv.org/abs/2111.14899}{{\ttfamily arXiv:2111.14899
  [hep-th]}}.

\bibitem{Schaub:2023scu}
V.~Schaub, ``{Spinors in (Anti-)de Sitter Space},''
  \href{http://arxiv.org/abs/2302.08535}{{\ttfamily arXiv:2302.08535
  [hep-th]}}.

\bibitem{Kitaev:2005dm}
A.~Kitaev and J.~Preskill, ``{Topological entanglement entropy},''
  \href{http://dx.doi.org/10.1103/PhysRevLett.96.110404}{{\em Phys. Rev. Lett.}
  {\bfseries 96} (2006) 110404},
  \href{http://arxiv.org/abs/hep-th/0510092}{{\ttfamily arXiv:hep-th/0510092}}.

\bibitem{Levin:2006zz}
M.~Levin and X.-G. Wen, ``{Detecting Topological Order in a Ground State Wave
  Function},'' \href{http://dx.doi.org/10.1103/PhysRevLett.96.110405}{{\em
  Phys. Rev. Lett.} {\bfseries 96} (2006) 110405},
  \href{http://arxiv.org/abs/cond-mat/0510613}{{\ttfamily
  arXiv:cond-mat/0510613}}.

\bibitem{Henneaux:1985nw}
M.~Henneaux, ``{QUANTUM GRAVITY IN TWO-DIMENSIONS: EXACT SOLUTION OF THE JACKIW
  MODEL},'' \href{http://dx.doi.org/10.1103/PhysRevLett.54.959}{{\em Phys. Rev.
  Lett.} {\bfseries 54} (1985) 959--962}.

\bibitem{Letsios:2020twa}
V.~A. Letsios, ``{The eigenmodes for spinor quantum field theory in global de
  Sitter space\textendash{}time},''
  \href{http://dx.doi.org/10.1063/5.0038651}{{\em J. Math. Phys.} {\bfseries
  62} no.~3, (2021) 032303}, \href{http://arxiv.org/abs/2011.07875}{{\ttfamily
  arXiv:2011.07875 [gr-qc]}}.

\bibitem{Vasiliev:1999ba}
M.~A. Vasiliev, ``{Higher spin gauge theories: Star product and AdS space},''
  \href{http://arxiv.org/abs/hep-th/9910096}{{\ttfamily arXiv:hep-th/9910096}}.

\bibitem{Seiberg:Fun}
N.~Seiberg, ``Fun with free field theory,'' 2015.
\newblock \url{https://www.youtube.com/watch?v=pqgNrVTQ4yM}.

\bibitem{Iliesiu:2019xuh}
L.~V. Iliesiu, S.~S. Pufu, H.~Verlinde, and Y.~Wang, ``{An exact quantization
  of Jackiw-Teitelboim gravity},''
  \href{http://dx.doi.org/10.1007/JHEP11(2019)091}{{\em JHEP} {\bfseries 11}
  (2019) 091}, \href{http://arxiv.org/abs/1905.02726}{{\ttfamily
  arXiv:1905.02726 [hep-th]}}.

\bibitem{Maldacena:2016upp}
J.~Maldacena, D.~Stanford, and Z.~Yang, ``{Conformal symmetry and its breaking
  in two dimensional Nearly Anti-de-Sitter space},''
  \href{http://dx.doi.org/10.1093/ptep/ptw124}{{\em PTEP} {\bfseries 2016}
  no.~12, (2016) 12C104}, \href{http://arxiv.org/abs/1606.01857}{{\ttfamily
  arXiv:1606.01857 [hep-th]}}.

\bibitem{Kitaev:2017awl}
A.~Kitaev and S.~J. Suh, ``{The soft mode in the Sachdev-Ye-Kitaev model and
  its gravity dual},'' \href{http://dx.doi.org/10.1007/JHEP05(2018)183}{{\em
  JHEP} {\bfseries 05} (2018) 183},
  \href{http://arxiv.org/abs/1711.08467}{{\ttfamily arXiv:1711.08467
  [hep-th]}}.

\bibitem{Iliesiu:2020zld}
L.~V. Iliesiu, J.~Kruthoff, G.~J. Turiaci, and H.~Verlinde, ``{JT gravity at
  finite cutoff},'' \href{http://dx.doi.org/10.21468/SciPostPhys.9.2.023}{{\em
  SciPost Phys.} {\bfseries 9} (2020) 023},
  \href{http://arxiv.org/abs/2004.07242}{{\ttfamily arXiv:2004.07242
  [hep-th]}}.

\bibitem{Moncrief:1978te}
V.~Moncrief, ``{Invariant States and Quantized Gravitational Perturbations},''
  \href{http://dx.doi.org/10.1103/PhysRevD.18.983}{{\em Phys. Rev. D}
  {\bfseries 18} (1978) 983--989}.

\bibitem{Moncrief:1979bg}
V.~Moncrief, ``{QUANTUM LINEARIZATION INSTABILITIES},''
  \href{http://dx.doi.org/10.1007/BF00756792}{{\em Gen. Rel. Grav.} {\bfseries
  10} (1979) 93--97}.

\bibitem{Chandrasekaran:2022cip}
V.~Chandrasekaran, R.~Longo, G.~Penington, and E.~Witten, ``{An algebra of
  observables for de Sitter space},''
  \href{http://dx.doi.org/10.1007/JHEP02(2023)082}{{\em JHEP} {\bfseries 02}
  (2023) 082}, \href{http://arxiv.org/abs/2206.10780}{{\ttfamily
  arXiv:2206.10780 [hep-th]}}.

\bibitem{Chakraborty:2023yed}
T.~Chakraborty, J.~Chakravarty, V.~Godet, P.~Paul, and S.~Raju, ``{The Hilbert
  space of de Sitter quantum gravity},''
  \href{http://arxiv.org/abs/2303.16315}{{\ttfamily arXiv:2303.16315
  [hep-th]}}.

\bibitem{Friedan:2012hi}
D.~Friedan and A.~Konechny, ``{Curvature formula for the space of 2-d conformal
  field theories},'' \href{http://dx.doi.org/10.1007/JHEP09(2012)113}{{\em
  JHEP} {\bfseries 09} (2012) 113},
  \href{http://arxiv.org/abs/1206.1749}{{\ttfamily arXiv:1206.1749 [hep-th]}}.

\bibitem{Liu:1987nz}
J.~Liu and J.~Polchinski, ``{Renormalization of the Mobius Volume},''
  \href{http://dx.doi.org/10.1016/0370-2693(88)91566-3}{{\em Phys. Lett. B}
  {\bfseries 203} (1988) 39--43}.

\bibitem{Maldacena:2016hyu}
J.~Maldacena and D.~Stanford, ``{Remarks on the Sachdev-Ye-Kitaev model},''
  \href{http://dx.doi.org/10.1103/PhysRevD.94.106002}{{\em Phys. Rev. D}
  {\bfseries 94} no.~10, (2016) 106002},
  \href{http://arxiv.org/abs/1604.07818}{{\ttfamily arXiv:1604.07818
  [hep-th]}}.

\bibitem{Anninos:2016szt}
D.~Anninos, T.~Anous, and F.~Denef, ``{Disordered Quivers and Cold Horizons},''
  \href{http://dx.doi.org/10.1007/JHEP12(2016)071}{{\em JHEP} {\bfseries 12}
  (2016) 071}, \href{http://arxiv.org/abs/1603.00453}{{\ttfamily
  arXiv:1603.00453 [hep-th]}}.

\bibitem{Anninos:2017cnw}
D.~Anninos, T.~Anous, and R.~T. D'Agnolo, ``{Marginal deformations
  \textbackslash{}\& rotating horizons},''
  \href{http://dx.doi.org/10.1007/JHEP12(2017)095}{{\em JHEP} {\bfseries 12}
  (2017) 095}, \href{http://arxiv.org/abs/1707.03380}{{\ttfamily
  arXiv:1707.03380 [hep-th]}}.

\bibitem{Yoon:2017nig}
J.~Yoon, ``{SYK Models and SYK-like Tensor Models with Global Symmetry},''
  \href{http://dx.doi.org/10.1007/JHEP10(2017)183}{{\em JHEP} {\bfseries 10}
  (2017) 183}, \href{http://arxiv.org/abs/1707.01740}{{\ttfamily
  arXiv:1707.01740 [hep-th]}}.

\bibitem{Gonzalez:2018enk}
H.~A. Gonz\'alez, D.~Grumiller, and J.~Salzer, ``{Towards a bulk description of
  higher spin SYK},'' \href{http://dx.doi.org/10.1007/JHEP05(2018)083}{{\em
  JHEP} {\bfseries 05} (2018) 083},
  \href{http://arxiv.org/abs/1802.01562}{{\ttfamily arXiv:1802.01562
  [hep-th]}}.

\bibitem{Holstein:1940zp}
T.~Holstein and H.~Primakoff, ``{Field dependence of the intrinsic domain
  magnetization of a ferromagnet},''
  \href{http://dx.doi.org/10.1103/PhysRev.58.1098}{{\em Phys. Rev.} {\bfseries
  58} (1940) 1098--1113}.

\bibitem{Fischler:2000}
W.~Fischler, ``Taking de sitter seriously. talk given at role of scaling laws
  in physics and biology (celebrating the 60th birthday of geoffrey west).''
  Unpublished, 2000.

\bibitem{Banks:2006rx}
T.~Banks, B.~Fiol, and A.~Morisse, ``{Towards a quantum theory of de Sitter
  space},'' \href{http://dx.doi.org/10.1088/1126-6708/2006/12/004}{{\em JHEP}
  {\bfseries 12} (2006) 004},
  \href{http://arxiv.org/abs/hep-th/0609062}{{\ttfamily arXiv:hep-th/0609062}}.

\bibitem{Bousso:2000nf}
R.~Bousso, ``{Positive vacuum energy and the N bound},''
  \href{http://dx.doi.org/10.1088/1126-6708/2000/11/038}{{\em JHEP} {\bfseries
  11} (2000) 038}, \href{http://arxiv.org/abs/hep-th/0010252}{{\ttfamily
  arXiv:hep-th/0010252}}.

\bibitem{Parikh:2004wh}
M.~K. Parikh and E.~P. Verlinde, ``{De Sitter holography with a finite number
  of states},'' \href{http://dx.doi.org/10.1088/1126-6708/2005/01/054}{{\em
  JHEP} {\bfseries 01} (2005) 054},
  \href{http://arxiv.org/abs/hep-th/0410227}{{\ttfamily arXiv:hep-th/0410227}}.

\bibitem{Kitaev:2017hnr}
A.~Kitaev, ``{Notes on $\widetilde{\mathrm{SL}}(2,\mathbb{R})$
  representations},'' \href{http://arxiv.org/abs/1711.08169}{{\ttfamily
  arXiv:1711.08169 [hep-th]}}.

\bibitem{bielski2013orthogonality}
S.~Bielski, ``Orthogonality relations for the associated legendre functions of
  imaginary order,'' {\em Integral transforms and special functions} {\bfseries
  24} no.~4, (2013) 331--337.

\end{thebibliography}\endgroup

\end{document}